\documentclass[reqno,a4paper,11pt]{article}
\pdfoutput=1

\usepackage{ogonek}
\usepackage{xcolor}
\usepackage{graphicx}
\usepackage[textwidth = 430 pt, textheight = 630 pt]{geometry}

\usepackage{epsfig,rotating}
\usepackage{amsmath,amssymb}
\usepackage{amsfonts}
\usepackage{mathrsfs}
\usepackage{bbm}
\usepackage[normalem]{ulem}

\usepackage{latexsym}
\usepackage{amsthm}
\usepackage{thmtools}
\usepackage[all,knot]{xy}
\xyoption{arc}

\usepackage[utf8x]{inputenc}
\usepackage{orcidlink}

\definecolor{MyDarkBlue}{rgb}{0.15,0.25,0.45}

\usepackage{hyperref}
\hypersetup{
	hypertexnames=false,
	colorlinks=true,
	citecolor=MyDarkBlue,
	linkcolor=MyDarkBlue,
	urlcolor=MyDarkBlue,
	pdfauthor={Hyungrok Kim,Christian S\"amann},
	pdftitle={},
	pdfsubject={hep-th math-ph},
	breaklinks=true
}
\usepackage[nameinlink,noabbrev]{cleveref}

\usepackage{tikz}
\usetikzlibrary{matrix,cd,arrows}
\usepackage{mathtools}
\usepackage[all,knot]{xy}
\xyoption{arc}

\usepackage[toc,page]{appendix}

\usepackage[vcentermath]{youngtab}

\let\fn\footnote
\renewcommand{\footnote}[1]{\linespread{1.1}\fn{#1}\linespread{1.29}}

\declaretheorem[numberwithin=section,name=Theorem,refname={theorem,theorems},Refname={Theorem,Theorems}]{theorem}
\declaretheorem[sibling=theorem,name=Lemma,refname={lemma,lemmas},Refname={Lemma,Lemmas}]{lemma}
\declaretheorem[sibling=theorem,name=Definition,refname={definition,definitions},Refname={Definition,Definitions}]{definition}
\declaretheorem[sibling=theorem,name=Definition/Theorem,refname={definition,definitions},Refname={Definition,Definitions}]{definitiontheorem}
\declaretheorem[sibling=theorem,name=Proposition,refname={proposition,propositions},Refname={Proposition,Propositions}]{proposition}

\declaretheorem[sibling=theorem,name=Remark,refname={remark,remarks},Refname={Remark,Remarks}]{remark}

\makeatletter
\renewcommand{\section}{\@startsection
	{section}{1}{\z@}{-3.5ex plus -1ex minus
		-.2ex}{2.3ex plus .2ex}{\mathversion{bold}\bf }}
\renewcommand{\subsection}{\@startsection{subsection}{2}{\z@}{-3.25ex
		plus -1ex minus
		-.2ex}{1.5ex plus .2ex}{\mathversion{bold}\bf }}
\renewcommand{\subsubsection}{\@startsection{subsubsection}{3}{-2.45ex}{-3.25ex
		plus -1ex minus -.2ex}{1.5ex plus .2ex}{\it }}
\renewcommand\paragraph{\@startsection{paragraph}{4}{\z@}%
	{3.25ex \@plus1ex \@minus.2ex}%
	{-1em}%
	{\normalfont\normalsize\bfseries\mathversion{bold}}}

\renewcommand{\thesection}{\arabic{section}}
\renewcommand{\thesubsection}{\arabic{section}.\arabic{subsection}}
\renewcommand{\@seccntformat}[1]{\@nameuse{the#1}.~~}

\renewcommand{\theequation}{\thesection.\arabic{equation}}
\makeatletter \@addtoreset{equation}{section}
\def\Ddots{\mathinner{\mkern1mu\raise\p@
		\vbox{\kern7\p@\hbox{.}}\mkern2mu
		\raise4\p@\hbox{.}\mkern2mu\raise7\p@\hbox{.}\mkern1mu}}
\renewcommand{\appendices}{
	\section*{Appendix}\label{appendices}\setcounter{subsection}{0}
	\addcontentsline{toc}{section}{Appendix}
	\setcounter{equation}{0}
	\renewcommand{\theequation}{\Alph{subsection}.\arabic{equation}}
	\renewcommand{\thesubsection}{\Alph{subsection}}
	\@addtoreset{equation}{subsection}
	\crefalias{subsection}{appendix}
}

\makeatother

\newcommand{\makecommand}[3]{%
	\foreach \i in #3 {%
		\expandafter\xdef\csname #1\i\endcsname{\noexpand#2{\unexpanded\expandafter{\i}}}%
	}%
}

\makecommand{fr}{\mathfrak}{{c,der,g,h,A,X,gl,sl,o,so,osp,u,su,sp,spin,string,Poisson,td,tb}}

\makecommand{sf}{\mathsf}{{id,String,Lie,Hom,SU,SO,Cat,Set,Spin,Pin,inv,CE,Diff,HSymp,TD,TB,pr,Aut,GL,SL,IO,Mat,Sym,Inn,Out,AUT,hom}}

\makecommand{rm}{\mathrm}{{Ham,diag,im,deg,id,Ad}}

\newcommand{\latinalphabet}{A,a,B,b,C,c,d,D,E,e,F,f,G,g,H,h,I,i,J,j,K,k,L,l,M,m,N,n,O,o,P,p,Q,q,R,r,S,s,T,t,U,u,V,v,W,w,X,x,Y,y,Z,z}
\newcommand{\caplatinalphabet}{A,B,C,D,E,F,G,H,I,J,K,L,M,N,O,P,Q,R,S,T,U,V,W,X,Y,Z}
\makecommand{I}{\mathbbm}{\latinalphabet}
\makecommand{bf}{\mathbf}{\latinalphabet}
\makecommand{bm}{\bm}{\latinalphabet}
\makecommand{ca}{\mathcal}{\caplatinalphabet}
\makecommand{fr}{\mathfrak}{\latinalphabet}
\makecommand{rm}{\mathrm}{\latinalphabet}
\makecommand{sf}{\mathsf}{\latinalphabet}
\makecommand{sc}{\mathscr}{\latinalphabet}

\newcommand{\sfGO}{\sfG\sfO}
\newcommand{\scTD}{\scT\!\scD}

\newcommand{\scGO}{\scG\!\scO}
\newcommand{\acton}{\triangleright}

\def\slasha#1{\setbox0=\hbox{$#1$}#1\hskip-\wd0\hbox to\wd0{\hss\sl/\/\hss}}

\def\periodb#1{\setbox0=\hbox{$#1$}#1\hskip-\wd0\hbox to\wd0{-}}

\newcommand{\unit}{\mathbbm{1}}   			

\newcommand{\CatCat}{\mathsf{Cat}}

\newcommand{\comment}[1]{}     				
\def\tyng(#1){\raisebox{0.05cm}{\hbox{\tiny$\yng(#1)$}}}			
\def\tyoung(#1){\hbox{\tiny$\young(#1)$}}			

\newcommand{\eand}{~~~\mbox{and}~~~}

\newcommand{\beq}{\begin{eqnarray}}
	\newcommand{\eeq}{\end{eqnarray}}

\newcommand\tildeotimes{\mathbin{\tilde\otimes}}
\newcommand\sslash{\mathbin{/\!/}}

\let\oldref\ref
\AtBeginDocument{\renewcommand{\ref}[1]{\cref{#1}}}
\AtBeginDocument{\renewcommand{\eqref}[1]{{\rm (\oldref{#1})}}}
\tikzset{Rightarrow/.style={double equal sign distance,>={Implies},->},
	triple/.style={-,preaction={draw,Rightarrow}},
	quadruple/.style={preaction={draw,Rightarrow,shorten >=0pt},shorten >=1pt,-,double,double
		distance=0.2pt}}

\usepackage{stmaryrd}

\begin{document}
    \begin{titlepage}
        \begin{flushright}
            EMPG--22--05
        \end{flushright}
        \vskip0.6cm
        \begin{center}
            {\LARGE \bf
                Non-Geometric T-Duality as\\[0.5cm]
                Higher Groupoid Bundles with Connections
            }
            \vskip0.5cm
            {\Large Hyungrok Kim${}^1$ and Christian Saemann${}^2$}
            \setcounter{footnote}{0}
            \renewcommand{\thefootnote}{\arabic{thefootnote}}
            \vskip0.5cm
            {\em 
                ${}^1$Centre for Theoretical Physics Research\\
                Department of Physics, Astronomy, and Mathematics\\
                University of Hertfordshire
                
                \
                
                ${}^2$Maxwell Institute for Mathematical Sciences\\
                Department of Mathematics, Heriot--Watt University\\
                Colin Maclaurin Building, Riccarton, Edinburgh EH14 4AS, U.K.}\\[0.5cm]
            {Email: {\ttfamily h.kim2@herts.ac.uk~,~c.saemann@hw.ac.uk}}
        \end{center}
        \vskip0.4cm
        \begin{center}
            {\bf Abstract}
        \end{center}
        \begin{quote}
            We propose a description of T-duality between general geometric and non-geometric backgrounds as higher groupoid bundles with connections. Our description extends the previous observation by Nikolaus and Waldorf that the topological aspects of geometric and half-geometric T-dualities can be described in terms of higher geometry. We extend their construction in two ways. First, we endow the higher geometries with adjusted connections, which allow us to discuss explicit formulas for the metric and the Kalb--Ramond field of a T-background. Second, we extend the principal 2-bundles to principal 2-groupoid bundles, which accommodate the scalar fields arising in T-dualities along two directions as well as $Q$-fluxes. Our proposals reproduce key examples from the literature. They are manifestly covariant under the full T-duality group $\sfGO(n,n;\IZ)$ and have interesting physical and mathematical implications. Eventually, we also comment on the case of T-duality in the presence of scalar fluxes.
        \end{quote}
                
    \end{titlepage}
    
    \tableofcontents
    
    \newpage
    
    \section{Introduction and results}
    
    \subsection*{Background}
    
    T-duality~\cite{Giveon:1994fu} is a crucial feature of string theory which sets it apart from field theories of point particles. In its simplest form, T-duality relates two string theories whose target spaces are of the form $X\times S^1$ for some Lorentzian manifold $X$, interchanging the momentum modes and the winding modes along the circle $S^1$. More complicated is the case in which a T-duality involves a non-trivial circle bundle that carries in addition a Kalb--Ramond two-form field $B$ describing a topologically non-trivial gerbe. Here, a T-duality can link target spaces with different topologies~\cite{Giveon:1993ph,Bouwknegt:2003vb}. Even more generally, we can consider T-dualities along several circle directions. Such T-dualities can link fully geometric target spaces to non-geometric target spaces. A class of mildly non-geometric target spaces is known as T-folds~\cite{Hull:2004in}. These are still locally geometric, but their local data are glued together by an element in the T-duality group $\sfO(n,n;\IZ)$. T-dualities can, however, also produce $R$-spaces for which there is not even a local geometric description~\cite{Shelton:2005cf,Shelton:2006fd} (see~\cite{Plauschinn:2018wbo} for a review). It is clear that a complete understanding of string theory requires an understanding of these non-geometric backgrounds. Given that T-duality is a duality between topologically non-trivial target spaces, it is particularly important not to work merely locally, as done in much of the literature. One of the aims of this paper is to provide a proposal for a clean mathematical description of some of these non-geometric backgrounds arising in the context of non-trivial topologies.
    
    By now, T-duality has attracted considerable mathematical interest due to its relation to a number of important mathematical constructions such as mirror symmetry and the Fourier--Mukai transform. The observation that T-duality can change the topology of the target space was linked, in a formalism usually called topological T-duality, to the existence of a Gysin sequence~\cite{Bouwknegt:2003vb,Bouwknegt:2003zg}. The latter provides an explicit relation between different topological classes, e.g.~between the first Chern class of a torus fibration and the Dixmier--Douady class of a gerbe on its total space. Subsequent works have found interpretations of non-geometric backgrounds in terms of non-commutative~\cite{Mathai:2004qq} and non-associative geometries~\cite{Bouwknegt:2004ap}. 
    
    A precise mathematical formalization of T-duality called T-correspondences was given in~\cite{Bunke:2005sn,Bunke:2005um}. Here, a T-background is defined as a torus bundle $P$ over a base manifold $X$ together with the Dixmier--Douady class $H$ of an abelian gerbe over $P$. A T-duality between two such pairs $(\check P, \check H)$ and $(\hat P,\hat H)$ is then formulated as a relation between the pullbacks of $\check H$ and $\hat H$ to the correspondence space $\check P\times_X\hat P$; the data $(\check P, \check H)$, $(\hat P,\hat H)$, and the relation are collectively called a T-duality triple. More recently, it was observed that T-backgrounds and indeed full T-duality triples can be represented by 2-stacks~\cite{Nikolaus:2018qop}. That is, a geometric T-background can be equivalently seen as a principal 2-bundle or gerbe with a particular structure 2-group $\sfTB^\text{F2}_n$ (see \ref{def:TBF2}) that encodes both the torus directions as well as the gerbe data. The same holds for a T-duality triple between geometric T-backgrounds, where the structure 2-group is denoted by $\sfTD_n$ (see \ref{def:TDn}). This structure group comes with a natural projection to $\sfTB^\text{F2}_n$ as well as an action of the T-duality group $\sfO(n,n;\IZ)$. The latter allows for constructing a second projection, consisting of the composition of the flip transformation in $\sfO(n,n;\IZ)$ and the projection. Together with the original projection, these maps induce on principal 2-bundles the data of a T-duality triple. Interestingly, even half-geometric T-dualities, which link geometric backgrounds with T-folds, can be captured in terms of principal 2-bundles. This opens up the exciting possibility that non-commutative and perhaps even non-associative geometries can be resolved into ordinary but higher geometrical structures, which would clearly constitute a simplification: higher geometry enters the description of geometric backgrounds anyway in the form of gerbes, and higher geometric objects are more readily derived than their non-commutative and, in particular, non-associative counterparts.
    
    The appealing picture obtained in~\cite{Nikolaus:2018qop} poses four evident questions:
    \begin{enumerate}
        \item[(i)] Can we extend the topological constructions to a more complete picture by adding a differential refinement in the form of connections?
        \item[(ii)] Can we extend the half-geometric correspondences of~\cite{Nikolaus:2018qop} further to the more general cases of T-dualities between T-folds?
        \item[(iii)] Can this be further extended to $R$-spaces?
        \item[(iv)] Is there a fully $\sfO(n,n;\IZ)$-covariant formulation that manifests the action of the T-duality group on the components of this description?
    \end{enumerate}
    In this paper, we answer questions (i), (ii), and (iv) affirmatively by proposing a concrete, differentially refined, and $\sfO(n,n;\IZ)$-covariant description of T-duality between T-folds. The answer to question (iii) is only partially positive: we can capture a simple situation, namely T-duality with scalar fluxes, but not the most general T-dualities between $R$-spaces.
    
    \subsection*{Results}
    
    \paragraph{Result 1: generalization of $\sfTD_n$ to $\sfTD_n^\ltimes$.} The principal torus bundles captured by $\sfTD_n$ are not sufficient for general abelian T-duality, which requires a generalization to affine torus bundles. We present a natural generalized 2-group $\sfTD_n^\ltimes$ that achieves just this for affine torus bundles in the sense of~\cite{Baraglia:1105.0290}.
    
    \paragraph{Result 2: explicit T-duality 2-group and action on $\sfTD_n$.} We formulate an explicit and useful form $\scGO(n,n;\IZ)$ of the automorphism 2-group of $\sfTD_n$ and give explicit expressions for the data of an action of this 2-group on $\sfTD_n$. This allows us to render all our constructions explicitly $\scGO(n,n;\IZ)$-covariant, i.e.~all constructions carry manifest and expected actions of the T-duality group.
    
    \paragraph{Result 3: differential refinement of principal $\sfTD_n$-bundles.} A differential refinement of the description of geometric T-duality triples in terms of principal 2-bundles should be straightforward by abstract nonsense. Unfortunately, this is complicated by the fact that the connections on principal 2-bundles as conventionally defined in the literature are either too general, see e.g.~\cite{Breen:math0106083} and~\cite{Aschieri:2003mw}, or too restrictive because they imply a particular flatness condition on the curvature, see e.g.~\cite{Baez:0511710}.  Instead, one has to work with adjusted connections as developed in~\cite{Saemann:2019dsl,Kim:2019owc,Borsten:2021ljb,Rist:2022hci}, see also~\cite{Sati:2008eg} for an earlier treatment of a special case (and further references to the literature in \ref{app:higher_principal_bundles}). In particular, the finite form of differentially refined, adjusted cocycles was only identified very recently~\cite{Rist:2022hci}. Using this technology, we construct the relevant adjustment and endow the principal 2-bundles describing geometric T-duality correspondences with adjusted connections. Recently, it was shown that our differential refinement reproduces the Buscher rules locally~\cite{Waldorf:2022tib}.
    
    \paragraph{Result 4: Proposal for an extension to the half-geometric case.} A differential refinement in the half-geometric case, however, requires more work. Recall that T-duality can be interpreted as a Kaluza--Klein reduction of the correspondence space, cf.~\cite{Berman:2019biz,Alfonsi:2019ggg,Alfonsi:2020nxu,Alfonsi:2021ymc}. A Kalb--Ramond $B$-field on the correspondence space will thus give rise to scalar fields on the base space $X$ after T-duality, or dimensional reduction along two directions. These scalar fields then take values in the Narain moduli space~\cite{Narain:1985jj}, which is a coset space by the T-duality group. The physical requirement that there must be scalar degrees of freedom demands that the principal 2-bundles used in~\cite{Nikolaus:2018qop} need to be extended to principal 2-groupoid bundles with structure 2-groupoid $\scTD_n$ given by the 2-group $\sfTD_n$ fibered over the Narain moduli space $M_n$. Our proposal reproduces the half-geometric T-dualities of~\cite{Nikolaus:2018qop}.

    \paragraph{Result 5: Key example.} For all cases, we work out the details of a key example familiar from the string theory literature: the three T-duals for the 3-torus $T^3$ carrying a gerbe with non-trivial Dixmier--Douady class or ``$H$-flux.''
    
    \paragraph{Result 6: Extension to T-duality with scalar fluxes.}
    While our construction cannot capture the most general case of $R$-spaces, there is an evident continuation of our picture to T-duality with scalar fluxes: the requirement for $0$-form curvatures corresponding to (non-existing) $(-1)$-forms suggests augmenting the 2-groupoid $\scTD_n$ in the simplicial sense to an augmented 2-quasi-groupoid $\scTD^\text{aug}_n$. The relevant space of $(-1)$-simplices is then identified from the observation that $R$-fluxes are related to particular embedding tensors in gauged supergravity~\cite{Aldazabal:2011nj,Geissbuhler:2011mx,Grana:2012rr}, and we show that these moduli match the expectations from the literature. Again, the relevant cocycles can be written down, and we obtain a description of T-duality correspondences involving scalar fluxes.\footnote{The slides~\cite{Waldorf:2019aa} mention a possible extension of the framework of~\cite{Waldorf:2022lzs} to $R$-spaces by using Lie 2-groupoids, but they do not give details; in particular, augmentation, which we see as necessary for a proper description of $R$-fluxes, is not mentioned.}
    
    \
    
    We note that the principal $\scTD^\text{aug}_n$-bundles naturally contain the principal $\scTD_n$-bundles describing T-dualities with T-folds, which, in turn, contain the principal $\sfTD_n$-bundles describing geometric T-dualities. Moreover, all constructions are manifestly $\scGO(n,n;\IZ)$-covariant: the action of our T-duality 2-group $\scGO(n,n;\IZ)$ is always explicit. In this sense, our approach is similar in spirit to double field theory.
    
    Our constructions only capture fields that have trivial dependence on the T-duality directions. This is simply due to the fact that we interpret T-duality as a Kaluza--Klein reduction from the correspondence space and ignore any massive modes arising therefrom. The latter encode the part of the geometry and dynamics that is not invariant under translation along the fibers, and these fields do not introduce new gauge symmetries. Since the compactification is toroidal, the truncation to massless fields is consistent.
    
    Our constructions are also based on situations in which the T-duality directions and their duals can be regarded as fibers of an affine torus bundle. This immediately excludes the general case of $R$-spaces, in which the fibers can be e.g.~pairs of nilmanifolds.

    \subsection*{Outline}
    
    We begin with a brief review of T-duality as a basis for our further discussion in \ref{sec:review}. We then give an explicit construction of the 2-group of automorphisms $\scGO(n,n;\IZ)$ of $\sfTD_n$ in \ref{sec:higher_groups} together with the corresponding semidirect product 2-group $\scGO(n,n;\IZ)\ltimes \sfTD_n$. The 2-group $\scGO(n,n;\IZ)$ is equivalent as a 2-group to the automorphism 2-group of $\sfTD_n$ constructed in~\cite{Waldorf:2022lzs}. 
    
    Next, we provide an explicit description of geometric T-dualities in terms of principal 2-bundles in \ref{sec:geometric_t-duality}, extending the topological picture of~\cite{Nikolaus:2018qop} to general torus bundles and providing a differential refinement. We explicitly show how to treat the well-known case of the three-dimensional nilmanifold and how to recover the individual T-dual geometries from the principal 2-bundle data.
    
    This picture is further extended in \ref{sec:T-folds} to a description of T-duality between T-folds. Concretely, we construct a Lie 2-groupoid $\scTD_n$ acting as the key structure 2-groupoid in a span of principal 2-groupoid bundles. An explicit description of the cocycles of principal $\scTD_n$-bundles is given, and we discuss an explicit example of a T-fold in this context. We also show how the half-geometric T-dualities of~\cite{Nikolaus:2018qop} are subsumed in our construction.
    
    Finally, we turn to a more speculative extension of our construction to T-dualities with scalar fluxes in \ref{sec:non_geometric}. To arrive at this picture, we recall that certain non-geometric fluxes are related to the embedding tensor in supergravity. This allows us to identify an interesting representation of the scalar fluxes, which we then adjoin as $(-1)$-simplices to the simplicial form of the Lie 2-groupoid $\scTD_n$. The result is the augmented Lie 2-quasi-groupoid $\scTD^\text{aug}_n$, and it is not hard to write down the explicit cocycles for the principal $\scTD^\text{aug}_n$-bundles describing general T-dualities. We then use these cocycles to classify the branes arising in toroidal compactifications of string theory. Finally, we comment on explicit examples of T-duality with scalar fluxes from this perspective.
    
    \subsection*{Open questions}
    
    There are a number of open questions arising from our constructions. 
    \begin{itemize}
        \item[(i)] First of all, our proposals for the non-geometric cases clearly need to be compared in detail to the various expectations from the string theory and mathematical literature, in particular to~\cite{Mathai:2004qq,Bouwknegt:2004ap}.
        \item[(ii)] We observe that the T-duality group $\sfGO(n,n;\IZ)$ does not act on the 2-group $\sfTD_n$, while the extension $\scGO(n,n;\IZ)$ does. This leads to additional moduli in our description, which are then canceled by condition~\eqref{eq:z-rel} arising from demanding the existence of adjusted curvatures. Similarly, our cocycles for principal $\scTD^\text{aug}_n$-bundles impose topological restrictions on the set of $Q$- and $R$-fluxes. It would be useful to understand both of these from a physical perspective. 
        \item[(iii)] It would be very interesting to relate our constructions much more closely to double field theory, in particular to the global constructions of~\cite{Deser:2018flj} based on the formalism of~\cite{Deser:2016qkw}.
        \item[(iv)] It may be possible to use our framework to make progress with the definition and the understanding of non-abelian T-duality as well as Poisson--Lie T-duality; an interesting perspective on the latter has recently been given in~\cite{Arvanitakis:2021lwo}. The issue here is that with the inclusion of non-abelian gauge groups, the relevant 2-groups including the gauge potential become increasingly complicated, cf.~\cite{Rist:2022hci}.
        \item[(v)] In a related vein, while in this paper we restrict to the case of ungauged (super)gravities, it may be feasible to generalize our results to the gauged case with more non-trivial tensor hierarchies. 
        \item[(vi)] Finally, all our constructions readily lift, in principle, to U-duality; see also~\cite{Alfonsi:2021ymc,Sati:2021rsd} for related work.
    \end{itemize}
    
    \section{Lightning review: T-duality}\label{sec:review}
    
    In the following, we collect some basic facts about T-duality from the literature; helpful reviews for further reading include~\cite{Giveon:1994fu,Plauschinn:2018wbo}.
    
    \subsection{Topological T-duality}
    
    We start with a brief review of topological T-duality~\cite{Bouwknegt:2003vb,Bouwknegt:2003zg} with an emphasis on the T-correspondences of~\cite{Bunke:2005sn,Bunke:2005um}. 
    
    \paragraph{Geometric T-backgrounds.} The low-energy sector of a geometric string theory background, or an $\caN=0$ supergravity background, is given by a smooth Riemannian manifold $(M,g)$ that carries an abelian gerbe $\scG$, whose connective structure provides the Kalb--Ramond field $B$~\cite{Gawedzki:1987ak,Freed:1999vc}.\footnote{Technically, the $\caN=0$ supergravity background also includes the dilaton $\phi$. Its T-duality transformation, however, is trivial: the rescaled combination $\exp(-2\phi)\sqrt{|\det g|}$ remains invariant. We therefore neglect it in this work. It will, however, become important in the extension to U-duality~\cite{Borsten:2022ab}.} Recall that abelian gerbes can be described in a geometrically appealing fashion as bundle gerbes~\cite{Murray:9407015,Murray:2007ps} or as central groupoid extensions; here, however, we will be using the equivalent but simpler Hitchin--Chatterjee gerbes~\cite{Hitchin:1999fh,Chatterjee:1998}. For us, a topological abelian gerbe is thus simply a cocycle in Čech cohomology $h_{\rm top}\in \rmH^3(M,\IZ)$. It becomes differentially refined, i.e.~equipped with a connection, if this Čech cocycle is extended to a cocycle in Deligne cohomology $h_{\rmD}\in \rmH^3_{\rmD}(M,\IZ)$. The cohomology class of $h_{\rm top}$ is called the Dixmier--Douady class of the gerbe; if the gerbe carries a connection with 2-form potential $B$, then the image of $[h_D]$ in de~Rham cohomology is the cohomology class of the 3-form curvature $H\in \Omega^3(M)$ of the gerbe, where $H=\rmd B$ locally.
    
    Most commonly, T-duality is defined for string theory backgrounds with a circle or, more generally, a torus action\footnote{or, even more generally, a $\sfGL(n;\IZ)\ltimes\IT^n$-action on a certain $|\sfGL(n;\IZ)|$-fold cover of a possibly unoriented $n$-torus bundle; see \cref{sec:geometric_t-duality}} that preserves the metric and the curvature 3-form of the gerbe. We therefore focus on backgrounds containing a number of 1-cycles that are fibered as a torus bundle $M=P$ over a base manifold $X$. Recall that principal torus bundles are always oriented; we want to explicitly permit unoriented affine torus bundles as considered in~\cite{Baraglia:1105.0290}. As an additional geometric datum, there is an abelian gerbe $\scG$ on the total space of this bundle. We call the triple $(X,P,\scG)$ a \emph{geometric T-background}; if both $P$ and $\scG$ carry connections and $X$ carries a Riemannian structure, we speak of a \emph{differentially refined geometric T-background}, cf.~\cite{Bunke:2005um,Nikolaus:2018qop}. Note that a T-background is not necessarily a consistent background of supergravity or string theory, in the sense 
    that the involved data need not satisfy the relevant equations of motion.

    \paragraph{Topological geometric T-duality.} Let us consider the case of geometric T-duality in more detail and focus on the purely topological aspect. Topological T-duality~\cite{Bouwknegt:2003vb,Bouwknegt:2003zg} is based on the existence of the \emph{Gysin sequence}~\cite{Gysin:1941:61-122}, see also~\cite[Prop.~14.33]{Bott:1982aa}. Given a principal circle bundle $\:\check P\rightarrow X$ with first Chern class $\check F\in \rmH^2(X,\IZ)$, the following sequence is exact:
    \begin{equation}\label{eq:Gysin_sequence}
        \ldots~\xrightarrow{~~~}~\rmH^k(X,\IZ)~\xrightarrow{~\check \pi^*~}~\rmH^k(\check P,\IZ)~\xrightarrow{~\check \pi_*~}~\rmH^{k-1}(X,\IZ)~\xrightarrow{~\check F\,\smile~}~\rmH^{k+1}(X,\IZ)~\xrightarrow{~~~}~\ldots
    \end{equation}
    For topological T-duality, we are interested in this segment for $k=3$. Any 3-form $\check H\in \rmH^3(P,\IZ)$ comes with an associated element $\hat F\coloneqq \check\pi_* \check H\in \rmH^2(X,\IZ)$ with $\check F\smile \hat F=0$ in $\rmH^4(X,\IZ)$. We can now consider a second circle bundle $\hat \pi\colon\hat P\rightarrow X$ with first Chern class $\hat F$. Because $\check F\smile \hat F=\hat F\smile \check F=0$, exactness of the Gysin sequence with hatted maps now shows that there is an $\hat H$ such that $\hat\pi_*\hat H=\check F$. \emph{Topological geometric T-duality} is then the transition from $(\check P,\check H)$ to $(\hat P,\hat H)$. As shown in~\cite{Bouwknegt:2003vb}, this construction matches the various expectations from string theory considerations. As explained in~\cite{Baraglia:1105.0290}, it also extends to the case of affine torus bundles, and there is a corresponding Gysin sequence.
    
    \paragraph{T-duality correspondence.} We can arrive at a more geometric picture if we include the correspondence space $\check P\times_X \hat P$ and regard $\check H$ and $\hat H$ as the Dixmier--Douady classes of some bundle gerbes $\check \scG$ and $\hat \scG$, respectively. This then leads to the commutative diagram
    \begin{equation}\label{eq:T-duality_correspondence}
        \begin{tikzcd}[column sep=1cm, row sep=0.8cm]
            & & \scG_\rmC=\check\sfp^*\check\scG\otimes \hat \sfp^*\hat \scG^{-1} \arrow[d]& & \\
            & & \arrow[ld,"\check \sfp",swap] \check P\times_X\hat P \arrow[rd,"\hat \sfp"]& & \\
            \check\scG \arrow[r] & \check P \arrow[rd,"\check \pi"] & & \hat P \arrow[ld,"\hat \pi",swap] & \hat \scG \arrow[l]\\
            & & X & &             
        \end{tikzcd}
    \end{equation}
    which is crucial in the definition of topological T-duality in terms of T-duality triples~\cite{Bunke:2005um}. Such a T-duality triple is given by the data $((\check P,\check H),(\hat P,\hat H), u)$, where $u$ is a trivialization of the gerbe~$\scG_\rmC$, relating it to the Poincar\'e bundle over the correspondence space, cf.~also~\cite[Rem.~6.3]{Fiorenza:2016oki} for a string theoretic interpretation.
    
    \paragraph{Towards a full description of T-duality.} In order to describe a geometric T-back\-ground $(X,P,\scG)$ completely, we need to provide a Riemannian metric on $X$, together with connections on $P$ and $\scG$. 
    
    Recall that a connection on $P$ gives rise to the \emph{Kaluza--Klein metric} on $P$ as follows. We can describe the connection on the principal $\sfU(1)^n$-bundle $P$ as a principal $G$-connection $\theta$. Recall that such a connection is a $\frt^n\coloneqq \sfLie(\sfU(1)^n)\cong \IR^n$-valued $1$-form $\theta\in\Omega^1(P,\frt^n)$ such that, for any fundamental vector field $X_\xi\in \Gamma(TX)$ of $\xi\in \frt^n$, the $1$-form $\theta$ is equivariant, $\caL_{X_\xi}\theta=0$, and reproduces $\xi$ in the sense that $\iota_{X_\xi}\theta=\xi$.
    
    Together with a Riemannian metric $g$ on $X$, the connection $\theta$ induces the Kaluza--Klein metric $\tilde g$ on $P$ defined by\footnote{The scale of the fibered part of the metric is usually set by the dilaton, which we suppress here.}
    \begin{equation}\label{eq:KK_metric}
        \tilde g\coloneqq \pi^*g+\theta^i\otimes \theta^i~.
    \end{equation}
    In the case of an affine torus bundle, we use the connection on the corresponding principal $(\sfGL(n;\IZ)\ltimes \sfU(1)^n)$-bundle, which corresponds to $\fru(1)^n$-valued vector fields defined locally and up to invertible integer linear transformations. This ambiguity, however, drops out of~\eqref{eq:KK_metric}. Throughout this paper, the Riemannian metrics on $\check P$ and $\hat P$ are always coming from the Kaluza--Klein metric, i.e.~from a Riemanninan metric on $X$ and connections on the involved principal bundles.
    
    We then have to explain how differential refinements for $(X,\check P,\check \scG)$ and $(X,\hat P,\hat \scG)$ are linked over the correspondence space into a T-duality correspondence and locally, we expect to reproduce the Buscher rules~\cite{Buscher:1987sk,Buscher:1987qj}. We will be using the picture of~\cite{Nikolaus:2018qop} and replace T-correspondence with spans of principal 2-bundles. Our proposal will then be: a full geometric T-duality amounts to a differential refinement for the principal 2-bundle sitting on top of this span, together with a reconstruction prescription for the differentially refined T-backgrounds. 
    
    \subsection{The T-duality group}\label{ssec:T-duality_group}
    
    Let us briefly motivate the appearance of the group $\sfGO(n,n;\IZ)$, a central actor in our discussion.
    
    \paragraph{The group $\sfO(n,n;\IZ)$.} T-duality is often presented as an involution given by a $\IZ_2$-action. On a string background, this action maps the radius of the involved circle direction $R$ to the inverse radius\footnote{We put $\alpha'=1$.} $\frac{1}{R}$ and interchanges the momentum and the winding modes of the string. There is an additional freedom of reversing the sign in the latter interchange so that the full T-duality group for T-duality along a circle direction should be identified with $\IZ_2\times \IZ_2\cong\sfO(1,1;\IZ)$.
    
    For an $n$-dimensional torus $\IT^n$, this group is enlarged to the group $\sfO(n,n;\IZ)$, see~\cite{Giveon:1988tt,Shapere:1988zv}. Elements $g$ of $\sfO(n,n;\IZ)$ are $2n\times2n$ integer matrices that leave the form 
    \begin{equation}\label{eq:Onn-metric}
        \eta\coloneqq \begin{pmatrix} 0 & \unit_n \\ \unit_n & 0\end{pmatrix}
    \end{equation}
    invariant in the sense that $g^\rmT \eta g=\eta$, which in components becomes 
    \begin{equation}\label{eq:g-parameterization}
        \begin{gathered}
            g=\begin{pmatrix} A & B \\ C & D \end{pmatrix}~,~~~A,B,C,D\in \sfMat(n;\IZ)~,
            \\
            A^\rmT C+C^\rmT A=B^\rmT D+D^\rmT B=0~,~~~A^\rmT D+C^\rmT B=\unit_n~.
        \end{gathered}
    \end{equation}
    
    The group $\sfO(n,n;\IZ)$ is a subgroup of the larger group\footnote{where the nontrivial element of $\IZ_2$ acts as conjugation by $(\begin{smallmatrix}0&\unit_n\\-\unit_n&0\end{smallmatrix})$} $\sfG\sfO(n,n;\IZ)\coloneqq \sfO(n,n;\IZ)\rtimes \IZ_2$ originally defined in~\cite{Mathai:2004qc,Mathai:2005fd}, which becomes relevant for T-duality with general torus bundles. For $n>0$, this group can be identified with the $2n\times2n$ integer matrices that leave $\eta$ invariant up to sign in the sense that $g^\rmT \eta g=\pm\eta$, which in components becomes
    \begin{equation}\label{eq:def_GO}
        A^\rmT C+C^\rmT A=B^\rmT D+D^\rmT B=0~,~~~A^\rmT D+C^\rmT B=\pm\unit_n~.
    \end{equation}
    For $n=0$, we have $\sfG\sfO(0,0;\IZ)\cong\IZ_2$. For future convenience, we introduce the indicator function $|-|\colon\sfG\sfO(n,n;\IZ)\rightarrow \{0,1\}$, which is simply the projection onto the $\IZ_2$ component. In particular, $(-1)^{|g|}=+1$ for all $g\in \sfO(n,n;\IZ)$.

    \paragraph{Local description.} Given a T-background, we can locally\footnote{as well as in the case in which the Kalb--Ramond $B$-field is globally defined} combine the Kalb--Ramond $B$-field with the metric $g$ into the tensor $\caE\coloneqq g+B$. We now have a Möbius-like, non-linear action of the group $\sfGO(n,n;\IZ)$ on $\caE$, which for an element $g\in \sfGO(n,n;\IZ)$ of the form~\eqref{eq:g-parameterization} is given by
    \begin{equation}\label{eq:Mobius_action}
        \tilde \caE=g\acton \caE\coloneqq \frac{A \caE+B}{C \caE+D}~.
    \end{equation}
    
    In order to render the above transformation linear, we can switch to the \emph{generalized metric}~\cite{Shapere:1988zv,Giveon:1988tt,Maharana:1992my,Gualtieri:2003dx}
    \begin{equation}\label{eq:generalized_metric}
        \caH\coloneqq \begin{pmatrix}
            g-Bg^{-1} B & B g^{-1}
            \\
            -g^{-1} B & g^{-1}
        \end{pmatrix}~,
    \end{equation}
    satisfying $\caH^{-1}=\eta \caH \eta$. The action~\eqref{eq:Mobius_action} is then translated into the simple adjoint action of $\sfGO(n,n;\IZ)$ on $\caH$,
    \begin{equation}\label{eq:trafo_generalized_metric}
        \tilde \caH=g\acton \caH\coloneqq g\caH g^\rmT ~.
    \end{equation}
    
    \paragraph{Global description.} In the case in which the $B$-field is not globally defined, we still have the group $\sfGO(n,n;\IZ)$ as the relevant group of T-dualities. Its action on the data making up the differential refinement, however, is more complicated. In particular, the combination $\caE\coloneqq g+B$ can exist only locally (for T-folds) or not exist at all (for $R$-spaces). For further details, see e.g.~\cite{Hull:2006qs,Belov:2007qj}. In all our constructions, we will always have an explicit action of the group $\sfGO(n,n;\IZ)$, or rather of a 2-group generalization $\scGO(n,n;\IZ)$ on the higher bundles forming T-duality correspondences according to our proposal.

    \paragraph{Subgroups of $\sfGO(n,n;\IZ)$.} It is useful to introduce the following subgroups of the T-duality group $\sfGO(n,n;\IZ)$, which together generate the entirety of $\sfGO(n,n;\IZ)$, cf.~\cite{Giveon:1994fu}:
    \begin{itemize}
        \item[$A$)] The subgroup $\sfGL(n;\IZ)\subset \sfO(n,n;\IZ)$ of \emph{$A$-transformations} consists of group elements
        \begin{equation}\label{eq:trafos_A}
            g_A=\begin{pmatrix} A & 0 \\ 0 & (A^\rmT )^{-1} \end{pmatrix}~~~\mbox{with}~~~A\in \sfGL(n;\IZ)~.
        \end{equation}
        These transformations are simply the automorphism $\sfAut(\IT^n)\cong \sfGL(n;\IZ)$ of the $n$-dimensional torus $\IT^n$ forming the fibers of the torus bundle, and this subgroup is therefore also sometimes called the \emph{geometric (sub)group}.
        \item[$B$)] The abelian torsion-free subgroup $\fro(n;\IZ)\subset \sfO(n,n;\IZ)$ of \emph{$B$-transformations} consists of group elements
        \begin{equation}\label{eq:trafos_B}
            g_B=\begin{pmatrix} \unit_n & B  \\ 0 & \unit_n \end{pmatrix}~~~\mbox{with}~~~B\in\{A\in \sfMat(n;\IZ)~|~A^\rmT =-A\}~.
        \end{equation}
        We note that if we tensor this subgroup with functions along the $n$-torus, then certain $B$-transformations are naturally identified with the 2-form $\rmd \Lambda$ for a 1-form $\Lambda$ along the torus direction. The corresponding $B$-transformations then describe gauge transformations, as familiar from the Courant algebroid description, cf.~e.g.~\cite{Deser:2018flj}.
        \item[$\beta$)] The abelian torsion-free subgroup $\fro(n;\IZ)\subset \sfO(n,n;\IZ)$ of \emph{$\beta$-transformations} consists of group elements
        \begin{equation}\label{eq:trafos_beta}
            g_B=\begin{pmatrix} \unit_n & 0 \\ \beta & \unit_n \end{pmatrix}~~~\mbox{with}~~~\beta\in\{A\in \sfMat(n;\IZ)~|~A^\rmT =-A\}~.
        \end{equation}
        These will arise e.g.~in T-dualities relating geometric backgrounds to T-folds.
        \item[$T_k$)] The abelian torsion subgroup $\sfO(1,1;\IZ)^{n}\cong (\IZ_2)^{2n}\subset \sfO(n,n;\IZ)$ of the \emph{factorized dualities} is generated by group elements
        \begin{equation}\label{eq:trafos_fac_dual}
            g^\pm_{T_k}=\begin{pmatrix} \unit_n-1_k & \pm 1_k \\ \pm 1_k & \unit_n-1_k \end{pmatrix}~~~\mbox{with}~~~1_k=\rmdiag(\underbrace{0,\ldots,0}_{k-1},1,\underbrace{0,\ldots,0}_{n-k-1})~.
        \end{equation}
        These transformations can be identified with the involutions that are T-dualities along the $k$th circle direction. Clearly, $(g^\pm_{T_k})^2=\unit_{2n}$. Note that these factorized dualities $T_k$ reproduce the Buscher rules~\cite{Buscher:1987sk,Buscher:1987qj} for the transformations of the metric and $B$-field along a circle direction when inserted into~\eqref{eq:Mobius_action} or~\eqref{eq:trafo_generalized_metric}.

        \item[$G$)] The abelian torsion subgroup $\IZ_2\times\IZ_2\subset\sfGO(n,n;\IZ)$ consists of group elements
        \begin{equation}
            g_G^{s_1,s_2}=\begin{pmatrix}
                s_1\unit_n & 0 \\
                0 & s_2\unit_n
            \end{pmatrix}~~~\mbox{with}~~~s_1,s_2\in\{\pm1\}~.
        \end{equation}
        Note that $(-1)^{|g_G^{s_1,s_2}|}=s_1s_2$. In other words, when $s_1s_2=-1$, then $g_G^{s_1,s_2}\not\in\sfO(n,n;\IZ)$.
    \end{itemize}
    We note that the general form of the generalized metric~\eqref{eq:generalized_metric} for integer $B$ as in~\eqref{eq:trafos_B} is obtained from a diagonal form by a $B$-transformation, hence the name:
    \begin{equation}
        \caH=\begin{pmatrix} \unit_n & B  \\ 0 & \unit_n \end{pmatrix}\begin{pmatrix} g & 0  \\ 0 & g^{-1}\end{pmatrix}\begin{pmatrix} \unit_n & B  \\ 0 & \unit_n \end{pmatrix}^\rmT ~.
    \end{equation}

    \paragraph{$\sfGO(n,n;\IZ)$ versus $\sfO(n,n;\IZ)$.} Throughout this paper, we will work with the larger T-duality group $\sfGO(n,n;\IZ)$ instead of $\sfO(n,n;\IZ)$. The difference between the two groups is that the former contains the additional generator $g_G^{-+}=\rmdiag(-\unit_n,\unit_n)$. We will see in \cref{ssec:2-group_action} that this group element flips the sign of the Kalb--Ramond field along the fiber directions of $\check P\times_X\hat P\rightarrow X$. Identifying gerbes of opposite orientation becomes a necessity because working with general torus bundles implies that we also identify principal bundles with opposite orientation. As an example, consider a principal $\sfU(1)$-bundle $P$ regarded as a general circle bundle, i.e.~a principal $\sfO(2)$-bundle. A constant coboundary equal to the additional $\IZ_2$-factor in $\sfO(2)$ over $\sfU(1)\cong \sfSO(2)$ now flips the orientation of $P$, rendering $P$ and its dual isomorphic. It is well-known that T-duality can interchange the topological invariant of the torus bundle with the topological invariant of the gerbe. Thus, working with general torus bundles implies that we have to enlarge the T-duality group from $\sfO(n,n;\IZ)$ to $\sfGO(n,n;\IZ)$. For further discussion, see also~\cite{Mathai:2004qc,Mathai:2005fd}.
    
    \subsection{Non-geometric backgrounds}\label{ssec:non_geometric_backgrounds}
    
    One of the key realizations has been that T-dualities can not only change the topology of a space, but also map a geometric T-background into a much more general object. These are usually called non-geometric backgrounds, and there are essentially two types or levels of these backgrounds: T-folds and $R$-spaces. Studying these in detail usually requires generalized geometry or even double field theory. For introductory reviews, see e.g.~\cite{Plauschinn:2018wbo,Szabo:2018hhh}.
    
    \paragraph{T-folds.} T-folds\footnote{also called monodrofolds in~\cite{Hellerman:2002ax}; see also references therein for earlier suggestions for extending the monodromy group}~\cite{Hull:2004in,Hull:2006va,Belov:2007qj} are T-backgrounds, most importantly torus fibrations with $B$-field, that are locally geometric, but whose local data in the form of metric and $B$-field are glued together by general elements of the T-duality group~$\sfGO(n,n;\IZ)$. Therefore, T-folds always have a global double geometry~\cite{Hull:2006va}. Because some T-folds arise as T-duals of geometric T-backgrounds, it is natural to propose that string theory is well-defined on these and, in particular, that they have to be included in the space of possible string backgrounds. There is, in fact, a constrained sigma model description of certain T-fold backgrounds~\cite{Hull:2004in}. We also note that a higher geometric local model of T-folds was given in~\cite{Fiorenza:2016oki}.
    
    T-folds are locally geometric, and we can still apply the Buscher rules locally. We note that if the local data of a T-fold is glued together with elements of the geometric subgroup of the T-duality group, then the T-fold specializes to a usual geometric T-background. 
    
    There is a new type of curvature or topological invariant, called $Q$-flux, that plays a similar role to the first Chern classes of the circle fibrations in the T-backgrounds and the Dixmier--Douady class of the gerbe, which are often called \emph{$f$-} and \emph{$H$-flux} in the string theory literature. T-folds are correspondingly sometimes called $Q$-spaces. Note that the T-folds in this paper can be endowed with non-trivial versions of all three types of topological invariants.
    
    As mentioned above, T-folds can be studied using generalized geometry. An alternative interpretation, at least in some cases, has been given in terms of non-commutative geometry~\cite{Mathai:2004qq}. In this paper, we essentially follow the former approach as, roughly speaking, the total space of the principal 2-bundle forming a non-geometric T-correspondence contains a doubled principal torus bundle with an evident connection to generalized geometry.
    
    \paragraph{$R$-spaces.} A further generalization is given by the so-called $R$-spaces~\cite{Shelton:2005cf,Wecht:2007wu} which may not even locally admit a geometric description. That is, there may no longer be a local description in terms of a metric and a $B$-field, and correspondingly, the Buscher rules may  no longer be applicable. The new and additional type of curvature that arises in an $R$-space (analogously to the $f$-, $H$-, and $Q$-fluxes of T-folds) is usually called $R$-flux~\cite{Shelton:2005cf,Shelton:2006fd}, whence the name $R$-space.
    
    These $R$-spaces can be studied within generalized geometry or double field theory, and they can be regarded as certain non-commutative geometries, see~\cite{Bouwknegt:2004ap,Ellwood:2006my,Blumenhagen:2011ph} as well as~\cite{Szabo:2018hhh,Plauschinn:2018wbo} for helpful reviews. Again, we will take a perspective that is close to double field theory, but ours is, in principle, capable of capturing arbitrary topologies.
    
    Note that in our nomenclature, $R$-spaces contain T-folds as special cases, which in turn contain geometric T-backgrounds as special cases.
    
    \paragraph{Classification.} There is now a useful classification of T-backgrounds that allows us to make statements about their T-duals. The Serre spectral sequence associated to the fibration $\pi\colon P\rightarrow X$ defines a filtration
    \[
    \pi^*\rmH^k(X)\eqqcolon F^k \subset F^{k-1} \subset \dotsb \subset F^0 \coloneqq\rmH^k(P)
    \]
    relating the cohomologies of the base $X$ and the total space $P$, cf.~\cite{Bunke:2005um,Baraglia:1105.0290}.
    In particular, the Dixmier--Douady class $h\in\rmH^3(P,\IZ)$ of a gerbe lies within the filtration $F^3\subset F^2\subset F^1\subset F^0$. If the gerbe carries a connection with curvature $H$, then $\scG$ belongs to $F^i$ if some contractions of $H$ with $3-i$ vector fields along the fiber directions are non-trivial. We say that a T-background is \emph{(of type) $F^i$} or simply an \emph{$F^i$-background} if its Čech cocycle lies in $F^i$ but not in $F^{i+1}$.
    
    This classification now allows us to make clear statements about the image of a T-background $(X,\check P,\check \scG)$ under T-duality along fiber directions.
    \begin{itemize}\itemsep-2pt
        \item[$F^3$:] The gerbe $\check\scG$ is the pullback of a gerbe on $X$ along $\check \pi\colon\check P\rightarrow X$, and T-duality maps the T-background $(X,\check P,\check \scG)$ to itself.
        \item[$F^2$:] As shown in~\cite{Bunke:2005um}, this is the minimum requirement for having a \emph{geometric T-dual}. A \emph{geometric T-duality} relates a geometric T-background to another geometric T-background, $(X,\check P,\check \scG)\mapsto (X,\hat P,\hat \scG)$, preserving the total dimension of $\check P$ but generically not the topologies of $\check P$ or $\check \scG$, cf.~e.g.~\cite{Bouwknegt:2003vb}. In particular, if $\check P$ is a principal $\IT^n\cong \sfU(1)^n$-bundle, then so is $\hat P$.
        \item[$F^1$:] T-duality along the fibers maps such a toric T-background to a non-geometric background~\cite{Mathai:2004qc,Mathai:2005fd,Hull:2006qs}. Locally, an $F^1$-background is always $F^2$, and it has local T-duals, which can then be glued together into a T-fold. This was in particular the perspective adopted in~\cite{Nikolaus:2018qop}. 
        \item[$F^0$:] The T-dual of an $F^0$-background is not even locally geometric, and the resulting image is an $R$-space.
    \end{itemize}
    
    Let us briefly consider the simple case of a trivial torus bundle $P=X\times \IT^n$ together with a gerbe $\scG$ with curvature 3-form $H$. In this case, the Dixmier--Douady 3-class $h$ and, correspondingly, the curvature $H$ naturally decompose into four parts:
    \begin{equation}\label{eq:direct_product_cohomology_decomposition}
        \begin{aligned}
            H\in\rmH^3(X\times\IT^n)&\cong\underbrace{\rmH^3(X)\oplus\underbrace{(\rmH^2(X))^{\oplus n}\oplus\underbrace{(\rmH^1(X))^{\oplus\binom n2}\oplus\underbrace{(\rmH^0(X))^{\binom n3}}_{F^3}}_{F^2}}_{F^1}}_{F^0}~,
            \\
            H&=~H^{(3)}~\,+~\,\sum_{i=1}^nH^{(2)}_i~\,\,+~\,\sum_{i,j=1}^nH^{(1)}_{ij}~~\,+~\sum_{i,j,k=1}^nH^{(0)}_{ijk}~.
        \end{aligned}
    \end{equation}
    In this decomposition, the 2-classes $H^{(2)}_1,\dotsc,H^{(2)}_n\in\rmH^2(X)$ dualize, under T-duality, to the Chern classes of a non-trivial torus bundle on $X$, thus to a geometric background.
    The 1-classes $H^{(1)}_{ij}$ correspond under T-duality to $Q$-fluxes. That is, the T-dual is formed by starting with a geometric universal cover and taking a possibly non-geometric quotient given by $\sfO(n,n;\IZ)$-transformations encoded by $H^{(1)}_{ij}$ to form T-folds.
    Finally, the 0-classes $H^{(0)}_{ijk}$ correspond to $R$-fluxes, which encode the degree to which even local geometry fails to exist.
    
    \paragraph{Key example.} An important and guiding example of a T-duality with non-geometric backgrounds, which we will also be using in exemplifying our proposal throughout the paper, is the following chain of three T-dualities~\cite{Hull:2004in,Shelton:2005cf,Dabholkar:2005ve}. Consider a T-background $\caT=(T^3,H)$ given by the 3-torus $T^3$ and an abelian gerbe with curvature or $H$-flux $H$ on this manifold. For a T-duality along any cycle in the torus, $\caT$ is an $F^2$-background, and performing such a T-duality yields a nilmanifold\footnote{See \ref{ssec:ex_top_T_nilmanifolds} for a detailed discussion, also from our new perspective.} $N$, i.e.~a principal circle bundle over $T^2$ with first Chern class or $f$-flux $F=H$. This can be regarded as a flux in the Riemannian metric, regarded as the Kaluza--Klein metric for the principal circle bundle.
    
    A further T-duality along a cycle in the base $T^2$ of the nilmanifold amounts to a T-duality along two cycles of the T-background $\caT$, for which $\caT$ is an $F^1$-background. Performing this T-duality now leads to a T-fold in which the local forms of the metric and the Kalb--Ramond $B$-field are glued together by a $\beta$-transformation that is given by the $Q$-flux $H$. This T-fold can be regarded as a bundle of non-commutative 2-tori with base $S^1$, see e.g.~\cite{Mathai:2004qq}.
    
    There is one last cycle left in $T^3$, the ``base'' of the T-fold, and a T-duality along this direction amounts to a T-duality along all three cycles in $T^3$ of the T-background $\caT$, for which this background is an $F^0$-background. Performing this T-duality to the extent that it has been explored in the string theory literature leads to what is interpreted as a non-geometric background without a local geometric description. The $Q$-flux $H$ here is turned into an $R$-flux of the same quantity.

    \section{Higher geometric groups for T-duality}\label{sec:higher_groups}
    
    We now begin the main discussion by defining the structure group of the principal 2-bundles relevant for T-duality. We start by reviewing the 2-group $\underline{\sfTD}_n$ of~\cite{Nikolaus:2018qop}, which we then extend to a 2-group $\underline{\sfTD}_n^\ltimes$ suitable for the description of affine torus bundles. We also discuss the automorphism group of $\underline{\sfTD}_n$, which will govern the action of the T-duality group, and compute explicit data describing this action.
    
    \subsection{The Lie 2-group \texorpdfstring{$\underline{\sfTD}_n$}{TDn}}\label{ssec:Onn_subgroup_and_actions}
    
    Recall that T-duality correspondences exist for T-backgrounds of type $F^2$. As shown in~\cite{Nikolaus:2018qop}, there is a strict Lie 2-group $\sfTB^\text{F2}_n$ that represents such backgrounds. In other words, we can regard a toric T-background as a principal 2-bundle, i.e.~a higher or categorified principal bundle with $\sfTB^\text{F2}_n$ as its structure 2-group. An interesting observation of~\cite{Nikolaus:2018qop} is that not only the T-backgrounds but also the correspondence space and the gerbe $\scG_\rmC$ over it can be replaced by a principal 2-bundle. A T-duality correspondence then amounts to a double fibration or span of principal 2-bundles which is induced by an underlying span of Lie 2-groups. 
    
    It is clear that the principal 2-bundle taking over the role of the correspondence space should describe the bundle $\check P\times_X \hat P\rightarrow X$, so the structure group should contain the abelian group $\sfU(1)^{2n}$. As observed in~\cite{Nikolaus:2018qop}, this group needs to be extended to a categorical torus, see~\cite{Ganter:2014zoa}, denoted by $\sfTD_n$. This means extending $\sfU(1)^{2n}$ by an additional factor of $\sfB\sfU(1)=\sfU(1)\rightrightarrows *$, which will turn the torus bundle into a torus bundle carrying a (slightly twisted) abelian gerbe or principal $\sfB\sfU(1)$-bundle. This component will ensure the correct relation between the abelian gerbes in the two $T$-backgrounds, cf.~\eqref{eq:T-duality_correspondence}.
    
    We will mostly use crossed modules of Lie groups in order to describe Lie 2-groups. Some background material and further pointers to the literature on higher groups and bundles, as well as our conventions, are found in \ref{app:higher_groups,app:higher_principal_bundles}.
    
    We regard the abelian group $\sfU(1)^{2n}$ as the quotient $\IR^{2n}/\IZ^{2n}$ and extend the corresponding action groupoid by the above mentioned factor of $\sfB\sfU(1)$. We will always use additive notation for elements in $\IR/\IZ\cong \sfU(1)$. 
    \begin{definition}[\cite{Nikolaus:2018qop}]\label{def:TDn}
        The (strict) Lie 2-group $\underline{\sfTD}_n$ is given by the Lie groupoid
        \begin{subequations}\label{eq:TD_n_2-group}
            \begin{equation}
                \begin{gathered}
                    \begin{tikzcd}
                        \IR^{2n}\times \IZ^{2n}\times \sfU(1)\arrow[r,shift left] 
                        \arrow[r,shift right] & \IR^{2n}
                    \end{tikzcd}~,~
                    \begin{tikzcd}[column sep=2.0cm,row sep=large]
                        \phantom{\sft(h)} \xi & \xi-m_1\arrow[l,bend left,swap,out=-20,in=200]{}{(\xi,m_1,\phi_1)}& \xi-m_1-m_2\arrow[l,bend left,swap,out=-20,in=200]{}{(\xi-m_1,m_2,\phi_2)}\arrow[ll,bend right,out=20,in=-200]{}{(\xi,m_1+m_2,\phi_1+\phi_2)}
                    \end{tikzcd}~,
                    \\
                    \sfid_\xi\coloneqq (\xi,0,0)~,~~~(\xi,m,\phi)^{-1}\coloneqq (\xi-m,-m,-\phi)~,
                \end{gathered}
            \end{equation}    
            together with the monoidal structure and inverse functor defined by
            \begin{equation}
                \begin{aligned}
                    (\xi_1,m_1,\phi_1)\otimes (\xi_2,m_2,\phi_2)&\coloneqq(\xi_1+\xi_2,m_1+m_2,\phi_1+\phi_2-\langle \xi_1,m_2\rangle)~,
                    \\
                    \sfinv(\xi,m,\phi)&\coloneqq (-\xi,-m,-\phi-\langle \xi,m\rangle)
                \end{aligned}
            \end{equation}    
        \end{subequations}    
        for $\xi,\xi_{1,2}\in \IR^{2n}$, $m,m_{1,2}\in \IZ^{2n}$ and $\phi,\phi_{1,2}\in \sfU(1)$. Here, the binary bracket $\langle-,-\rangle$ is defined by
        \begin{equation}
            \langle \xi_1,\xi_2\rangle=\xi^\rmT _1
            \begin{pmatrix} 
                0 & 0 \\ 
                \unit_n &0 
            \end{pmatrix}
            \xi_2
            ~~~\mbox{or}~~~
            \left\langle
            \begin{pmatrix}
                \hat \xi_1
                \\
                \check \xi_1
            \end{pmatrix},
            \begin{pmatrix}
                \hat \xi_2
                \\
                \check \xi_2
            \end{pmatrix}\right\rangle=\check \xi_1 \hat \xi_2
        \end{equation}
        for $\xi_{1,2}\in \IR^{2n}$.
    \end{definition}
    We note that the Lie 2-group $\underline{\sfTD}_n$ corresponds to the following crossed module of Lie groups:
    \begin{equation}\label{eq:def_TD_n}
        \begin{gathered}
            \sfTD_n~\coloneqq~\big(\IZ^{2n}\times\sfU(1) \xrightarrow{~\sft~}\IR^{2n}\big)~,
            \\
            \sft(m,\phi)\coloneqq m~,
            \\
            \xi\acton(m,\phi)\coloneqq (m,\phi-\langle \xi,m\rangle)
        \end{gathered}
    \end{equation}
    with abelian and evident group products in $\IZ^{2n}\times\sfU(1)$ and $\IR^{2n}$. 
    
    The Lie 2-algebra $\frtd_n$, in the form of a crossed module of Lie algebras, is given by 
    \begin{equation}\label{eq:def_TD_n_alg}
        \begin{gathered}
            \frtd_n~=~\big(\IR\xrightarrow{~\sft~}\IR^{2n}\big)~,
            \\
            \sft(y)=0~,~~~
            \xi\acton y=0
        \end{gathered}
    \end{equation}
    for $y\in \IR$ and $\xi\in \IR^{2n}$. Weak morphisms of Lie 2-groups describing automorphisms of $\underline{\sfTD}_n$ naturally translate to invertible Lie 2-algebra morphisms $\phi\colon\frtd_n\rightarrow \frtd_n$ as defined in \ref{app:higher_Lie_algebras}. 
    
    \subsection{A group of automorphisms of \texorpdfstring{$\underline{\sfTD}_n$}{TDn}}
    
    In~\cite{Nikolaus:2018qop}, the authors announced that in the group of isomorphism classes of objects of $\sfTD_n$, the group $\underline{\pi_0}(\sfAut(\underline{\sfTD}_n))$ in the 2-group of automorphisms given by crossed intertwiners $\sfAut(\underline{\sfTD}_n)$ is isomorphic to the group $\sfG\sfO(n,n;\IZ)$ defined in~\eqref{eq:def_GO}; the proof was given recently in~\cite{Waldorf:2022lzs}. 
    
    In order to prepare the discussion of automorphisms of $\underline{\sfTD}_n$ and to set up some helpful notation to be used later, we begin by studying a group of automorphisms $\Phi\colon\underline{\sfTD}_n\rightarrow \underline{\sfTD}_n$ in detail. These morphisms can equivalently be regarded as invertible weak 2-functors between the corresponding one-object 2-groupoids $\Phi\colon\sfB\underline{\sfTD}_n\rightarrow \sfB\underline{\sfTD}_n$ as defined in~\ref{app:2-groupoid_basics}. Such a 2-functor consists of a functor $\Phi_1\colon\underline{\sfTD}_n\rightarrow \underline{\sfTD}_n$ and a natural transformation given by a map $\Phi_2\colon\IR^{2n}\times \IR^{2n}\rightarrow \IR^{2n}\times\IZ^{2n}\times \sfU(1)$ that satisfies the naturality and coherence conditions listed in~\eqref{eq:weak_morphism}. 
    
    We begin by defining a useful set of maps on $\sfGO(n,n;\IZ)$ and listing their properties.
    \begin{lemma}\label{lem:def_maps_rho_sigma}
        Consider the group $\sfGO(n,n;\IZ)$ as a matrix group. We define the map $\rho\colon\sfGO(n,n;\IZ)\rightarrow \frgl(2n;\IZ)$ by 
        \begin{equation}
            \begin{aligned}
                \rho(g)&\coloneqq 
                g^\rmT \begin{pmatrix} 0_n & 0_n \\ \unit_n & 0 \end{pmatrix}g
                -(-1)^{|g|}\begin{pmatrix} 0_n & 0_n \\ \unit_n & 0 \end{pmatrix}
                \\
                &=\begin{pmatrix}
                    C^\rmT A & C^\rmT  B 
                    \\
                    D^\rmT A -(-1)^{|g|}\unit_n & D^\rmT  B
                \end{pmatrix}=
                \begin{pmatrix}
                    C^\rmT A & C^\rmT  B 
                    \\
                    -B^\rmT C & D^\rmT  B
                \end{pmatrix}
                \\
                &=-\rho^\rmT (g)~,
            \end{aligned}
        \end{equation}
        $g\in \sfGO(n,n;\IZ)$. We further decompose this matrix into its lower triangular part $\rho_L(g)$ and its transpose, defining a second map $\rho_L\colon\sfGO(n,n;\IZ)\rightarrow \frgl(2n;\IZ)$ by 
        \begin{equation}\label{eq:def_rho_L}
            \rho(g)\eqqcolon\rho_L(g)-\rho_L(g)^\rmT~.
        \end{equation}    
        For all $g_{1,2}\in \sfGO(n,n;\IZ)$, we have 
        \begin{equation}
            g^\rmT _2\rho(g_1)g_2+(-1)^{|g_1|}\rho(g_2)=\rho(g_1g_2)~.
        \end{equation}
        An analogous relation does not hold for $\rho_L$, and we can describe this failure by the map $\sigma_L\colon\sfGO(n,n;\IZ)\times\sfGO(n,n;\IZ)\rightarrow\sfSym(2n;\IZ)$ with 
        \begin{equation}\label{eq:def_sigma_L}
            \sigma_L(g_1,g_2)\coloneqq g^\rmT _2\rho_L(g_1)g_2+(-1)^{|g_1|}\rho_L(g_2)-\rho_L(g_1g_2)~.
        \end{equation}
        Here, $\sfSym(2n;\IZ)$ denotes the set of $2n\times 2n$-dimensional symmetric matrices. For all $g_{1,2,3}\in \sfGO(n,n;\IZ)$, we then have
        \begin{equation}\label{eq:sigma_relation}
            -g_3^\rmT \sigma_L(g_1,g_2)g_3+\sigma_L(g_1,g_2g_3)+(-1)^{|g_1|}\sigma_L(g_2,g_3)-\sigma_L(g_1g_2,g_3)=0~.
        \end{equation}
    \end{lemma}
    \begin{proof}
        By direct computation.\footnote{Many proofs in this section are straightforward but tedious; we therefore suggest using a computer algebra system with a suitable term substitution engine to complete these.}
    \end{proof}
    \begin{lemma}\label{lem:subgroupsOK}
        For two elements $g_1,g_2\in \sfGO(n,n;\IZ)$ that are both in the geometric subgroup, or both in the subgroup of $\beta$-transformations, or both in the subgroup of $B$-transformations, we have 
        \begin{equation}
			\sigma_L(g_1,g_2)=0~.
		\end{equation}
    \end{lemma}
    \begin{proof}
        By direct computation.
    \end{proof}
    
    Next, we establish a useful group of (2-group) automorphisms on $\underline{\sfTD}_n$.
    \begin{theorem}\label{theorem:action}
        Let $\widehat{\sfGO}(n,n;\IZ)$ be the group with underlying set $\sfGO(n,n;\IZ)\times \sfSym(2n;\IZ)$, group product 
        \begin{equation}\label{eq:group_product}
            (g_1,\zeta_1)\times(g_2,\zeta_2)=\big(g_1g_2,g_2^\rmT \zeta_1g_2+(-1)^{|g_1|}\zeta_2+\sigma_L(g_1,g_2)\big)
        \end{equation}    
        for $g_{1,2}\in \sfGO(n,n;\IZ)$ and $\zeta_{1,2}\in \sfSym(2n;\IZ)$, and evident unit and inverses. There is a group of automorphisms (given by invertible 2-functors) $\Phi\colon\underline{\sfTD}_n\rightarrow \underline{\sfTD}_n$ that is isomorphic to $\widehat{\sfGO}(n,n;\IZ)$. 
        Explicitly, $(g,\zeta)\in \sfGO(n,n;\IZ)\times \sfSym(2n;\IZ)$ parameterizes the 2-functor 
        \begin{equation}\label{eq:automorphisms_TD_n}
            \begin{aligned}
                \Phi^{g,\zeta}_1(\xi,m,\phi)&=\big(g\xi,gm,(-1)^{|g|}\phi+m^\rmT (\rho_L(g)+\zeta)\xi\big)~,
                \\
                \Phi^{g,\zeta}_2(\xi_1,\xi_2)&=(g(\xi_1+\xi_2),0,\xi_1^\rmT (\rho_L(g)+\zeta)\xi_2)~.
            \end{aligned}
        \end{equation}
        Composition of two such 2-functors yields the group product~\eqref{eq:group_product}, and the identity in the group parameterizes the identity automorphism.
    \end{theorem}
    \begin{proof}
        The fact that~\eqref{eq:group_product} is an associative product compatible with unit and inverses follows from~\eqref{eq:def_sigma_L} and~\eqref{eq:sigma_relation} by direct computation. Once the parameterization~\eqref{eq:automorphisms_TD_n} is established, the compatibility of composition of 2-functors with the group product on $\widehat{\sfGO}(n,n;\IZ)$ is also readily verified by computation.
        
        In fact, the group $\widehat{\sfGO}(n,n;\IZ)$ parameterizes a large set of all automorphisms $\underline{\sfTD}_n\rightarrow \underline{\sfTD}_n$. Let us therefore present also a more detailed derivation. 
        
        We start from the most general ansatz for such an automorphism,
        \begin{equation}
            \begin{aligned}
                \Phi_1(\xi,m,\phi)&=(\Phi_1^0(\xi,m,\phi),\Phi_1^1(\xi,m,\phi),\Phi_1^2(\xi,m,\phi))~,
                \\
                \Phi_2(\xi_1,\xi_2)&=(\Phi_2^0(\xi_1,\xi_2),\Phi_2^1(\xi_1,\xi_2),\Phi_2^2(\xi_1,\xi_2))~,
            \end{aligned}
        \end{equation}
        where all components are evident smooth maps. The naturality condition reads as
        \begin{equation}\label{eq:TD_naturality}
            \begin{aligned}
                \Phi_2(\xi_1,\xi_2)\circ &(\Phi_1(\xi_1,m_1,\phi_1)\tildeotimes \Phi_1(\xi_2,m_2,\phi_2))
                \\
                &=\Phi_1(\xi_1+\xi_2,m_1+m_2,\phi_1+\phi_2-\langle \xi_1,m_2\rangle)\circ \Phi_2(\xi_1-m_1,\xi_2-m_2)~,
            \end{aligned}
        \end{equation}
        and the coherence condition is
        \begin{equation}\label{eq:TD_coherence}
            \begin{aligned}
                \Phi_2&(\xi_1+\xi_2,\xi_3)\tilde \circ(\Phi^0_2(\xi_1,\xi_2)+\Phi_1^0(\xi_3),\Phi^1_2(\xi_1,\xi_2),\Phi^2_2(\xi_1,\xi_2))
                \\
                &=
                \Phi_2(\xi_1,\xi_2+\xi_3)\tilde \circ (\Phi_1^0(\xi_1)+\Phi_2^0(\xi_2,\xi_3),\Phi_2^1(\xi_2,\xi_3),\Phi_2^2(\xi_2,\xi_3)-\langle \Phi_1^0(\xi_1),\Phi_2^1(\xi_2,\xi_3)\rangle)~.
            \end{aligned}
        \end{equation}
        Because $\Phi_1$ is a functor, we have 
        \begin{equation}
            \begin{aligned}
                \Phi_1^0(\xi,m,\phi)&=\Phi_1^0(\xi)~,
                \\
                \Phi_1^1(\xi,m,\phi)&=\Phi_1^0(\xi)-\Phi_1^0(\xi-m)~,
                \\
                \Phi_1^2(\xi,0,0)&=0~,
                \\
                \Phi_1^2(\xi,m_1+m_2,\phi_1+\phi_2)&=\Phi^2_1(\xi,m_1,\phi_1)+\Phi^2_1(\xi-m_1,m_2,\phi_2)~.
            \end{aligned}
        \end{equation}    
        Applying the target map to both sides of~\eqref{eq:TD_naturality} implies that $\Phi_2^0(\xi_1,\xi_2)=\Phi_1^0(\xi_1+\xi_2)$. The composition on the left-hand side of~\eqref{eq:TD_naturality} implies that $\Phi^1_2$ measures the failure of $\Phi_1^0$ to be additive:
        \begin{equation}
            \Phi^1_2(\xi_1,\xi_2)=\Phi_1^0(\xi_1+\xi_2)-\Phi_1^0(\xi_1)-\Phi_1^0(\xi_2)~.
        \end{equation}
        The other composition is automatically satisfied, and the sources of both sides of~\eqref{eq:TD_naturality} match. 
        
        We now restrict ourselves to a particular class of morphisms in which $\Phi_1^0$ is a group isomorphism on objects. This amounts to
        \begin{equation}
            \Phi_1^0(\xi)=g\xi~,~~~\Phi_1^1(\xi,m,\phi)=gm~,~~~\Phi_2^1=0
        \end{equation}
        for $g\in \sfGL(2n;\IZ)$, and we further restrict $g$ to be an element in $\sfGO(n,n;\IZ)$. We note that the restriction to crossed intertwiners in~\cite{Nikolaus:2018qop} certainly implies this restriction. The coherence condition then reduces to
        \begin{equation}
            \Phi_2^2(\xi_1,\xi_2)+\Phi_2^2(\xi_1+\xi_2,\xi_3)=\Phi_2^2(\xi_1,\xi_2+\xi_3)+\Phi_2^2(\xi_2,\xi_3)~,
        \end{equation}
        which implies 
        \begin{equation}
            \Phi_2^2(\xi_1,0)=\Phi_2^2(0,\xi_2)=\Phi_2^2(0,0)~.
        \end{equation}
        Also, the naturality condition~\eqref{eq:TD_naturality} for $\xi_1=m_1=\phi_2=0$ reduces to
        \begin{equation}
            \Phi_1^2(\xi_2,m_2,\phi_1)=\Phi_1^2(\xi_2,m_2,0)+\Phi_1^2(0,0,\phi_1)~,
        \end{equation}
        allowing us to split $\Phi_1^2$ into two components,
        \begin{equation}
            \Phi_1^2(\xi_2,m_2,\phi_1)=\Phi_1^{21}(\xi_2,m_2)+\Phi_1^{22}(\phi_1)
        \end{equation}
        with $\Phi_1^{21}(0,0)=0$. Naturality for $\xi_1=m_1=0$ then implies linearity of $\Phi_1^{22}$.
        
        Following~\cite{Nikolaus:2018qop} further, we restrict to morphisms with $\Phi_1^{21}(\xi,m)=\Phi_2^2(m,\xi)$ and assume $\Phi_2^2$ to be bilinear. This now completely solves the coherence relation~\eqref{eq:TD_coherence}. We further set $\Phi_1^2(\phi)=(-1)^{|g|}\phi$, reducing the naturality condition~\eqref{eq:TD_naturality} to
        \begin{equation}
            \Phi_2^2(\xi_1,m_2)-\Phi_2^2(m_2,\xi_1)
            =\langle g \xi_1,g m_2\rangle-(-1)^{|g|}\langle \xi_1,m_2\rangle~.
        \end{equation}
        For an element $g$ parameterized as in~\eqref{eq:g-parameterization}, we have
        \begin{equation}
            \begin{aligned}
                \Phi_2^2(\xi,m)-\Phi^2_2(m,\xi)&=\xi^\rmT 
                \begin{pmatrix}
                    C^\rmT A & C^\rmT  B 
                    \\
                    D^\rmT A -(-1)^{|g|}\unit_n & D^\rmT  B
                \end{pmatrix}m
                \\
                &=\xi^\rmT 
                \begin{pmatrix}
                    C^\rmT A & C^\rmT  B 
                    \\
                    (-1)^{|g|}\unit_n-(-1)^{|g|}\unit_n-B^\rmT C & D^\rmT  B
                \end{pmatrix}m~,
            \end{aligned}
        \end{equation}
        and, using the notation of \ref{lem:def_maps_rho_sigma}, 
        \begin{equation}
            \Phi_2^2(\xi_1,\xi_2)\coloneqq \xi_1^\rmT \big(\rho_L(g)+\zeta\big) \xi_2
        \end{equation}
        for $\zeta\in \sfSym(2n;\IZ)$. Altogether, we have identified a subset of automorphisms of $\underline{\sfTD}_n$ which is parameterized by $\widehat{\sfGO}(n,n;\IZ)$ according to~\eqref{eq:automorphisms_TD_n}.
    \end{proof}
    
    The group $\widehat{\sfGO}(n,n;\IZ)$ can further be lifted to a 2-group, by adding natural 2-transfor\-mations. However, the above structures will suffice for our purposes; moreover, the automorphism 2-group of $\underline{\sfTD}_n$ was computed in~\cite{Waldorf:2022lzs}.
    
    \subsection{2-group action on \texorpdfstring{$\underline{\sfTD}_n$}{TDn}}\label{ssec:2-group_action}
    
    The automorphisms 2-group of $\underline{\sfTD}_n$ will replace the usual T-duality group $\sfGO(n,n;\IZ)$ in our proposal for describing T-duality. Its action on $\underline{\sfTD}_n$ will induce the familiar $A$-, $B$-, $\beta$- and T-duality transformations on the 2-bundles.
    
    The action of a 2-group on a 2-group is readily defined, see \ref{app:higher_groups}. It is now interesting to observe that $\sfGO(n,n;\IZ)$ is not trivially a subgroup of $\widehat{\sfGO}(n,n;\IZ)$ due to the product listed in~\ref{theorem:action}. We therefore do not expect an action of $\sfGO(n,n;\IZ)$ on $\underline{\sfTD}_n$. We could consider an action of $\widehat{\sfGO}(n,n;\IZ)$ on $\underline{\sfTD}_n$, but, as mentioned above, we have the following proposition.
    \begin{proposition}[{\cite[Theorem 5.2.6]{Waldorf:2022lzs}}]\label{prop:Waldorf1}
        The 2-group $\IA_n\cong \sfAut(\underline{\sfTD}_n)$ is a non-central extension
        \begin{equation}
            1\rightarrow \IZ^{2n}\rightarrow \IA_n\rightarrow \sfGO(n,n;\IZ)\rightarrow 1~.
        \end{equation}
    \end{proposition}
    Let us therefore define a Lie 2-group of this form, making all structure maps explicit. To do so, we first note the following.
    \begin{lemma}\label{lem:diag_Z}
        Consider the map $\rmdiag\colon\sfSym(2n;\IZ)\rightarrow \IZ^{2n}$ that extracts the diagonal vector of an element in $\sfSym(2n;\IZ)$, regarded as a matrix. Then 
        \begin{equation}
            \tfrac12(g_3^\rmT \rmdiag(\zeta)-\rmdiag(g_3^\rmT \zeta g_3))\in \IZ^{2n}
        \end{equation}
        for $\zeta\in\sfSym(2n;\IZ)$.
    \end{lemma}
    \begin{proof}
        This follows from the fact that all terms proportional to off-diagonal elements of $\zeta$ appear twice, and all terms proportional to diagonal elements appear with the even factor of the form $(g_3)_{ii}((g_3)_{ii}-1)$.
    \end{proof}
    
    \begin{definitiontheorem}
        There is a Lie 2-group $\scGO(n,n;\IZ)$ with underlying Lie group\-oid
        \begin{subequations}\label{eq:def_scGOnnZ}
            \begin{equation}
                \begin{gathered}
                    \begin{tikzcd}
                        \sfGO(n,n;\IZ)\times \IZ^{2n}\arrow[r,shift left] 
                        \arrow[r,shift right] & \sfGO(n,n;\IZ)
                    \end{tikzcd}~,~~
                    \begin{tikzcd}[column sep=2.0cm,row sep=large]
                        g & g \arrow[l,bend left,swap,out=-20,in=200]{}{(g,z_1)}& g\arrow[l,bend left,swap,out=-20,in=200]{}{(g,z_2)}\arrow[ll,bend right,out=20,in=-200]{}{(g,z_1+z_2)}
                    \end{tikzcd}~,
                    \\
                    \sfid_g=(g,0)~,~~~(g,z)^{-1}=(g,-z)~,
                \end{gathered}
            \end{equation}   
            monoidal product and corresponding inverse given by 
            \begin{equation}
                \begin{aligned}
                    (g_1,z_1)\otimes (g_2,z_2)&=(g_1g_2,z_1+g_1z_2)
                    \eand
                    \sfinv(g,z)&=(g^{-1},-g^{-1}z)
                \end{aligned}
            \end{equation}
            for all $g,g_{1,2}\in \sfGO(n,n;\IZ)$ and $z,z_{1,2}\in \IZ^{2n}$, and associator
            \begin{equation}\label{eq:assoc_GOnnZ}
                \begin{aligned}
                    \sfa(g_1,g_2,g_3)&=\Big(g_1g_2g_3,\frac{(-1)^{|g_1g_2g_3|}}{2}g_1g_2g_3\eta\times\big(g_3^\rmT \rmdiag(\sigma_L(g_1,g_2))
                    \\
                    &\hspace{3cm}+\rmdiag\big(\sigma_L(g_1g_2,g_3)-\sigma_L(g_1,g_2g_3)-(-1)^{|g_1|}\sigma_L(g_2,g_3)\big)\Big)
                    \\
                    &=\Big(g_1g_2g_3,\frac{(-1)^{|g_1g_2g_3|}}{2}g_1g_2g_3\eta~\big(g_3^\rmT \rmdiag(\sigma_L(g_1,g_2))-\rmdiag(g_3^\rmT \sigma_L(g_1,g_2)g_3)\big)\Big)
                \end{aligned}
            \end{equation}
            for $g_{1,2,3}\in \sfGO(n,n;\IZ)$, where $\eta$ is the usual $\sfO(n,n;\IZ)$-invariant metric defined in~\eqref{eq:Onn-metric}.
        \end{subequations}    
    \end{definitiontheorem}
    \begin{proof}
        The associator is well-defined because of \ref{lem:diag_Z}. The remaining axioms for a 2-group are directly verified using~\eqref{eq:sigma_relation} in the computations. 
    \end{proof}
    
    Interestingly, further computation shows that the form of the associator is essentially fixed by demanding an action $\scGO(n,n;\IZ)\curvearrowright\underline{\sfTD}_n$. Moreover, the associator is fully encoded in a normalized cocycle $\sfGO(n,n;\IZ)^3\rightarrow \IZ^{2n}$: in particular, $\sfa(1,g_2,g_3)$, $\sfa(g_1,1,g_3)$, and $\sfa(g_1,g_2,1)$ are trivial, and 
    \begin{equation}
        \begin{aligned}
            \sfid_{g_1}\otimes \sfa(g_2,g_3,g_4)+\sfa(g_1,g_2g_3,g_4)+\sfa(g_1,g_2,g_3)=\sfa(g_1,g_2,g_3g_4)+\sfa(g_1g_2,g_3,g_4)~.
        \end{aligned}
    \end{equation}   
    
    The 2-group $\scGO(n,n;\IZ)$ is thus a special Lie 2-group in the sense of~\cite{Baez:0307200}. In particular, it is skeletal, i.e.~isomorphic objects are equal. Altogether, it meets all the expectations of the 2-group $\IA_n$ of \ref{prop:Waldorf1}.
    
    We also have the explicit form of an action.
    \begin{theorem}\label{thm:action_GO_on_TD}
        The following data defines an action (cf.~\ref{app:higher_groups}) $\scGO(n,n;\IZ)\curvearrowright\underline{\sfTD}_n$: the unital bifunctor
        \begin{subequations}\label{eq:GO-action}
            \begin{equation}
                \begin{aligned}
                    \acton\colon\scGO(n,n;\IZ)\times \underline{\sfTD}_n&\rightarrow \underline{\sfTD}_n~,
                    \\
                    (g,z)\times(\xi,m,\phi)&\mapsto (g\xi,gm,(-1)^{|g|}\phi+m^\rmT \rho_L(g)\xi+z^\rmT \eta g \xi)~,
                \end{aligned}
            \end{equation}
            the natural transformation
            \begin{equation}
                \begin{aligned}
                    \Upsilon_{\scGO(n,n;\IZ)}\colon (g_1g_2)\acton \xi &\xrightarrow{~\cong~} g_1\acton(g_2\acton \xi)~,
                    \\
                    \Upsilon_{\scGO(n,n;\IZ)}(g_1,g_2,\xi)&\coloneqq (g_1g_2\xi,0,\tfrac12 \xi^\rmT \sigma_L(g_1,g_2)\xi+\tfrac12\rmdiag(\sigma_L(g_1,g_2))^\rmT \xi)~,
                \end{aligned}
            \end{equation}
            and the natural transformation
            \begin{equation}
                \begin{aligned}
                    \Upsilon_{\sfTD_n}\colon g\acton(\xi_1+\xi_2)&\xrightarrow{~\cong~} (g\acton \xi_1)+(g\acton \xi_2)~,
                    \\
                    \Upsilon_{\sfTD_n}(g,\xi_1,\xi_2)&\coloneqq (g(\xi_1+\xi_2),0,-\xi_1^\rmT \rho_L(g)\xi_2)~.
                \end{aligned}
            \end{equation}
        \end{subequations}    
    \end{theorem}
    \begin{proof}
        One directly verifies that these data satisfy all the relations required for a 2-group action. In particular, the functors $\Upsilon_{\sfGO(n,n;\IZ)}$ and $\Upsilon_{\sfTD_n}$ indeed satisfy the required coherence relations found in~\cite[Prop.~3.2]{Garzn:2001aa}. In the underlying computations, we have to use the fact that
        \begin{equation}
            \tfrac12 m^\rmT \zeta m+\tfrac12\rmdiag(\zeta)^\rmT m
        \end{equation}
        is an integer for all $\zeta\in \sfSym(2n;\IZ)$ and $m\in \IZ^{2n}$ from \ref{lem:diag_Z}. 
    \end{proof}
    
    We note that this action matches closely the parameterization of automorphisms that we obtained in \ref{theorem:action}, up to the discrepancy in $\zeta$- and $z$-terms. Let us now trivially embed any element in $\sfO(n,n;\IZ)$ into $\sfGO(n,n;\IZ)\times \IZ^{2n}$ by $g\mapsto(g,0)$ and consider the automorphism on $\underline{\sfTD}_n$ induced by the action bifunctor $\acton((g,0),-)$. Again, these automorphisms are of the form~\eqref{eq:automorphisms_TD_n}, and we can compute them for elements of the various subgroups of $\sfGO(n,n;\IZ)$ that we reviewed in \ref{ssec:T-duality_group}.\footnote{Note that, as explained above, the full group $\sfGO(n,n;\IZ)$ does not act on $\underline{\sfTD}_n$ without the extension to $\scGO(n,n;\IZ)$. Some of the subgroups in isolation, however, do act on $\underline{\sfTD}_n$.}
    \begin{subequations}
        \begin{itemize}
            \item[$A)$]  $A$-transformations are parameterized by  elements $A\in \sfGL(n;\IZ)\subset \sfO(n,n;\IZ)$, and the corresponding automorphisms are strict:
            \begin{equation}
                \begin{gathered}
                    \Phi^A_1(\xi,m,\phi)\coloneqq \left(\begin{pmatrix} A & 0 \\ 0 & (A^\rmT )^{-1}  \end{pmatrix} \xi ~,~ \begin{pmatrix} A & 0 \\ 0 & (A^\rmT )^{-1}  \end{pmatrix} m~,~\phi\right)~,
                    \\
                    \Phi^A_2(\xi_1,\xi_2)\coloneqq \left(\begin{pmatrix} A & 0 \\ 0 & (A^\rmT )^{-1}  \end{pmatrix} (\xi_1+\xi_2)~,~0~,~0\right)~.
                \end{gathered}
            \end{equation}
            \item[$B)$] $B$-transformations are parameterized by an antisymmetric, integer-valued matrix $B$. The corresponding automorphisms read as
            \begin{equation}
                \begin{gathered}
                    \Phi^B_1(\xi,m,\phi)\coloneqq \left(\begin{pmatrix} \unit_n & -B  \\ 0 & \unit_n \end{pmatrix} \xi ~,~ \begin{pmatrix} \unit_n & -B  \\ 0 & \unit_n \end{pmatrix} m~,~\phi-\check m^\rmT  B_L \check \xi\right)~,
                    \\
                    \Phi^B_2(\xi_1,\xi_2)\coloneqq \left(\begin{pmatrix} \unit_n & -B  \\ 0 & \unit_n \end{pmatrix} (\xi_1+\xi_2)~,~0~,~-\check \xi^\rmT _1 B_L \check \xi_2\right)~,
                \end{gathered}
            \end{equation}
            where $B=B_L-B_L^\rmT $. This action was also defined, with minor differences in $\Phi_2$, in~\cite[Sect.~4.1]{Nikolaus:2018qop}.
            \item[$\beta)$] $\beta$-transformations are parameterized by an antisymmetric, integer-valued matrix $\beta$. The corresponding automorphisms read as
            \begin{equation}
                \begin{gathered}
                    \Phi^\beta_1(\xi,m,\phi)\coloneqq \left(\begin{pmatrix} \unit_n & 0  \\ -\beta & \unit_n \end{pmatrix} \xi ~,~ \begin{pmatrix} \unit_n & 0  \\ -\beta & \unit_n \end{pmatrix} m~,~\phi+\hat m^\rmT  \beta_L \hat \xi\right)~,
                    \\
                    \Phi^\beta_2(\xi_1,\xi_2)\coloneqq \left(\begin{pmatrix} \unit_n & 0  \\ -\beta & \unit_n \end{pmatrix} (\xi_1+\xi_2)~,~0~,~\hat \xi^\rmT _1 \beta_L \hat \xi_2\right)~,
                \end{gathered}
            \end{equation}
            where $\beta=\beta_L-\beta_L^\rmT $.
            \item[$T_k$)] The generators of factorized dualities are parameterized by $k\in\{1,\ldots,n\}$ and a sign. The corresponding automorphisms read as
            \begin{equation}\label{eq:Tk-morphisms}
                \begin{gathered}
                    \Phi^{T_k}_1(\xi,m,\phi)\coloneqq \left( \begin{pmatrix} \unit_n-1_k & \pm 1_k \\ \pm 1_k & \unit_n-1_k \end{pmatrix} \xi ~,~ \begin{pmatrix} \unit_n-1_k & \pm 1_k \\ \pm 1_k & \unit_n-1_k \end{pmatrix} m,\phi-\langle m ,1_k \xi\rangle\right)~,
                    \\
                    \Phi^{T_k}_2(\xi_1,\xi_2)\coloneqq \left(\begin{pmatrix} \unit_n-1_k & \pm 1_k \\ \pm 1_k & \unit_n-1_k \end{pmatrix} (\xi_1+\xi_2)~,~0~,~-\langle \xi_1,1_k\xi_2\rangle\right)~.
                \end{gathered}
            \end{equation}
            \item[$G$)] The $G$-transformations are parameterized by $s_{1,2}\in\{\pm1\}$, and the corresponding automorphisms are strict:
            \begin{equation}
                \begin{gathered}
                    \Phi^G_1(\xi,m,\phi)\coloneqq \left( \begin{pmatrix} s_1\unit_n & 0 \\ 0 & s_2\unit_n \end{pmatrix} \xi ~,~ \begin{pmatrix} s_1\unit_n & 0 \\ 0 & s_2\unit_n \end{pmatrix} m,s_1s_2\phi\right)~,
                    \\
                    \Phi^G_2(\xi_1,\xi_2)\coloneqq \left(\begin{pmatrix} s_1\unit_n & 0 \\ 0 & s_2\unit_n \end{pmatrix} (\xi_1+\xi_2)~,~0~,~0\right)~.
                \end{gathered}
            \end{equation}
        \end{itemize}
    \end{subequations}    
    
    Particularly important will be the following map, already contained in~\cite{Nikolaus:2018qop}.
    \begin{definition}
        The \emph{flip morphism} is simply the action of the concatenation of all factorized dualities $T^+_1\circ \cdots \circ T^+_n$, i.e.~the $\sfGO(n,n;\IZ)$-transformation $g=\eta$. The corresponding automorphism reads as
        \begin{equation}\label{eq:flip-plain}
            \begin{gathered}
                \Phi_1(\xi,m,\phi)\coloneqq \left( \eta \xi ~,~ \eta m~,~\phi-\langle m,\xi\rangle \right)~,
                \\
                \Phi_2(\xi_1,\xi_2)\coloneqq \left(\eta (\xi_1+\xi_2)~,~0~,~-\langle \xi_1,\xi_2\rangle\right)~,
            \end{gathered}
        \end{equation}
        cf.~\eqref{eq:Tk-morphisms}.
    \end{definition}
    At the level of the crossed module of Lie algebras, $\Phi$ induces the following endomorphism $\phi$ on $\frtd_n$:
    \begin{equation}\label{eq:infinitesimal_flip_morphism}
        \begin{gathered}
            \phi^\text{flip}_0(\xi)=
            \begin{pmatrix} 
                0 & \unit_n
                \\
                \unit_n & 0         
            \end{pmatrix}
            \xi
            ~,~~~
            \phi^\text{flip}_1(y)= y~,
            \\
            \phi^\text{flip}_2(\xi_1,\xi_2)=\langle \xi_2,\xi_1\rangle-\langle \xi_1,\xi_2\rangle~.
        \end{gathered}
    \end{equation}
    
    \begin{remark}\label{rem:action_of_subgroups}
        Recall from \ref{lem:subgroupsOK} that for elements $g_1,g_2$ in either the geometric subgroup or the subgroups of $B$- and $\beta$-transformations, the expression $\sigma_L(g_1,g_2)$ vanishes. This implies that the associator in $\scGO(n,n;\IZ)$ vanishes, and there is a 2-group embedding of these subgroups, trivially regarded as 2-groups, into $\scGO(n,n;\IZ)$. Furthermore, for $g_{1,2}$ in these subgroups, the natural transformation $\Upsilon_{\scGO(n,n;\IZ)}(g_1,g_2,\xi)$ is trivial. Altogether, we obtain a direct action of the subgroups on $\sfTD_n$ by the above embedding.
    \end{remark}

    \subsection{The Lie 2-group \texorpdfstring{$\underline{\sfTD}^\ltimes_n$}{TDnx}}    
    
    In this paper, we want to capture T-dualities involving general affine torus bundles, which, following~\cite{Baraglia:1105.0290}, we regard as principal $(\sfGL(n;\IZ)\ltimes \sfU(1)^n)$-bundles. We therefore extend the Lie 2-group $\underline{\sfTD}_n$ by the action of the geometric subgroup $\sfGL(n;\IZ)\subset \sfGO(n,n;\IZ)$ to a semidirect product $\sfGL(n;\IZ)\ltimes \underline{\sfTD}_n$, where the action of $\sfGL(n;\IZ)$ is that of the geometric subgroup on $\sfGO(n,n;\IZ)$. Recall that the geometric subgroup $\sfGL(n;\IZ)$ indeed acts on $\underline{\sfTD}_n$ by \ref{rem:action_of_subgroups}.
    
    We also note that T-duality merely allows for an extension by the group $\sfGL(n;\IZ)\subset \sfGO(n,n;\IZ)$ and not the perhaps expected group $\sfGL(2n;\IZ)$. This amounts to a link between the orientations of the torus bundles $\check P$ and $\hat P$ in~\eqref{eq:T-duality_correspondence}.
    
    \begin{definition}
        The Lie 2-group $\underline{\sfTD}^\ltimes_n$ is given by the Lie groupoid
        \begin{subequations}\label{eq:TD_n_hat}
            \begin{equation}
                \begin{gathered}
                    \begin{tikzcd}
                        \sfGL(n;\IZ)\times \IR^{2n}\times \IZ^{2n}\times \sfU(1)\arrow[r,shift left] 
                        \arrow[r,shift right] & \sfGL(n;\IZ)\times \IR^{2n}
                    \end{tikzcd}~,
                    \\
                    \begin{tikzcd}[column sep=2.0cm,row sep=large]
                        (g,\xi) & (g,\xi-m_1)\arrow[l,bend left,swap,out=-20,in=200]{}{(g,\xi,m_1,\phi_1)}& (g,\xi-m_1-m_2)\arrow[l,bend left,swap,out=-20,in=200]{}{(g,\xi-m_1,m_2,\phi_2)}\arrow[ll,bend right,out=20,in=-200]{}{(g,\xi,m_1+m_2,\phi_1+\phi_2)}
                    \end{tikzcd}~,
                    \\
                    \sfid_{(g,\xi)}=(g,\xi,0,0)~,~~~(g,\xi,m,\phi)^{-1}=(g,\xi-m,-m,-\phi)
                \end{gathered}
            \end{equation}    
            together with the monoidal structure and inverse functor
            \begin{equation}
                \begin{aligned}
                    (g_1,\xi_1,m_1,\phi_1)\otimes (g_2,\xi_2,m_2,\phi_2)&\coloneqq(g_1g_2,\xi_1+g_1\xi_2,m_1+g_1m_2,\phi_1+\phi_2-\langle \xi_1,g_1m_2\rangle)~,
                    \\
                    \sfinv(g,\xi,m,\phi)&\coloneqq (g^{-1},-g^{-1}\xi,-g^{-1}m,-\phi-\langle \xi,g^{-1}m\rangle)
                \end{aligned}
            \end{equation}    
        \end{subequations}    
        for $g\in \sfGL(n;\IZ)\subset \sfO(n,n;\IZ)$, $\xi,\xi_{1,2}\in \IR^{2n}$, $m,m_{1,2}\in \IZ^{2n}$, and $\phi,\phi_{1,2}\in \sfU(1)$. 
    \end{definition}
    
    Just as $\underline{\sfTD}_n$, also $\underline{\sfTD}^\ltimes_n$ is a strict Lie 2-group, and the corresponding crossed module of Lie groups is 
    \begin{equation}\label{eq:def_TD_n_hat}
        \begin{gathered}
            \sfTD^\ltimes_n~\coloneqq~\big(\IZ^{2n}\times\sfU(1) \xrightarrow{~\sft~}\sfGL(n;\IZ)\ltimes\IR^{2n}\big)~,
            \\
            \sft(m,\phi)=(1,m)~,
            \\
            (g,\xi)\acton(m,\phi)=(gm,\phi-\langle \xi,gm\rangle)~.
        \end{gathered}
    \end{equation}
    The associated crossed module of Lie algebras is $\frtd_n$.
    
    We note that the action of automorphisms of $\underline{\sfTD_n}$ trivially extends to automorphisms of $\underline{\sfTD}^\ltimes_n$ because the group $\sfGL(n,\IZ)$ in the semidirect product is to be seen as a subgroup of $\sfGO(n,n;\IZ)$. For example, the action of the flip morphisms~\eqref{eq:flip-plain} on $\underline{\sfTD}^\ltimes_n$ is given by the automorphism
    \begin{equation}
        \begin{gathered}
            \Phi_1(g,\xi,m,\phi)\coloneqq \left( \eta g\eta~,~\eta \xi ~,~ \eta m~,~\phi-\langle m,\xi \rangle \right)~,
            \\
            \Phi_2(g_1,\xi_1;g_2,\xi_2)\coloneqq \left(\eta (\xi_1+\xi_2)~,~0~,~-\langle \xi_1,\xi_2\rangle\right)~.
        \end{gathered}
    \end{equation}
    
    \subsection{From the origins of \texorpdfstring{$\sfTD_n$}{TDn} to principal 2-groupoid bundles}\label{ssec:origins_of_TDn}
    
    In this section, we abandon rigor and sketch the origins of the 2-group $\sfTD_n$, based on material available in the literature. This will naturally motivate the use of principal 2-groupoid bundles in \ref{sec:T-folds} as well as augmented principal 2-groupoid bundles in \ref{sec:non_geometric}.
    
    On general grounds, it is clear that T-duality is intimately related to Kaluza--Klein reduction: the definition of topological T-duality via the Gysin sequence relies on fiber integration, and the metric on the total space of the principal torus bundle is defined as the Kaluza--Klein metric. For a more detailed discussion of this point with regards to double field theory, see also~\cite{Berman:2019biz}. For a related but slightly different perspective that interprets double field theory as a Kaluza--Klein theory, see also~\cite{Alfonsi:2019ggg,Alfonsi:2020nxu,Alfonsi:2021ymc}.
    
    In a Kaluza--Klein reduction we reduce a geometric structure on a principal bundle $P$ over a manifold $X$ to a geometric structure on the manifold $X$. Usually, $P$ is a principal torus bundle. Mathematically, most geometric structures we want to reduce (such as Riemannian metrics, gerbes, and principal bundles with connections) are given by functors that are represented by a classifying space $\caC$. An example would be principal $\sfG$-bundles for $\sfG$ some topological group, which are maps from $P$ to $\caC=\sfB\sfG$. In the following, we will focus on topological aspects, which suffice for the present section.
    
    If $P=X\times \IT^n$ is topologically trivial, then we have the usual currying relation
    \begin{equation}
        C^0(X\times \IT^n,\caC)=C^0(X,C^0(\IT^n,\caC))~,
    \end{equation}
    where $C^0(A,B)$ denotes the space of continuous maps from $A$ to $B$.\footnote{For this to hold, one must work in a Cartesian-closed category of topological spaces. The category of all topological spaces and continuous maps fails to be Cartesian-closed, but there are well-known fixes for this, e.g.\ working with compactly generated weakly Hausdorff spaces. For physical purposes, one might want to work with smooth maps rather than continuous ones; in that case one can use e.g.\ diffeological spaces, cf.~\cite{Baez:0807.1704} for a detailed discussion. Here, we neglect such technical details.
    } This is due to the functors $\IT^n\times-$ and $C^0(\IT^n,-)$ forming an adjunction in a Cartesian-closed category. Taking homotopy classes,
    \begin{equation}
        [X\times \IT^n,\caC]=[X,C^0(\IT^n,\caC)]~,
    \end{equation}
    we see that $C^0(\IT^n,\caC)$ classifies $\caC$-objects on a trivial $n$-torus bundle.
    Note that, for $n=1$, we obtain the maps from $X$ into the free loop space $L\caC\coloneqq C^0(S^1,\caC)$ of $\caC$.
    
    If $P$ is non-trivially fibered over $X$, then the above discussion holds only locally. In particular, the fibers can only be identified with $\IT^n$ up to an action of $\sfU(1)^n$, and we replace the mapping space $[\IT^n,\caC]$ with the homotopy quotient space\footnote{This is not an ordinary quotient since the action of $\sfU(1)^n$ has fixed points (on constant maps). Technically, such homotopy quotients can be realized in the category of topological spaces by using (topologically realized) classifying spaces to remove such fixed points. These details do not concern us, however: for our purposes, it is much more natural to model them in terms of higher group(oid)s as we explain.}
    \begin{equation}\label{eq:gen_cyclic_loop_space}
        C^0(\IT^n,\caC)\sslash\sfU(1)^n~.
    \end{equation}
    In the case $n=1$, this quotient is also called the \emph{cyclic loop space}, and the mapping
    \begin{equation}\label{eq:reduction_functor}
        [P,\caC]\rightarrow [X,C^0(\IT^n,\caC)\sslash\sfU(1)^n]
    \end{equation}
    is also called \emph{double dimensional reduction}; see~\cite{Fiorenza:2016ypo,Fiorenza:2016oki} and also the corresponding $n$Lab page\footnote{\url{https://ncatlab.org/nlab/show/geometry+of+physics+--+fundamental+super+p-branes}} for further details. We note that there is again an adjunction between the reduction functor~\eqref{eq:reduction_functor} and the corresponding oxidation functor. 
    
    Here, we restrict ourselves to the zero modes along the fibers, as this allows for much computational simplification. For one-dimensional T-duality, we are interested in the case $n=1$ and $\caC=\sfB\sfB\sfU(1)$, the classifying space for abelian bundle gerbes\footnote{This is simply the strict 2-category with a single object, a single 1-cell and $\sfU(1)$ as its 2-cells.}. Recall that for any higher group $\sfG$, there is a homotopy equivalence between $L\sfB\sfG$ and the homotopy quotient $\sfB\sfG\sslash\sfG$, which can be modeled by the corresponding action groupoid. In the case of $\sfG=\sfB\sfU(1)$, we thus identify\footnote{Here, $\big(\sfU(1)\times \sfU(1)\rightrightarrows \sfU(1)\rightrightarrows *\big)$ stands for some bicategory with a single object/0-cells, morphisms/1-cells $\sfU(1)$, and 2-morphisms/2-cells $\sfU(1)\times \sfU(1)$.} 
    \begin{equation}
        \begin{aligned}
            L\sfB\sfB\sfU(1)&\cong \sfB\sfB\sfU(1)\times \sfB\sfU(1)
            \\
            &\cong \big(\sfU(1)\times \sfU(1)\rightrightarrows \sfU(1)\rightrightarrows *\big)~.
        \end{aligned}
    \end{equation}
    The cyclic loop space $L\sfB\sfB\sfU(1)\sslash\sfU(1)$ is again a homotopy quotient, and we arrive at
    \begin{equation}
        \begin{aligned}
            \big(L\sfB\sfB\sfU(1)\sslash\sfU(1)\big)&\cong \sfB\sfU(1)\times L\sfB\sfB\sfU(1)
            \\
            &\cong \big(\sfU(1)\times\sfU(1)\times \sfU(1)\rightrightarrows \sfU(1)\times \sfU(1)\rightrightarrows *\big)~.
        \end{aligned}
    \end{equation}
    We note that the latter space looks like the classifying space of a smooth Lie 2-group $\scG$,
    \begin{equation}
        \big(L\sfB\sfB\sfU(1)\sslash\sfU(1)\big)\cong \sfB\scG~,
    \end{equation}
    where the underlying Lie groupoid (we ignore the monoidal product) is given by
    \begin{equation}
        \scG=\sfB\sfU(1)\times \sfU(1)\times \sfU(1)~.
    \end{equation}
    Replacing the groups $\sfU(1)$ with 2-groups $\IR\times \IZ\rightrightarrows \IR$, we arrive at $\underline{\sfTD}_1$ as a Lie groupoid 
    \begin{equation}
        \underline{\sfTD}_1=\left(\IR^2\times\IZ^{2}\times \sfU(1)\rightrightarrows \IR^2\right)~.
    \end{equation}
    We note that more work is certainly necessary to derive the monoidal product. Here, we are just interested in a motivational sketch. 
    
    We can now iterate the above procedure. We regard $\scG$ as a Cartesian product of Lie groupoids and do this for each factor separately. Above we saw that, in each step, we make a replacement
    \begin{equation}
        \sfB\sfB\sfU(1)\rightarrow \sfB(\sfB\sfU(1)\times \sfU(1)\times \sfU(1))~.
    \end{equation}
    Similarly, it is easy to see that we have the replacement
    \begin{equation}
        \sfB\sfU(1)\rightarrow \sfB\sfU(1)\times \sfU(1)\times\sfB\sfU(1)~,
    \end{equation}
    where the last factor, coming from the $\sfU(1)$-action of the cyclification, acts on the second, producing an image in the first. We can consistently truncate to the first factor $\sfB\sfU(1)$ in order to retain a group. Iterating this procedure $n$ times and replacing $\sfU(1)$-factors with $\IZ\rightarrow \IR$, we arrive at the Lie groupoid underlying $\underline{\sfTD}_n$,
    \begin{equation}
        \underline{\sfTD}_n=\left(\IR^{2n}\times\IZ^{2n}\times \sfU(1)\rightrightarrows \IR^{2n}\right)~.
    \end{equation}
    Again, deriving the correct monoidal product requires substantially more work.
    
    Note that in the iteration procedure above, we truncated the part obtained from $\sfB\sfU(1)$-factors to preserve the 2-group structure. After two dimensional reductions, however, we ought to keep these groupoid parts. This is intuitively clear as a 2-form $B$-field, dimensionally reduced twice, will give rise to scalar fields, which should take values in the space of objects of this groupoid. We will develop this point in \ref{sec:T-folds}. A further dimensional reduction can then be captured by an augmented groupoid, and we will discuss this later in \ref{sec:non_geometric}.
    
    \section{Geometric T-duality with principal 2-bundles}\label{sec:geometric_t-duality}
    
    In this section, we generalize the principal 2-bundles of~\cite{Nikolaus:2018qop} to ones suitable for describing topological T-dualities with affine torus bundles as discussed in~\cite{Baraglia:1105.0290}. We also present a natural differential refinement of the underlying cocycles together with the details for describing T-dualities using these structures. We then show that in the case of nilmanifolds, our description matches all expectations\footnote{We note that after a first version of this paper appeared, it was shown in~\cite{Waldorf:2022tib} that our proposal indeed reproduces the Buscher rules locally.}.
    
    \subsection{Topological T-duality correspondences as principal 2-bundles}\label{ssec:t-duality_as_principal_2-bundle}
    
    As mentioned before, it has been shown in~\cite{Nikolaus:2018qop} that geometric T-duality correspondences for principal circle bundles can be formulated as spans of principal 2-bundles $\check \scP$, $\hat \scP$, and $\scP_\rmC$ over $X$,
    \begin{equation}\label{eq:geometric_T_duality_span}
        \begin{tikzcd}[column sep=1cm, row sep=0.8cm]
            & \arrow[ld,"\check \sfp",swap] \scP_\rmC \arrow[rd,"\hat \sfp"]& & \\
            \check \scP & & \hat \scP
        \end{tikzcd}
    \end{equation}
    which are induced by correspondences of Lie 2-groups. In the following, we generalize this picture to affine torus bundles, giving details for the cocycle description as well as the projections $\check \sfp$ and $\hat \sfp$ between them. Throughout, we consider principal 2-bundles subordinate to a surjective submersion $Y\rightarrow X$.
    
    We start with the generalization of $\scP_\text{C}$, which is a principal $\sfTD^\ltimes_n$-bundle. Correspondingly, the general cocycle relations for principal 2-bundle~\eqref{eq:adjusted_cocycles} specialize as follows:
    \begin{subequations}\label{eq:TD-top-cocycles}
        \begin{equation}
            \begin{gathered}
                h=(m_{ijk},\phi_{ijk})\in C^\infty(Y^{[3]},\IZ^{2n}\times\IR/\IZ)~,\\
                g=(g_{ij},\xi_{ij})\in C^\infty(Y^{[2]},\sfGL(n;\IZ)\times\IR^{2n})
            \end{gathered}
        \end{equation}
        with $(ij)\in Y^{[2]}$ and $(ijk)\in Y^{[3]}$, which satisfy
        \begin{equation}
            \begin{aligned}
                \phi_{ikl}+\phi_{ijk}&=\phi_{ijl}+\phi_{jkl}-\langle \xi_{ij},g_{ij}m_{jkl}\rangle~,
                \\
                m_{ikl}+m_{ijk}&=m_{ijl}+g_{ij}m_{jkl}~,
                \\
                g_{ik}&=g_{ij}g_{jk}~,
                \\
                \xi_{ik} &= m_{ijk}+\xi_{ij}+g_{ij}\xi_{jk}
            \end{aligned}
        \end{equation}
        on $Y^{[4]}$ and $Y^{[3]}$, respectively.
    \end{subequations}    
    
    Two such cocycles $(g,h)$ and $(\tilde g,\tilde h)$ are considered equivalent if they are related by a coboundary consisting of maps 
    \begin{subequations}\label{eq:TD-top-coboundaries}
        \begin{equation}
            \begin{gathered}
                (m_{ij},\phi_{ij})\in C^\infty(Y^{[2]},\IZ^{2n}\times \IR/\IZ)~,\\
                (g_i,\xi_i)\in C^\infty(Y,\sfGL(n;\IZ)\times\IR^{2n})~,
            \end{gathered}
        \end{equation}
        that link the cocycles by the relations
        \begin{equation}
            \begin{aligned}
                \phi_{ik}+\phi_{ijk}&=\tilde \phi_{ijk}-\langle \xi_i,g_i\tilde m_{ijk}\rangle+\phi_{ij}+\phi_{jk}-\langle \xi_{ij},g_{ij}m_{jk}\rangle~,
                \\
                m_{ik}+m_{ijk}&=g_i\tilde m_{ijk}+m_{ij}+g_{ij}m_{jk}~,
                \\
                \tilde g_{ij}&=g_i^{-1}g_{ij}g_j~,
                \\
                \xi_i +g_i\tilde \xi_{ij}&=m_{ij}+\xi_{ij}+g_{ij}\xi_j
            \end{aligned}
        \end{equation}
        over $Y^{[3]}$ and $Y^{[2]}$, cf.~the general formulas~\eqref{eq:adjusted_coboundaries}.
    \end{subequations}    
    
    Next, we come to the two principal 2-bundles $\check\scP$ and $\hat \scP$, whose structure 2-group we introduce first.
    \begin{definition}\label{def:TBF2}
        We define the crossed module of Lie groups
        \begin{equation}\label{eq:def_TB_F2}
            \begin{gathered}
                \sfTB^\ltimes_n~=~\big(\IZ^n\times C^\infty(\IT^n,S^1)\xrightarrow{~\sft~}\sfGL(n;\IZ)\ltimes\IR^n\big)~,
                \\
                \sft(m,f)=m~,
                \\
                (g,\xi)\acton(m,f)=(gm,c\mapsto f(c-g\xi))=(gm,f\circ \sfs_{g\xi})
            \end{gathered}
        \end{equation}
        for all $g\in \sfGL(n;\IZ)$, $\xi\in \IR^n$, $m\in \IZ^n$, and $f\in C^\infty(\IT^n,S^1)$, where $\sfs_\xi$ denotes the function $\sfs_\xi t\mapsto t-\xi$. We denote the corresponding strict Lie 2-group by $\underline{\sfTB^\ltimes_n}$.
    \end{definition}
    \noindent We note that the crossed module $\sfTB^{\mathrm{F2}}_n$ defined in~\cite{Nikolaus:2018qop} is obtained by restricting to $g=\unit$. 
    
    The cocycles describing principal $\sfTB^\ltimes_n$-bundles then consist of the following data:
    \begin{subequations}\label{eq:TBF2-cocycles}
        \begin{equation}
            \begin{gathered}
                h=(m_{ijk},f_{ijk})\in C^\infty\big(Y^{[3]},\IZ^{n}\times C^\infty(\IT^n,S^1)\big)~,\\
                g=(g_{ij},\xi_{ij})\in C^\infty(Y^{[2]},\sfGL(n;\IZ)\times\IR^{n})~,
            \end{gathered}
        \end{equation}
        which satisfy
        \begin{equation}
            \begin{aligned}
                f_{ikl}+f_{ijk}&=f_{ijl}+f_{jkl}\circ \sfs_{g_{ij}\xi_{ij}}~,
                \\
                m_{ikl}+m_{ijk}&=m_{ijl}+g_{ij}m_{jkl}~,
                \\
                g_{ik} &= g_{ij}g_{jk}~,
                \\
                \xi_{ik} &= m_{ijk}+\xi_{ij}+g_{ij}\xi_{jk}~.
            \end{aligned}
        \end{equation}
    \end{subequations}    
    Two such cocycles $(g,h)$ and $(\tilde g,\tilde h)$ are considered equivalent if they are related by a coboundary consisting of maps 
    \begin{subequations}\label{eq:TBF2-coboundaries}
        \begin{equation}
            \begin{gathered}
                (m_{ij},f_{ij})\in C^\infty(Y^{[2]},\IZ^n\times C^\infty(\IT^n,S^1))~,
                \\
                (g_i,\xi_i)\in C^\infty(Y,\sfGL(n;\IZ)\times\IR^{n})~,
            \end{gathered}
        \end{equation}
        that link the cocycles by the relations
        \begin{equation}
            \begin{aligned}
                f_{ik}+f_{ijk}&=\tilde f_{ijk}\circ \sfs_{g_i\xi_i}+f_{ij}+f_{jk}\circ\sfs_{g_{ij}\xi_{ij}}~,
                \\
                m_{ik}+m_{ijk}&=g_i\tilde m_{ijk}+m_{ij}+g_{ij}m_{jk}~,
                \\
                \tilde g_{ij}&=g_i^{-1}g_{ij}g_j~,
                \\
                \xi_i +g_i\tilde \xi_{ij}&=m_{ij}+\xi_{ij}+g_{ij}\xi_j~.
            \end{aligned}
        \end{equation}
    \end{subequations}    
    
    We can now generalize the 2-group homomorphism denoted by $ri\ell e'$ in~\cite{Nikolaus:2018qop}\footnote{Note that we interchanged right and left as compared to~\cite{Nikolaus:2018qop}, so that the right-leg projection will lead to the left projection $\check \sfp$ in diagram~\eqref{eq:geometric_T_duality_span}.} to the following (strict) morphism of 2-groups:
    \begin{equation}
        \begin{gathered}
            \Psi : \underline{\sfTD}^\ltimes_n \rightarrow \underline{\sfTB}^\ltimes_n~,
            \\
            \Psi_1\left(
            \begin{pmatrix} 
                \hat g\\ \check g
            \end{pmatrix},
            \begin{pmatrix} 
                \hat \xi\\ \check \xi
            \end{pmatrix}
            ,
            \begin{pmatrix} 
                \hat m\\ \check m
            \end{pmatrix}
            ,\phi
            \right)=(\check g,\check \xi,\check m,c\mapsto \phi+\hat m^\rmT  \check gc)~,
        \end{gathered}
    \end{equation}
    where $\hat g$ and $\check g$ are not independent but parameterize the two block matrices appearing in the embedding of the geometric subgroup $\sfGL(n;\IZ)$ in $\sfGO(n,n;\IZ)$. That is, $\check g=(\hat g^T)^{-1}$.
    
    Postcomposing $\Psi$ with the 2-functor defining the 2-bundle $\scP_\text{C}$ in the sense of \ref{app:higher_principal_bundles}, we obtain the following morphisms of 2-bundles:
    \begin{equation}\label{eq:formula_geometric_pi-check}
        \begin{aligned}
            \check \sfp&:\scP_\text{C}\rightarrow \check \scP~,
            \\
            \check\sfp\colon(g_{ij},\xi_{ij},m_{ijk},\phi_{ijk})&=\left(
            \begin{pmatrix} 
                \hat g_{ij}\\ \check g_{ij}\end{pmatrix},
            \begin{pmatrix} 
                \hat \xi_{ij}\\ \check \xi_{ij}\end{pmatrix},
            \begin{pmatrix} 
                \hat m_{ijk}\\ \check m_{ijk}
            \end{pmatrix},\phi
            \right)
            \\
            &\mapsto (\check g_{ij},\check \xi_{ij},\check m_{ijk},c\mapsto \phi_{ijk}+\hat m_{ijk}^\rmT  \check g_{ij}c)~.
        \end{aligned}
    \end{equation}
    
    Another projection is obtained by precomposing $\Psi$ with the flip morphism defined in~\eqref{eq:flip-plain} and then postcomposing the resulting map with the 2-functor defining the 2-bundle $\scP_\text{C}$:
    \begin{equation}\label{eq:formula_geometric_pi-hat}
        \begin{aligned}
            \hat\sfp&:\scP_\text{C}\rightarrow \hat \scP~,
            \\
            \hat\sfp\colon(g_{ij},\xi_{ij},m_{ijk},\phi_{ijk})&=\left(
            \begin{pmatrix} 
                \hat g_{ij}\\ \check g_{ij}\end{pmatrix},
            \begin{pmatrix} 
                \hat \xi_{ij}\\ \check \xi_{ij}\end{pmatrix},
            \begin{pmatrix} 
                \hat m_{ijk}\\ \check m_{ijk}
            \end{pmatrix},
            \phi_{ijk}\right)
            \\
            &\mapsto \left(
            \hat g_{ij},
            \hat \xi_{ij}
            ,
            \hat m_{ijk}
            ,c\mapsto \phi_{ijk}+\check m_{ijk}^\rmT \hat g_{ij}c+\check \xi_{ij}^\rmT\hat \xi_{jk}   
            \right)~.
        \end{aligned}
    \end{equation}
    This completes our generalization\footnote{Strictly speaking, it is merely a proposal, as with all our other constructions. In future work, it will be necessary to match this proposal with various expectations, e.g.~those of~\cite{Baraglia:1105.0290}.} of topological T-duality given in~\cite{Nikolaus:2018qop} to the case of affine circle bundles: two principal $\sfTB^\ltimes_n$-bundles $\check \scP$ and $\hat \scP$ form a T-dual pair if there is a double fibration~\eqref{eq:geometric_T_duality_span} with projections~\eqref{eq:formula_geometric_pi-check} and~\eqref{eq:formula_geometric_pi-hat}.
    
    \begin{remark}\label{rem:TD_sufficient}
        We note that~\cite[Thm.~3.4.5]{Nikolaus:2018qop} shows that the left leg projection yields a bijection between isomorphism classes of principal $\sfTD_n$-bundles and principal $\sfTB^\text{F2}_n$-bundles. This bijection evidently extends to a bijection between isomorphism classes of principal $\sfTD_n^\ltimes$-bundles and principal $\sfTB_n^\ltimes$-bundles. In this sense, it is clear that no information is gained or lost by choosing to work with either $\check \scP$ or $\scP_\rmC$, at least for principal torus bundles. This point will be important below because a suitable differential refinement only exists on $\scP_\rmC$.
    \end{remark}
    
    Finally, let us give an explicit procedure for deriving the topological T-background from a principal $\sfTB^\ltimes_n$-bundle, in the form of a principal bundle $\check P$ and an abelian gerbe $\check \scG$ on the total space of $\check P$. Consider the image of the right leg projection $\check \sfp$ as given in~\eqref{eq:formula_geometric_pi-check}, defining the principal $\sfTB^\ltimes_n$-bundle $\check \scP$ subordinate to a surjective submersion $\sigma\colon Y\rightarrow X$. The triple $(\check g_{ij},\check \xi_{ij},\check m_{ijk})$ clearly defines an affine torus bundle, which we regard as a principal bundle with fibers $\sfGL(n;\IZ)\times (\IR/\IZ)^n$ over $X$. This principal bundle is given by the quotient 
    \begin{equation}
        \check P \coloneqq (V/\IZ^n)/\sim
        ~~~\mbox{with}~~~
        V\coloneqq Y\times \sfGL(n;\IZ)\times \IR^n~,
    \end{equation}
    where two points $(y,s,t)$, $(y',s',t')\in V$ are equivalent if and only if 
    \begin{equation}
        \sigma(y)=\sigma(y')~,~~~s=\check g(y,y')s'~,\eand t-t'=\check g(y,y')\check \xi(y,y')~.
    \end{equation}
    Correspondingly, we may cover $\check P$ by the induced surjective submersion $V\rightarrow \check P$. Consider now cocycles describing the pullback 2-bundle $\check \pi^* \check\scP$ subordinate to $V\rightarrow \check P$, where $\check \pi$ is the projection defined in~\eqref{eq:T-duality_correspondence}, and recall that any bundle naturally trivializes when pulled back over itself. We note that 
    \begin{equation}
        \begin{aligned}
            V^{[2]}&\coloneqq V\times_{\check P} V
            \\
            &=\{(y_i,y_j,s_i,s_j,t_i,t_j)\in Y^2\times\IR^{2n}\mid
            \\
            &\hspace{3cm}
            \sigma(y_i)=\sigma(y_j),\;
            s_i=\check g_{ij}s_j,\;
            t_i-t_j-\check g_{ij}\check \xi_{ij}\in \IZ^n
            \}~.
        \end{aligned}
    \end{equation}
    Correspondingly, we can define a coboundary $(g_i,\xi_i,m_{ij},f_{ij})$ as in~\eqref{eq:TBF2-coboundaries} by
    \begin{equation}
        \begin{aligned}
            g_i\coloneqq s_i~,~~~\xi_i\coloneqq t_i~,~~~m_{ij}\coloneqq t_i-t_j-\check g_{ij}\check \xi_{ij}~,~~~f_{ij}\coloneqq 0~.
        \end{aligned}
    \end{equation}
    This coboundary induces a 2-bundle isomorphism which trivializes the $\check P$ part in the cocycle:
    \begin{equation}\label{eq:recovering_gerbe_data_top}
        \begin{aligned}
            (\check g_{ij},\check \xi_{ij},\check m_{ijk}, f_{ijk})~~\xrightarrow{~\cong~}~~\big(\unit,0,0,c \mapsto f_{ijk}+\check m_{ijk}^\rmT g_i(c-t_i)\big)~,
        \end{aligned}
    \end{equation}
    and the $\sfU(1)$-cocycle given by the part $(f_{ijk}-\check m_{ijk}^\rmT g_i t_i)$ constant in $c$ defines an abelian gerbe subordinate to the cover $V\rightarrow \check P$. This is the abelian gerbe $\check \scG$ that, together with $\check P$, forms the T-background captured by $\check \scP$.

    \subsection{Differential refinement of \texorpdfstring{$\scP_\rmC$}{PC}}
    
    In order to define a sufficiently general connection on a principal 2-bundle, one generically needs to lift the conventional definition in the literature to that of an adjusted one, see also \ref{app:higher_principal_bundles}. In particular, one has to deform the flat gauge transformation of the 2-form component of the connection to ensure 1) a consistent composition of gauge transformations and 2) the expected globular structure\footnote{i.e.~gauge transformations related by higher gauge transformation have the same image} of higher gauge transformations. Both are required for consistency of the higher gauge structure. One can show that such a deformation is captured by an algebraic datum that is intrinsic to the structure Lie 2-group of the principal 2-bundle. For strict Lie 2-groups or, equivalently, crossed modules of Lie groups, this algebraic datum is given by the following.
    \begin{definition}
        An \emph{adjustment} for a crossed module of Lie groups $\caG=(\sfH\xrightarrow{~\sft~}\sfG,\acton)$ is a map $\kappa\colon\sfG\times \frg\rightarrow \frh$ linear in $\frg$, where $\frg$ and $\frh$ are the Lie algebras of $\sfG$ and $\sfH$, respectively, satisfying 
        \begin{subequations}\label{eq:adjustmentCondition_text}
            \begin{align}
                \kappa(\sft(h),X)&=h(X\acton h^{-1})~,\label{eq:alternativeAdjustmentCondition_a}
                \\
                \kappa(g_2g_1,X)&=g_2\acton\kappa(g_1,X)+\kappa\big(g_2,g_1Xg^{-1}_1-\sft(\kappa(g_1,X))\big)\label{eq:alternativeAdjustmentCondition_b}
            \end{align}
        \end{subequations}
        for all $g_{1,2}\in \sfG$, $h\in \sfH$, and $X\in \frg$.         
    \end{definition}
    Ideally, we would now have adjustments for $\sfTD^\ltimes_n$ and $\sfTB^\ltimes_n$, but we arrive at the following observations:
    \begin{theorem}
        An adjustment for $\sfTD^\ltimes_n$ is given by the map 
        \begin{equation}\label{eq:adjustment_TD}
            \kappa\colon \sfGL(n;\IZ)\times \IR^{2n}\times \IR^{2n}\rightarrow \IZ^{2n}\times \sfU(1)~,~~~(g,\xi;X)\mapsto (0,-\langle X,\xi\rangle)~.
        \end{equation}
        There is no adjustment for $\sfTB_n$ or $\sfTB^\ltimes_n$.
    \end{theorem}
    \begin{proof}
        Verifying~\eqref{eq:adjustmentCondition_text} for~\eqref{eq:adjustment_TD} is immediate. For $\sfTB_n$, we consider the first condition~\eqref{eq:alternativeAdjustmentCondition_a}, which reads as
        \begin{equation}
			\kappa(\sft(m,f),X)=f-f\circ \sfs_X~~\in C^\infty(\IT^n,\IR)
		\end{equation}
        for $(m,f)\in \IZ^n\times C^\infty(\IT^n,S^1)$ and $X\in \IR^n=\sfLie(\IR^n)$. Since $\sft(m,f)=m$, this reduces to the condition\footnote{Suggestively written, $\rmd f(c,X)=f(c)-f(c-X)$.}
        \begin{equation}
            \kappa(m,X)(c)= \rmd f(c,X)~,~~~c\in \IT^n~,
        \end{equation}
        and it is clear that such a map $\kappa$ does not exist for arbitrary $f\in C^\infty(\IT^n,S^1)$. This problem persists for the Lie 2-group $\sfTB^\ltimes_n$.
    \end{proof}
    
    One can now speculate about the reasons for the absence of an adjustment for $\sfTB^\ltimes_n$. It may be that the descriptions of adjustments developed so far in~\cite{Saemann:2019dsl,Kim:2019owc,Rist:2022hci} are incomplete, which may well be the case. Another reason may be the disconnected nature of the components of the 2-group. Regarding the situation at the level of Lie 2-algebras as done in~\cite{Saemann:2019dsl} and in particular in~\cite{Borsten:2021ljb}, we would still conjecture that any crossed module of Lie groups obtained by integration from a crossed module of Lie algebras admits an adjustment. 
    
    In any case, \ref{rem:TD_sufficient} allows us to focus on the principal 2-bundle $\scP_C$ and, hence, on the 2-group $\sfTD_n$. We will see, indeed, that the pair of T-dual T-backgrounds can be reconstructed purely from the differentially refined cocycles of $\scP_C$, which will therefore be fully sufficient for our purposes.
    
    The cocycle relations for the latter are readily found by making formulas~\eqref{eq:adjusted_cocycles} concrete for the adjustment map~\eqref{eq:adjustment_TD}. Beyond the topological cocycle data $(g_{ij},\xi_{ij},m_{ijk},\phi_{ijk})$, cf.~\eqref{eq:TD-top-cocycles}, we have the 1- and 2-forms
    \begin{subequations}\label{eq:diff_refined_cocycles}
        \begin{equation}
            \Lambda\in \Omega^1(Y^{[2]})~,~~~
            A\in \Omega^1(Y,\IR^{2n})~,~~~
            B\in \Omega^2(Y)
        \end{equation}
        satisfying the gluing relations
        \begin{equation}
            \begin{aligned}
                \Lambda_{ik} &= \Lambda_{jk} + \Lambda_{ij}+\rmd \phi_{ijk}-\langle A_i,m_{ijk}\rangle~,
                \\
                A_j &= g_{ij}^{-1}A_i + g_{ij}^{-1}\rmd \xi_{ij}~,
                \\
                B_j &= B_i + \rmd \Lambda_{ij} + \langle\rmd A_i,\xi_{ij}\rangle~.
            \end{aligned}
        \end{equation}
    \end{subequations}        
    The adjusted curvature of this connection on $\scP_\rmC$ is given by locally defined 2- and 3-forms
    \begin{equation}
        F=\rmd A\in \Omega^2(Y,\IR^{2n})
        \eand
        H=\rmd B+\langle\rmd A,A\rangle\in \Omega^3(Y)~.
    \end{equation}
    
    Two differentially refined cocycles $(g,h,A,\Lambda,B)$ and $(\tilde g,\tilde h,\tilde A,\tilde \Lambda,\tilde B)$ are equivalent if there is a differentially refined coboundary between them. Such a coboundary is given by the data $(\xi,m,\phi)$ of a topological coboundary, cf.~\eqref{eq:TD-top-coboundaries}, together with a 1-form 
    \begin{equation}
        \lambda\in \Omega^1(Y)
    \end{equation}
    such that
    \begin{equation}
        \begin{aligned}
            \tilde \Lambda_{ij}&= \Lambda_{ij}+\lambda_j-\lambda_i-\rmd \phi_{ij}-\langle A_i,m_{ij}\rangle~,
            \\
            \tilde A_i&=g_i^{-1}A_i+g_i^{-1}\rmd \xi_i~,
            \\
            \tilde B_i&=B_i +\mathrm d\lambda_i
            +\langle \rmd A_i,\xi_i\rangle~.
        \end{aligned}
    \end{equation}
    
    \subsection{Proposal for a differential refinement of topological T-duality}\label{ssec:differential_refinement}
    
    While we do not have a span of differentially refined principal 2-bundles, we still have a differential refinement of $\scP_\text{C}$ on which the T-duality 2-group $\scGO(n,n;\IZ)$ acts. In particular, we have an action of the flip morphisms encoding T-duality itself.    
    \begin{proposition}
        The flip morphism~\eqref{eq:flip-plain} acts on a differentially refined cocycle describing a principal $\sfTD^\ltimes_n$-bundle with connection as follows:
        \begin{equation}\label{eq:images_of_flip}
            \begin{aligned}
                \xi_{ij}&\mapsto \tilde \xi_{ij}=
                \begin{pmatrix}
                    \hat{\tilde{\xi}}_{ij}
                    \\
                    \check{\tilde{\xi}}_{ij}
                \end{pmatrix}=
                \begin{pmatrix}
                    \check \xi_{ij}
                    \\
                    \hat \xi_{ij}
                \end{pmatrix}~,
                ~~~&
                h_{ijk}&\mapsto \tilde h_{ijk}=h_{ijk}-\check \xi_{ij}\cdot \hat \xi_{ij}~,
                \\
                g_{ij}&\mapsto \tilde g_{ij}=
                \begin{pmatrix}
                    \hat{\tilde{g}}_{ij}
                    \\
                    \check{\tilde{g}}_{ij}
                \end{pmatrix}=
                \begin{pmatrix}
                    \check g_{ij}
                    \\
                    \hat g_{ij}
                \end{pmatrix}~,
                \\
                A_i&\mapsto \tilde A_i=
                \begin{pmatrix}
                    \hat{\tilde{A}}_i
                    \\
                    \check{\tilde{A}}_i
                \end{pmatrix}=
                \begin{pmatrix}
                    \check A_i
                    \\
                    \hat A_i
                \end{pmatrix}~,
                ~~~&
                \Lambda_{ij}&\mapsto \tilde\Lambda_{ij}=\Lambda_{ij}~,
                \\
                B_i&\mapsto \tilde B_i=B_i+\check A_i\wedge\hat A_i~.
            \end{aligned}
        \end{equation}
        As a consequence, the curvatures are mapped to
        \begin{equation}
            F_i\mapsto \tilde F_i=\begin{pmatrix}
                \hat{\tilde{F}}_i
                \\
                \check{\tilde{F}}_i
            \end{pmatrix}
            =\begin{pmatrix}
                \rmd \check A_i
                \\
                \rmd \hat A_i
            \end{pmatrix}
            \eand 
            H_i\mapsto \tilde H_i=\rmd \tilde B_i+(\rmd \check A_i)\wedge \hat A_i~,
        \end{equation}
        and we note that the 3-form part of the curvature remains invariant: $H_i=\tilde H_i$.
    \end{proposition}
    \begin{proof}
        We can simply use~\eqref{eq:cocycles_under_morphisms} to identify the images of the cocycles describing the principal 2-bundle $\scP_\rmC$ under flips, and the result is the above map.
    \end{proof}
    
    Due to the absence of an adjustment for $\sfTB^\ltimes_n$, however, we do not have analogues of the left- or right-leg projections that map to the principal 2-bundles $\check \scP$ and $\hat\scP$. Instead, we need to reconstruct a T-background directly from the principal $\sfTD_n^\ltimes$-bundle. We will explain this procedure in the following.
    
    Consider a differentially refined, $\sfTD_n^\ltimes$-valued cocycle describing a principal 2-bundle $\scP_\text{C}$ with connection. The cocycle data $(\check g_{ij},\check \xi_{ij},\check m_{ijk}^\IZ,\check A_i)$ and $(\hat g_{ij},\hat \xi_{ij},\hat m_{ijk}^\IZ,\hat A_i)$ contained in a cocycle describing $\scP_\rmC$ describe two affine torus bundles $\check P$ and $\hat P$ over $X$ equipped with connections. We already have the diagram~\eqref{eq:T-duality_correspondence} and all contained maps up to the two gerbes $\hat \scG$ and $\check \scG$.
    
    To this end, we pull back $\scP_\rmC$ along $\Pi=\check \pi\circ \check \sfp=\hat \pi\circ \hat \sfp$, cf.~\eqref{eq:T-duality_correspondence}, so that the part corresponding to the affine torus bundles trivializes. Let $\sigma\colon Y\rightarrow X$ be a surjective submersion. Then the correspondence space forms an affine torus bundle, which we regard as a fiber bundle with fibers $\sfGL(n;\IZ)\times \IT^{2n}$. This bundle can be identified with
    \begin{equation}
        \check P\times_X\hat P=(V/\IZ^{2n})/\sim
        ~~~\mbox{with}~~~
        V\coloneqq Y\times \sfGL(n;\IZ)\times \IR^{2n}~,
    \end{equation}
    where two points $(y,s,t)$, $(y',s',t')\in V$ are equivalent if and only if 
    \begin{equation}
        \sigma(y)=\sigma(y')~,~~~~s=g(y,y')s'~,\eand t-t'=g(y,y')\xi(y,y')~.
    \end{equation}
    The fibered product over the correspondence space is then given by 
    \begin{equation}
        \begin{aligned}
            V^{[2]}=\{(y_i,y_j,s_i,s_j,t_i,t_j)&\in Y^2\times\sfGL(n;\IZ)^2\times \IR^{4n}\mid
            \\
            &\sigma(y_i)=\sigma(y_j),\;
            s_i=g_{ij}s_j,\;
            t_i-t_j-g_{ij}\xi_{ij}\in \IZ^{2n}
            \}~,
        \end{aligned}
    \end{equation}
    and we introduce the coboundary $(g_i,\xi_i,m_{ij},\phi_{ij})$ with
    \begin{equation}\label{eq:coboundary1}
        \begin{aligned}
            g_i\coloneqq s_i~,~~~
            \xi_i\coloneqq t_i~,~~~m_{ij}\coloneqq t_i-t_j-g_{ij}\xi_{ij}~,~~~\phi_{ij}\coloneqq 0~.
        \end{aligned}
    \end{equation}
    This coboundary mostly trivializes the cocycle describing $\scP_\rmC$:
    \begin{equation}\label{eq:recovering_gerbe_data}
        \begin{aligned}
            (\xi_{ij},m_{ijk},\phi_{ijk})~~\xrightarrow{~\cong~}~~\big(\unit,0,0,\phi_{ijk}+\langle \xi_{ij},t_j-t_k-g_{ij}\xi_{jk}\rangle\big)~.
        \end{aligned}
    \end{equation}
    The latter expression contains the cocycle $(\phi_{ijk}+\langle \xi_{ij},t_j-t_k-g_{ij}\xi_{jk}\rangle)$ that defines an abelian gerbe subordinate to the cover $\check P\times_X\hat P$. We note that this expression does not depend on $\check t$; therefore, it is the pullback of a gerbe $\hat\scG$ on $\hat \scP$ along the map $\hat \pi$ in~\eqref{eq:T-duality_correspondence}.
    
    Let us now consider the differential refinement. Assume that $\scP_\text{C}$ was endowed with an adjusted connection. Then the pullback along $\Pi$ and the subsequent gauge transformation by the coboundary $(g_i,\xi_i,m_{ij},\phi_{ij},\lambda_i)$ given by~\eqref{eq:coboundary1} together with 
    \begin{subequations}\label{eq:recovery_differential_refinement}
        \begin{equation}
            \lambda_i=-\langle A_i,t_i\rangle
        \end{equation}
        leads to the cocycle data~\eqref{eq:recovering_gerbe_data} and 
        \begin{equation}
            (\tilde \Lambda,\tilde A,\tilde B)\in \Omega^1(V^{[2]})\oplus\Omega^1(V,\IR^{2n})\oplus\Omega^2(V),
        \end{equation}
        satisfying the relations
        \begin{equation}
            \begin{aligned}
                \tilde \Lambda_{ik} = \tilde \Lambda_{jk} + \tilde \Lambda_{ij}+\rmd \phi_{ijk}~,~~~\tilde A_j = \tilde A_i~,~~~\tilde B_j = \tilde B_i + \rmd \tilde \Lambda_{ij} ~,
            \end{aligned}
        \end{equation}
        and the $(\tilde \Lambda,\tilde B)$ form the differential refinement of the gerbe $\hat \pi^*\hat \scG$. We note that the global one-form $\hat A$ gives rise to the Kaluza--Klein metric on $\hat P$. Moreover, 
        \begin{equation}
            \tilde B_i=B_i+\langle A_i,\rmd t_i\rangle=B+\check A_i\wedge \rmd \hat t_i
        \end{equation}
    \end{subequations}    
    is the 2-form connection $\tilde B_i$ of the gerbe. As expected, the connection $\check A_i$ of the bundle $\check P$ relevant in the T-dual situation is obtained by dimensional reduction of the 2-form $\tilde B_i$ along $\hat t_i$.
    
    In order to recover the gerbe $\check \scG$, we apply first the flip morphism to the cocycle data and then go through the same procedure.    
    
    \subsection{Example: Geometric T-duality with nilmanifolds}\label{ssec:ex_top_T_nilmanifolds}
    
    While we do not provide a proof for the validity of our description of T-duality\footnote{As mentioned in the introduction, such a proof was provided in~\cite{Waldorf:2022tib}.}, we work out the instructive example of T-duality between three-dimensional nilmanifolds with $H$-fluxes, i.e.~abelian gerbes with 3-form curvature $H$. This is the first instance of T-duality in the key example of \ref{ssec:non_geometric_backgrounds}.
    
    Recall that a three-dimensional nilmanifold $N_k$ is a principal circle bundle over $\IT^2$ characterized by its first Chern number $k\in \rmH^2(\IT^2,\IZ)\cong\IZ$. We can describe it as a quotient of $\IR^3$ with coordinates $(x,y,z)$ by the relations 
    \begin{equation}\label{eq:def_nil_1}
        (x,y,z)\sim(x,y+1,z)\sim(x,y,z+1)\sim(x+1,y,z-ky)~,
    \end{equation}
    where $x$ and $y$ are coordinates on the base and $z$ is the fiber coordinate. Subordinate to the surjective submersion $\IR^2\rightarrow \IT^2$, we can describe the principal circle bundle $N\rightarrow \IT^2$ by a Čech cocycle given by the map $g\colon(\IR^2)^{[2]}\rightarrow \IR$ with\footnote{We note that $(\IR^2)^{[2]}=\IR^2\times_{\IT^2}\IR^2\cong \IR^2\times \IZ^2$ and $(\IR^2)^{[3]}=\IR^2\times_{\IT^2}\IR^2\times_{\IT^2}\IR^2\cong \IR^2\times \IZ^2\times \IZ^2$.}    
    \begin{equation}\label{eq:nilmanifold_bundle_cocycle}
        g(x,y;x',y')\coloneqq k(x'-x)y~.
    \end{equation}
    
    A gerbe $\scG_\ell$ on the nilmanifold $N$ is characterized by its Dixmier--Douady class $\ell\in \rmH^3(N,\IZ)\cong\IZ$; subordinate to the surjective submersion $\IR^3\rightarrow N$, we can describe it by a Čech cocycle $h\colon(\IR^2)^{[3]}\rightarrow \IR$ with
    \begin{equation}\label{eq:nilmanifold_gerbe_cocycle}
        h(x,y,z;x',y',z';x'',y'',z'')\coloneqq\ell (x-x')(y'-y'')z~.
    \end{equation}
    
    It is well-known that the T-background $(\IT^2,N_k,\scG_\ell)$ is an $F^2$-background, and (topological) geometric T-duality corresponds to the duality
    \begin{equation}\label{eq:top_T-duality}
        (\IT^2,N_k,\scG_\ell)~~\xleftrightarrow{~~~}~~(\IT^2,N_\ell,\scG_k)~.
    \end{equation}
    
    The connection on the principal circle bundle $N_k\rightarrow \IT^2$ is given by $1$-forms
    \begin{subequations}\label{eq:def_nil_2}
        \begin{equation}
            A(x,y)\coloneqq k x\,\rmd y \in \Omega^1(\IR^2)~,
        \end{equation}
        which are local pullbacks of the global 1-form $\rmd z+kx\,\rmd y$ on $N_k$. This leads to the Kaluza--Klein metric
        \begin{equation}
            g(x,y,z) = \rmd x^2+\rmd y^2+(\rmd z+kx\,\rmd y)^2
        \end{equation}
        on the total space of $N_k$. Moreover, the gerbe $\scG_\ell$ over $N_k$ is endowed with a connective structure given by the 2-form connection and the 1-form 
        \begin{equation}
            \begin{aligned}
                B(x,y,z) &= \ell x\,\rmd y\wedge\rmd z\in \Omega^2(\IR^3)~,
                \\
                \Lambda(x,y,z;x',y',z') &=\ell(x-x')y \rmd z\in \Omega^1(\IR^3\times\IR^3)~,
            \end{aligned}
        \end{equation}
        and these data satisfy the cocycle conditions~\eqref{eq:adjusted_cocycles} for the 2-group $\sfB\sfU(1)=(\sfU(1)\rightarrow *)$ in additive notation. The curvature of the gerbe $\scG_\ell$ is the image of its Dixmier--Douady class in de~Rham cohomology,
        \begin{equation}
            H=\ell \rmd x\wedge \rmd y\wedge \rmd z~.
        \end{equation}
    \end{subequations}    
    
    Let us now describe the above from our perspective, using a principal $\sfTD^\ltimes_n$-bundle with connection. There is only one T-duality direction, and hence $n=1$. We thus consider a principal $\sfTD^\ltimes_1$-bundle $\scP_\rmC$ over $\IT^2$ subordinate to the submersion $\IR^2\rightarrow\IT^2$, and the cocycles~\eqref{eq:TD-top-cocycles} specialize as follows:
    \begin{subequations}\label{eq:nilmanifold_cocycle}
        \begin{equation}
            \begin{aligned}
                g&=\begin{pmatrix}
                    \hat g,~\hat \xi
                    \\
                    \check g,~\check \xi
                \end{pmatrix}~,~~~\begin{aligned}
                    \hat g(x,y;x',y') &= \unit~, ~~~&\hat \xi(x,y;x',y') &= \ell(x'-x)y~, \\
                    \check g(x,y;x',y') &= \unit~,~~~&\check \xi(x,y;x',y') &= k(x'-x)y~,
                \end{aligned}
                \\
                m&= \begin{pmatrix}
                    \hat m
                    \\
                    \check m
                \end{pmatrix}
                ~,~~~
                \begin{aligned}
                    \hat m(x,y;x',y';x'',y'') &= -\ell(x''-x')(y'-y)~,\\
                    \check m(x,y;x',y';x'',y'') &= -k(x''-x')(y'-y)~,
                \end{aligned}
                \\
                \phi(x,y;x',y';x'',y'') &= \tfrac12k\ell\left(y'(xx''-xx'-x'x'')-(x''-x')(y'^2-y^2)x\right)~.
            \end{aligned}
        \end{equation}
        We see that the structure group can be reduced to $\sfTD_1$. The differential refinement~\eqref{eq:diff_refined_cocycles} is given by 
        \begin{equation}
            \begin{aligned}
                A&=\begin{pmatrix}
                    \check A \\
                    \hat A
                \end{pmatrix}=
                \begin{pmatrix}
                    k x\,\mathrm dy
                    \\
                    \ell x\,\mathrm dy
                \end{pmatrix}~,
                \\
                B(x,y) &= 0~,
                \\
                \Lambda(x,y;x',y') &= \tfrac12k\ell(xx'\,\rmd y+(xy+x'y'+y^2(x'-x))\,\rmd x)~.
            \end{aligned}
        \end{equation}
    \end{subequations}    
    We note that under the flip homomorphism, the roles of $k$ and $\ell$ are interchanged, as expected.
    
    In order to recover the full T-backgrounds $(\check P=N_k,\check \scG=\check \scG_\ell)$ and $(\hat P=N_\ell,\hat \scG=\check \scG_k)$ directly from the cocycle~\eqref{eq:nilmanifold_cocycle}, we readily extract the cocycle data for the two circle bundles $N_k$ and $N_\ell$ described by $(\check \xi, \check m, \check A)$ and $(\hat \xi, \hat m, \hat A)$, respectively. The gerbe cocycles are also extracted from~\eqref{eq:nilmanifold_cocycle} using formulas~\eqref{eq:recovering_gerbe_data} and~\eqref{eq:recovery_differential_refinement}. In the case of the gerbe $\scG_\ell$ over $N_k$, we obtain the 2-form
    \begin{equation}
        B=\langle A,\rmd \hat t\rangle=kx\,\rmd y\wedge \rmd \hat t~.
    \end{equation}
    Identifying $\hat t$ with $z$, this potential 2-form has curvature
    \begin{equation}
        H=k\rmd x\wedge \rmd y\wedge \rmd z~,
    \end{equation}
    which is indeed the curvature of the gerbe $\hat \scG_k$ on $\hat P=N_\ell$.
    
    Altogether, we find that our proposal for differentially refined T-duality agrees with all expectations in this example.
    
    Note that for the topological part, we also have a span of principal 2-bundles in this case. Since no such example was given in~\cite{Nikolaus:2018qop}, let us briefly list the details here. The image of the projection $\check \sfp\colon\scP_\rmC\rightarrow \check \scP$ in~\eqref{eq:geometric_T_duality_span} is obtained from formula~\eqref{eq:formula_geometric_pi-check}. It is given by the $\sfTB^{\rm F2}_n$-bundle $\check P$ over $\IT^2$ which, subordinate again to the surjective submersion $\IR^2\rightarrow \IT^2$, is described by the cocycle
    \begin{equation}\label{eq:top_part_nilmanifold_higher_cocycle}
        \begin{aligned}
            f(x,y;x',y';x'',y'') &=\Big(c\mapsto\tfrac12k\ell\left(y'(xx''-xx'-x'x'')-(x''-x')(y'^2-y^2)x\right)\\
            &\hspace{3cm}-\ell(x''-x')(y'-y)c\Big)~,
            \\
            m(x,y;x',y';x'',y'') &=-k(x''-x')(y'-y)~,
            \\
            \xi(x,y;x',y') &=k(x'-x)y~.
        \end{aligned}
    \end{equation}
    On the other hand, we have the projection $\hat \sfp\colon\scP_\rmC\rightarrow \hat \scP$ in~\eqref{eq:geometric_T_duality_span}, whose image is 
    \begin{equation}
        \begin{aligned}
            f(x,y;x',y';x'',y'') &=\Big(c\mapsto\tfrac12k\ell\left(y'(xx''-xx'-x'x'')-(x''-x')(y'^2-y^2)x\right)\\
            &\hspace{3cm}-k(x''-x')(y'-y)c+k\ell(x'-x)y(x''-x')y'\Big)~,
            \\
            m(x,y;x',y';x'',y'') &=-\ell(x''-x')(y'-y)~,
            \\
            \xi(x,y;x',y') &=\ell(x'-x)y~.
        \end{aligned}
    \end{equation}

    \section{T-duality with T-folds and principal 2-groupoid bundles}\label{sec:T-folds}
    
    Next, we come to the generalization of our geometric framework for T-duality to mildly non-geometric T-backgrounds, namely T-folds~\cite{Hull:2004in}, i.e.~T-backgrounds which are locally geometric but globally glued together by general elements of the T-duality group. In this section, we construct the 2-groupoid $\scTD_n$ which is a minimal extension of the 2-group $\sfTD_n$, trivially regarded as a one-object 2-groupoid, with expected properties. We then show that the corresponding principal $\scTD_n$-groupoid bundles capture T-folds.
    
    \subsection{The Lie groupoid \texorpdfstring{$\scTD_n$}{TDn}}\label{ssec:def_TDn_groupoid}
    
    Recall that in our sketch of the origin of the 2-group $\underline{\sfTD}_n$ in \ref{ssec:origins_of_TDn}, the dimensional reduction led to a Lie 2-groupoid whose manifold of objects is the target space of the additional scalar fields. These have a 1-form field strength and correspond to $0$-branes from a string theory perspective. The physical expectation is that these scalar fields take values in the Narain moduli space~\cite{Narain:1985jj}, the moduli space of the Riemannian metric and the Kalb--Ramond $B$-field on $\IT^n$. This space is given by
    \begin{equation}
        M_n=\sfO(n,n;\IZ)~\backslash~\sfO(n,n;\IR)~/~\big(\sfO(n;\IR)\times \sfO(n;\IR)\big)=\sfO(n,n;\IZ)~\backslash~Q_n~,
    \end{equation}
    with dimension $n^2$, where
    \begin{equation}
        Q_n\coloneqq \sfO(n,n;\IR)~/~\big(\sfO(n;\IR)\times \sfO(n;\IR)\big)~,
    \end{equation}
    and all group actions are given by left and right multiplication, regarding all groups as matrix groups and embedding $\sfO(n;\IR)\times \sfO(n;\IR)$ block-diagonally into $\sfO(n,n;\IR)$. As argued above, we have to replace the T-duality group $\sfO(n,n;\IZ)$ by $\sfGO(n,n;\IZ)$ to allow for general torus bundles. Correspondingly, we will work with the scalar orbifold
    \begin{equation}
        GM_n=\sfGO(n,n;\IZ)~\backslash~\sfO(n,n;\IR)~/~\big(\sfO(n;\IR)\times \sfO(n;\IR)\big)= \sfGO(n,n;\IZ)~\backslash~Q_n~.
    \end{equation}
    
    We now want to construct the simplest Lie 2-groupoid $\scTD_n$, which combines the Narain moduli space with the 2-group $\underline{\sfTD_n}$. To this end, we start by replacing the quotient $GM_n$ by its action groupoid,
    \begin{equation}\label{eq:action_groupoid}
        \sfGO(n,n;\IZ)\ltimes Q_n~\rightrightarrows~Q_n~.
    \end{equation}
    This description has the advantage that we have a nice manifold of objects: while $GM_n$ generically has non-trivial 1-cycles, $Q_n$ is contractible\footnote{This is due to $\sfO(n;\IR)\times \sfO(n;\IR)$ being a maximal compact subgroup of $\sfO(n,n;\IR)$; the topology of a simple Lie group is essentially that of its maximal compact subgroup.}. 
    
    The action groupoid~\eqref{eq:action_groupoid} also suggests a simple combination with the 2-group $\sfTD_n$: enrich the morphisms in~\eqref{eq:action_groupoid} by a factor of $\underline{\sfTD_n}$, extend $\sfGO(n,n;\IZ)$ to $\scGO(n,n;\IZ)$ and have this 2-group act diagonally on both $Q_n$ and the additional $\sfTD_n$-factor in the fibers. The result is the following. 
    \begin{definition}
        The Lie 2-groupoid $\scTD_n$ has the following 2-, 1-, and 0-cells:
        \begin{equation}
            \begin{aligned}
                (\scTD_n)_2&=\sfGO(n,n;\IZ)\times \IZ^{2n}\times\IR^{2n}\times \IZ^{2n}\times \sfU(1)\times Q_n~,
                \\
                (\scTD_n)_1&=\sfGO(n,n;\IZ)\times \IR^{2n}\times Q_n~,
                \\
                (\scTD_n)_0&=Q_n~.
            \end{aligned}
        \end{equation}
        The 2- and 1-morphisms read as
        \begin{equation}
            \begin{gathered}
                (g,\xi,q)\xLeftarrow{~(g,z,\xi,m,\phi,q)~}(g,\xi-m,q)~,
                \\
                (q)\xleftarrow{~(g,\xi,q)~}(g^{-1}\acton q)~,
            \end{gathered}
        \end{equation}
        and they compose vertically and horizontally according to
        \begin{equation}
            (g,z_1,\xi,m_1,\phi_1,q)\circ(g,z_2,\xi-m_1,m_2,\phi_2,q)\coloneqq(g,z_1+z_2,\xi,m_1+m_2,\phi_1+\phi_2,q)
        \end{equation}
        and\footnote{Here, $|-|$ denotes again the indicator function on $\sfGO(n,n;\IZ)$ introduced in \ref{ssec:T-duality_group}.}
        \begin{equation}
            \begin{aligned}
                &(g_1,z_1,\xi_1,m_1,\phi_1,q)\otimes(g_2,z_2,\xi_2,m_2,\phi_2,g_1^{-1}\acton q)
                \\
                &\hspace{2cm}\coloneqq\Big(g_1g_2,z_1+g_1z_2,\xi_1+g_1\xi_2,m_1+g_1m_2,
                \\
                &\hspace{3cm}\phi_1+(-1)^{|g_1|}\phi_2-\langle \xi_1,g_1m_2\rangle+m_2^\rmT \rho_L(g_1)\xi_2+z_1^\rmT \eta g_1\xi_2,q\Big)~.
            \end{aligned}
        \end{equation}
        Horizontal composition is unital, but not associative, and we have the associator
        \begin{equation}
            \begin{aligned}
                &\sfa(g_1,\xi_1,q_1;g_2,\xi_2,q_2;g_3,\xi_3,q_3)
                \\
                &\hspace{1cm}=\Big(\sfa(g_1,g_2,g_3),(\sfid_{\xi_1}\otimes\Upsilon_{\sfTD_n}^{-1}(g_1,\xi_2,g_2\xi_3))
                \\
                &\hspace{6cm}
                \circ(\sfid_{\xi_1}\otimes\sfid_{g_1\xi_2}\otimes \Upsilon_{\scGO(n,n;\IZ)}(g_1,g_2,\xi_3)),q_1\Big)
                \\
                &\hspace{1cm}=\Big(\sfa(g_1,g_2,g_3),
                (\xi_1+g_1(\xi_2+g_2\xi_3),0,\xi_2^\rmT \rho_L(g_1)g_2\xi_3)
                \\
                &\hspace{3cm}\circ(\xi_1+g_1\xi_2+g_1g_2\xi_3,0,\tfrac12\xi_3^\rmT \sigma_L(g_1,g_2)\xi_3+\tfrac12\rmdiag(\sigma_L(g_1,g_2))^\rmT \xi_3),q_1\Big)
                \\
                &\hspace{1cm}=\Big(\sfa(g_1,g_2,g_3),\xi_1+g_1(\xi_2+g_2\xi_3),0,
                \\
                &\hspace{4cm}\xi_2^\rmT \rho_L(g_1)g_2\xi_3+\tfrac12\xi_3^\rmT \sigma_L(g_1,g_2)\xi_3+\tfrac12\rmdiag(\sigma_L(g_1,g_2))^\rmT \xi_3),q_1\Big)~,            
            \end{aligned}
        \end{equation}
        where $\sfa(g_1,g_2,g_3)$ is the associator in $\scGO(n,n;\IZ)$ defined in~\eqref{eq:assoc_GOnnZ}, and $q_{i+1}=g_i^{-1} q_i$, cf.~\eqref{eq:associator_semidirect_product}. 
    \end{definition}
    
    A few remarks on the geometric interpretation of $\scTD_n$ are in order.
    \begin{remark}
    	The Lie 2-groupoid $\scTD_n$ is equivalent to a bundle of 2-groups with fiber $\sfTD_n$ on the orbifold $Q_n/\sfGO(n,n;\IZ)$, up to the additional copy of $\IZ^{2n}$ in $(\scTD_n)_2$. Around a non-contractible cycle labeled by $g\in\sfGO(n,n;\IZ)$, the fiber $\sfTD_n$ undergoes a monodromy given by $g$. This construction is the direct categorified analogue of a bundle of groups on an orbifold $\Sigma/\Gamma$. Such bundles arise, for example, in Yang--Mills-matter theories with gauge group $\sfG$ with a scalar field taking values in a manifold $\Sigma$, for which a discrete subgroup $\Gamma\subset\sfAut(\sfG)$ that acts on $\Sigma$ has been gauged.
    \end{remark}
    
    \begin{remark}
        In our Lie 2-groupoid $\scTD_n$, the T-duality group $\sfGO(n,n;\IZ)$ appears explicitly on par with the gauge group $\IR^{2n}/\IZ^{2n}$. The T-duality group therefore is to be regarded as a gauge group. Because the group is discrete, there are no associated gauge potentials, but there are associated 2-groupoid bundle isomorphisms, effectively quotienting the space of inequivalent principal 2-groupoid bundles, while at the same time giving rise to new, topologically non-trivial bundles.        
    \end{remark}
    
    \begin{remark}\label{rem:contained_2-group}
        The groupoid $\scTD_n$ is still an action groupoid for the 2-group $\underline{\widehat{\sfTD}_n}$ with underlying groupoid
        \begin{equation}
            \underline{\widehat{\sfTD}_n}=\left(\sfGO(n,n;\IZ)\times \IZ^{2n}\times \IR^{2n}\times \IZ^{2n}\times \sfU(1)\rightrightarrows \sfGO(n,n;\IZ)\times \IR^{2n}\right)
        \end{equation}
        and monoidal product etc.~obtained from restricting the structures on $\scTD_n$. The 2-group $\underline{\widehat{\sfTD}_n}$ is essentially a higher semidirect product, $\underline{\widehat{\sfTD}_n}=\scGO(n,n;\IZ)\ltimes \underline{\sfTD_n}$, with the action $\scGO(n,n;\IZ)\curvearrowright\underline{\sfTD_n}$ given in \ref{thm:action_GO_on_TD}. 
        
        Because of this, there is an evident action of $\scGO(n,n;\IZ)$ on $\underline{\widehat{\sfTD}_n}$, and, by extending the action encoded in the action groupoid $\scTD_n$, on the total Lie 2-groupoid $\scTD_n$ via the adjoint action of $\scGO(n,n;\IZ)$ on itself. In particular, there is an action of the flip morphism~\eqref{eq:flip-plain} on $\scTD_n$.
    \end{remark}
    
    \subsection{Principal \texorpdfstring{$\scTD_n$}{TDn}-bundles with adjusted connections}
    
    Next, we develop the cocycle description of a principal $\scTD_n$-bundle. Recall that the cocycles describing principal bundles can be defined in terms of functors, and this definition readily extends.
    \begin{definition}
        Let $M$ be a manifold and $\sigma\colon Y\rightarrow M$ a surjective submersion, and let $\check\scC(Y\rightarrow M)$ be the corresponding Čech groupoid, cf.~\eqref{eq:def_Cech_groupoid}, trivially regarded as a (strict) higher Lie groupoid. Let $\scG$ be a (higher) Lie groupoid, which we call the \uline{structure groupoid}. A \uline{(higher) groupoid bundle} over $M$ subordinate to the surjective submersion $\sigma$ with structure groupoid $\scG$ is then a (higher) functor 
        \begin{equation}
            \Phi\colon\check\scC(Y\rightarrow M)\rightarrow \scG~.
        \end{equation}
        Groupoid bundle isomorphisms are given by (higher) natural transformations between two such functors, and the higher isomorphisms are then identified with modifications and higher transfors. 
    \end{definition}
    \noindent Note that a groupoid bundle whose structure groupoid is the delooping $\sfB\sfG$ of a Lie group $\sfG$ is just a principal $\sfG$-bundle, a higher groupoid bundle with structure groupoid $\sfB\sfB\sfU(1)$ is an abelian gerbe, etc.
    
    The definition of higher functors for Lie $n$-groupoids for $n>2$ is technically very involved, and it is a good idea to switch to the perspective of quasi-groupoids defined in terms of Kan simplicial manifolds, cf.~\ref{app:quasi-groupoids} and e.g.~\cite{Jurco:2016qwv}. The definition of a differential refinement also becomes technically involved.
    
    We remark that from a physical perspective, ordinary groupoid bundles are the geometric structures underlying gauged sigma models. We thus deal with the kinematical data of a higher gauged sigma model, as expected, in particular, from the perspective of supergravity.
    
    The specialization to principal Lie 2-groupoid bundles with structure groupoid $\scTD_n$ is now straightforward; it is a groupoid extension of the discussion in~\cite{Jurco:2014mva}. A principal $\scTD_n$-bundle over a manifold $M$ subordinate to the surjective submersion $\sigma\colon Y\rightarrow M$ is a weak 2-functor\footnote{see~\ref{app:2-groupoid_basics} for definitions} $\Phi\colon\check\scC(Y\rightarrow M)\rightarrow \scTD_n$. Such a functor is encoded in the data
    \begin{equation}
        \begin{aligned}
            (g,z,\xi,m,\phi,q)
            &\in C^\infty(Y^{[3]},\sfGO(n,n;\IZ)\times \IZ^{2n}\times\IR^{2n}\times \IZ^{2n}\times \sfU(1)\times Q_n)~,
            \\
            (g,\xi,q)&\in C^\infty(Y^{[2]},\sfGO(n,n;\IZ)\times \IR^{2n}\times Q_n)~,
            \\
            q&\in C^\infty(Y,Q_n)~,
        \end{aligned}
    \end{equation}
    which define the natural isomorphism $\Phi_2$, the functor $\Phi_1$, and the function $\Phi_0$, respectively. On $Y^{[2]}$, we then have 
    \begin{equation}
        q_j=g_{ij}^{-1}\acton q_i
    \end{equation}
    with $(ij)\in Y^{[2]}$, while on $Y^{[3]}$, we deduce that 
    \begin{equation}
        e_{ijk}\circ (\sfid_{d_{ij}}\otimes \sfid_{d_{jk}})=\sfid_{d_{ik}}\circ e_{ijk}
    \end{equation}
    for $d_{ij}=(g_{ij},\xi_{ij},q_{i})$, $e_{ijk}=(g_{ik},z_{ijk},\xi_{ik},m_{ijk},\phi_{ijk},q_{i})$, and $(ijk)\in Y^{[3]}$ or, equivalently,
    \begin{equation}\label{eq:groupoid_cocycle_relations_Y3}
        \begin{aligned}
            g_{ik}&=g_{ij}g_{jk}~,
            \\
            \xi_{ik}&=m_{ijk}+\xi_{ij}+g_{ij}\xi_{jk}~.
        \end{aligned}
    \end{equation}    
    Finally, on $Y^{[4]}$, we have
    \begin{equation}
        e_{ikl}\circ (e_{ijk}\otimes \sfid_{d_{kl}})=e_{ijl}\circ (\sfid_{d_{ij}}\otimes e_{jkl})\circ \sfa(d_{ij},d_{jk},d_{kl})
    \end{equation}
    or, equivalently,
    \begin{equation}
        \begin{aligned}
            z_{ijk}+z_{ikl} &= z_{ijl}+g_{ij}z_{jkl}
            \\
            &\hspace{0.5cm}+\frac{(-1)^{|g_{ik}|}}{2}g_{ik}\eta\,\rmdiag(\sigma_L(g_{ij},g_{jk}))
            +\frac{(-1)^{|g_{il}|}}{2}g_{il}\eta\,\rmdiag(\sigma_L(g_{ik},g_{kl}))
            \\
            &\hspace{0.5cm}-\frac{(-1)^{|g_{il}|}}{2}g_{il}\eta\,\rmdiag(\sigma_L(g_{ij},g_{jl}))
            -\frac{(-1)^{|g_{jl}|}}{2}g_{il}\eta\,\rmdiag(\sigma_L(g_{jk},g_{kl}))~,
            \\
            m_{ijk}+m_{ikl} &= m_{ijl}+g_{ij}m_{jkl}~,
            \\
            \phi_{ijk} + \phi_{ikl} &=\phi_{ijl} + (-1)^{|g_{ij}|}\phi_{jkl}-\langle \xi_{ij}, g_{ij}m_{jkl}\rangle+
            m_{jkl}^\rmT \rho_L(g_{ij})\xi_{jl} +
            \xi_{jk}^\rmT \rho_L(g_{ij})g_{jk}\xi_{kl}
            \\
            &\hspace{0.5cm}+\tfrac12 \xi_{kl}^\rmT \sigma_L(g_{ij},g_{jk})\xi_{kl}+\tfrac12\rmdiag(\sigma_{L}(g_{ij},g_{jk}))^\rmT \xi_{kl}-z_{ijk}^\rmT \eta g_{ik}\xi_{kl}~,
        \end{aligned}
    \end{equation}
    where the functions $\rho_L$ and $\sigma_{L}$ were defined in~\eqref{eq:def_rho_L} and~\eqref{eq:def_sigma_L}, respectively. Note that the second equation is automatically satisfied due to~\eqref{eq:groupoid_cocycle_relations_Y3}.
    
    Let us now differentially refine this smooth\footnote{We note that the ``scalar part'' of a smooth groupoid bundle can already be considered as a part of the differential refinement. This is certainly more sensible from a physical perspective, where scalar fields arise from dimensionally reducing gauge potentials.} cocycle data, i.e.~add the datum of a higher connection. Generically, this is a difficult problem, see e.g.~the discussion for principal groupoid bundles~\cite{Fischer:2024vak}. Here, however, we are working with a higher action groupoid, cf.~\ref{rem:contained_2-group}, which makes the construction tractable. Therefore a principal $\scTD_n$-bundle is a principal $\underline{\widehat{\sfTD}_n}$-bundle, described by the cocycle data $(g,z,\xi,m,\phi)$, which, in turn, imposes a gluing condition on the local scalar fields $q$. As a consequence, the notion of adjustment trivially lifts from a principal $\underline{\widehat{\sfTD}_n}$-bundle to a principal $\scTD_n$-bundle.\footnote{We note, however, that following~\cite{Rist:2022hci}, it is not hard to establish a generalization of adjustment for connections on principal 2-groupoid bundles. It is simply not necessary here.}
    
    The notion of adjustment, which is required for a higher connection, is only available for structure 2-groups with trivial associator. For $\underline{\widehat{\sfTD}_n}$, we therefore have to go back to first principles. Recall from~\cite{Rist:2022hci} that an adjustment is a consistent deformation of the differentially refined coboundaries and cocycles such that globularity of the action groupoid is preserved\footnote{Concretely, in the case of principal 2-bundles, one has to require that higher gauge transformations do not affect the image of gauge transformations}. 
    
    A detailed search with a computer algebra system did not produce any suitable such deformation for the general case\footnote{We suspect that a more comprehensive framework as the one developed in~\cite{Fischer:2024vak} for 1-groupoids is required.}. We did, however, find such a deformation in the case in which the Čech cocycle, i.e.~the cocycle before differential refinement, satisfies the condition 
    \begin{subequations}\label{eq:diff_refined_cocycles2}
        \begin{equation}\label{eq:z-rel}
            z_{ijk}=\frac{(-1)^{|g_{ik}|}}{2}g_{ik}\eta\,\rmdiag(\sigma_L(g_{ij},g_{jk}))~.
        \end{equation}
        We leave the study of the mathematical reasons for this subtlety to future work. It will not be relevant in our further discussion: for the cases we are interested in, the condition~\eqref{eq:z-rel} is automatically satisfied. With the above constraint imposed, an adjusted cocycle contains the additional data
        \begin{equation}
            \Lambda\in \Omega^1(Y^{[2]})~,~~~
            A\in \Omega^1(Y,\IR^{2n})~,~~~
            B\in \Omega^2(Y)~,
        \end{equation}
        which satisfy the gluing relations
        \begin{equation}
            \begin{aligned}
                \Lambda_{ik} &= (-1)^{|g_{ij}|}\Lambda_{jk} + \Lambda_{ij}+\rmd \phi_{ijk}-\langle A_i,m_{ijk}\rangle
                \\
                &\hspace{1cm}+\tfrac12 \rmd \xi_{jk}\rho(g_{ij})\xi_{jk}-A_k^\rmT \eta g_{kj}\eta \rho(g_{ij})\xi_{jk}+\tfrac12\rmd\left(\xi_{jk}^\rmT \rho_L(g_{ij})\xi_{jk}\right)
                \\
                &= (-1)^{|g_{ij}|}\Lambda_{jk} + \Lambda_{ij}+\rmd \phi_{ijk}-\langle A_i,m_{ijk}\rangle
                \\
                &\hspace{1cm}+\rmd \xi_{jk}\rho_L(g_{ij})\xi_{jk}-A_k^\rmT \eta g_{kj}\eta \rho(g_{ij})\xi_{jk}~,
                \\
                A_j &= g_{ij}^{-1}A_i + g_{ij}^{-1}\rmd \xi_{ij}~,
                \\
                (-1)^{|g_{ij}|}B_j &= B_i + \rmd \Lambda_{ij} + \langle\rmd A_i,\xi_{ij}\rangle-\tfrac12A_j^\rmT \rho(g_{ij})A_j~.
            \end{aligned}
        \end{equation}
    \end{subequations}        
    We note that for $g_{ij}\in \sfGL(n;\IZ)\subset \sfO(n,n;\IZ)$, the relation~\eqref{eq:z-rel} is automatically satisfied for $z_{ijk}=0$, and we recover the cocycle relations for differentially refined principal $\sfTD^\ltimes_n$-bundles.
    
    \subsection{T-duality correspondences involving T-folds}
    
    We note that the flip morphism~\eqref{eq:flip-plain} acts on the cocycle data by postcomposing the 2-functor defining the principal $\scTD_n$-bundle with this flip morphism, acting on $\scTD_n$ as explained in \ref{rem:contained_2-group}. It is then tempting to see a $\scTD_n$-bundle as a correspondence and to try to reconstruct the left- and right-backgrounds, similarly to the construction in \ref{ssec:differential_refinement}. As we expect for T-folds, this is not always possible, but two statements can be made.
    
    \begin{proposition}
	   	A principal $\scTD_n$-bundle with adjusted connection defines locally a T-background. On overlaps of patches, the corresponding data is glued together by $\sfGO(n,n;\IZ)$-transformations, and thus we have a global T-fold.
    \end{proposition}
    \begin{proof}
    	It is clear that the local data of an $\scTD_n$-bundle is the same as the local data of a $\sfTD_n$-bundle plus the Narain moduli space. Locally, this data is therefore an ordinary T-background. At the global level, the local data is glued together on $Y^{[2]}$ in particular by the $g_{ij}\in \sfGO(n,n;\IZ)$, and such geometries are called T-folds.
    \end{proof}    
    
    The left T-background in the correspondence is still an ordinary T-background in the following special case:
    \begin{definition}\label{def:special_TD_n_bundle}
        A \uline{special principal $\scTD_n$-bundle} is a principal $\scTD_n$-bundle that is described by a cocycle in which all $(g_{ij})$ are purely $B$-transformations, i.e.~they are of the form
        \begin{equation}
	       	g_{ij}=\begin{pmatrix}
	       		\unit & b_{ij}
	       		\\
	       		0 & \unit
	       	\end{pmatrix}~,~~~b_{ij} \in \sfMat(n;\IZ)~,~~b_{ij}^\rmT=-b_{ij}~.
        \end{equation}
    \end{definition}
    For special $\scTD_n$-bundles, $\sigma_L(g_1,g_2)$ vanishes by \ref{lem:subgroupsOK} so that $z_{ijk}$ is put to zero by~\eqref{eq:z-rel}. Moreover, let us split the $q_i\in Q_n\cong \IR^{n^2}$ as 
    \begin{equation}
    	q_i=\left[
    	\begin{pmatrix}
    		\hat q_{i,1} & \hat q_{i,2}
    		\\
    		\check q_{i,1} & \check q_{i,2}
    	\end{pmatrix}
    	\right]~,
    \end{equation}
    where the square brackets denote equivalence classes of the $\sfO(n;\IR)\times\sfO(n;\IR)$ action on $\sfO(n,n;\IR)$. For special principal $\scTD_n$-bundles, we then have 
    \begin{equation}
    	\check q_{i,a} = \check q_{j,a}~.
    \end{equation}
    Similarly, the cocycle relations for the components $\check \xi_{ij}$, $\check m_{ijk}$, and $\check A_i$ in the cocycle data are precisely the same as those in the case of trivial $g_{ij}$.
    \begin{proposition}
    	Special $\scTD_n$-bundles define a correspondence between an ordinary T-background and a T-fold.
    \end{proposition}
    \begin{proof}
    	As the $\check q_{ia}$ and $\check A_i$ describing the left T-background are those of an ordinary $\sfTD_n$-bundle, they capture a geometric T-background. The right T-background, however, features $\sfGO(n,n;\IZ)$-valued transition functions and is therefore a T-fold.
    \end{proof}
    
    The above observations lead us to make the following proposal:
    
    \
    
    \emph{A T-duality between geometric T-backgrounds and T-folds of circle fibrations is described by a principal $\scTD_n$-bundle.}
    
    \

    Since there is no full theory of T-duality involving T-folds, we cannot fully verify our construction; we can merely show that it reproduces expected examples and that it comprises the half-geometric topological T-duality in the form described in~\cite{Nikolaus:2018qop}. We will do both in the following two sections.     

    \subsection{Example of a T-fold}
    
    Let us consider again the well-known example of the three-dimensional nilmanifold defined in~\eqref{eq:def_nil_1} and~\eqref{eq:def_nil_2} with $k=0$ and T-dualize along the $y$- and $z$-directions. In the T-correspondence, our base manifold $X$ is then simply the circle parameterized by $x$, and we consider a principal $\scTD_2$-bundle over $S^1$ subordinate to the cover $\IR\rightarrow S^1$. 
    
    Due to dimensionality, there can be no non-trivial triple intersections, and we can thus set $z_{ijk}=m_{ijk}=\phi_{ijk}=0$. Since principal torus bundles over a circle are topologically trivial, we can trivialize the $(\xi_{ij})$. This leaves us with the scalars $(q_i)\colon Y\to Q_2$ and the transition functions $(g_{ij})\colon Y^{[2]}\rightarrow \sfGO(n,n;\IZ)$ with the only non-trivial cocycle conditions being
    \begin{equation}
        g_{ij}\acton q_j=q_i\eand g_{ij}g_{jk}=g_{ik}
    \end{equation}
    for all $(ij)\in Y^{[2]}$ and $(ijk)\in Y^{[3]}$, together with the evenness condition
    \begin{equation}\label{eq:g-even}
        g_{ik}\eta\,\rmdiag(\sigma_L(g_{ij},g_{jk}))\in2\IZ^{2n}
    \end{equation}
    required for adjustment.
    Modulo this evenness condition, these data are the same as those defining a groupoid bundle over $S^1$ with structure groupoid the action groupoid corresponding to the action $\sfGO(2,2;\IZ)\curvearrowright Q_2$. This, in turn, is the same as a map of orbifolds $q\colon S^1\rightarrow GM_2$.
    
    Consider now the case of a special such $\scTD_2$-bundle. As observed above, \eqref{eq:g-even} holds automatically in this case. Moreover, $Q_2\coloneqq\sfO(2,2;\IR)/\sfO(2;\IR)^2\cong \IR^4$, and this manifold is identified with the four scalars arising from the dimensional reduction of the metric and the Kalb--Ramond field. Consider the first T-background in the key example of~\ref{ssec:non_geometric_backgrounds}, i.e.~a gerbe on the nilmanifold $N_0=T^3$ with metric and Kalb--Ramond field given by 
    \begin{equation}
        g(x,y,z)=\rmd x^2+\rmd y^2+\rmd z^2\eand B=\ell x \rmd y\wedge \rmd z~.
	\end{equation}
    The relevant scalar fields are now the metric components $g_{yy}=1$, $g_{yz}=0$, $g_{zz}=1$ as well as the component $B_{yz}=\ell x$. To identify the effect of successive T-duality transformations on these scalar fields $(q_i)$, we identify the space $Q_2$ with the space of generalized metrics~\eqref{eq:generalized_metric}, which becomes generically a local quantity, and thus functions on $\IR$. For our concrete example, we obtain
    \begin{equation}
        \caH_H(x)=\begin{pmatrix}
            1+B_{yz}^2 & 0 & 0 & B_{yz}
            \\
            0 & 1+B_{yz}^2 & -B_{yz} & 0
            \\
            0 & -B_{yz} & 1 & 0
            \\
            B_{yz} & 0 & 0 & 1
        \end{pmatrix}~,~~B_{yz}=\ell x~.
    \end{equation}
    The factorized duality $g^+_{T_1}$ and the flip transformation~\eqref{eq:flip-plain} $g^+_{T_2}\circ g^+_{T_1}$ produce the following generalized metrics:
    \begin{equation}
		\begin{aligned}
			\caH_f(x)&\coloneqq g^+_{T_1}\caH (g^+_{T_1})^T=
            \begin{pmatrix}
                1 & -B_{yz}^2 & 0 & 0 
                \\
                -B_{yz} & 1+B_{yz}^2 & 0 & 0
                \\
                0 & 0 & 1+B_{yz}^2 & B_{yz}
                \\
                0 & 0 & B_{yz} & 1
            \end{pmatrix}~,
            \\
			\caH_Q(x)&\coloneqq g^+_{T_2}\circ g^+_{T_1}\caH (g^+_{T_2}\circ g^+_{T_1})^T=
            \begin{pmatrix}
                1 & 0 & 0 & -B_{yz}
                \\
                0 & 1 & B_{yz} & 0
                \\
                0 & B_{yz} & 1+B_{yz}^2 & 0
                \\
                -B_{yz} & 0 & 0 & 1+B_{yz}^2
            \end{pmatrix}~.
        \end{aligned}
	\end{equation}
    Comparing with the general form~\eqref{eq:generalized_metric} of the generalized metric, we clearly identify the following situations: 
    \begin{itemize}
        \item[$\caH_H)$] The metric is trivial (from the lower right block), and the $B$-field is non-trivial (from the off-diagonal blocks), inducing a non-trivial $H$-flux. This describes the 3-torus $T^3$ with a gerbe of curvature $H=\ell \rmd x \wedge \rmd y \wedge \rmd z$.
        \item[$\caH_f)$] The $B$-field is trivial (vanishing off-diagonal blocks), and there is a non-vanishing gauge potential (the component $g_{yz}$ from the lower right block). This describes the nilmanifold with curvature $F=\ell \rmd x \wedge \rmd y$.
        \item[$\caH_Q)$] The metric is not globally defined, but $x$-dependent (lower right block). This describes a T-fold.
    \end{itemize}
    
    As parts $(q_x)$ of the $\scTD_2$-bundle cocycles, the local metrics are glued together by elements in $\sfGO(n,n;\IZ)$, e.g.~
    \begin{equation}
        \caH_Q(x+i)=g^{-1}_{x+i,x}\caH_Q(x)(g^{-1}_{x+i,x})^T
	\end{equation}
    with 
    \begin{equation}\label{eq:T-fold_transition}
        g_{x+i,x}=\begin{pmatrix}
            1 & 0 & 0 & 0 
            \\
            0 & 1 & 0 & 0 
            \\
            0 & -i\ell & 1 & 0
            \\
            +i\ell & 0 & 0 & 1
        \end{pmatrix}~,~~~i\in \IZ~,
    \end{equation}
    and the non-trivial cocycle relations 
    \begin{equation}
	g_{x+i+j,x}=g_{x+i+j,x+j}g_{x+j,x}
    \eand
    q_{x+i}=g_{x+i,x}^{-1} q_x	
	\end{equation}
    are indeed satisfied for $i,j\in \IZ$. As for all special $\scTD_n$-bundles, $\sigma_L(g_{ij},g_{jk})=0$, and hence the condition~\eqref{eq:diff_refined_cocycles2} for the existence of a differential refinement is satisfied for $z_{ijk}=0$.
    
    The $\scTD_2$-bundle encoded in the cocycle data
    \begin{equation}
        (g_{ij},z_{ijk},\xi_{ij},m_{ijk},\phi_{ijk},q_i,\Lambda_{ij},A_i,B_i)=(g_{x+i,x},0,0,0,0,\caH_Q(x),0,0,0)
	\end{equation}
    thus describes the T-duality with the left T-background being the torus with $H$-flux, and the right T-background being the expected T-fold with the scalar fields in $\caH_Q(x)$ glued together by the $\beta$-trans\-formation~\eqref{eq:T-fold_transition}. 
    
    We can also introduce a third projection by considering only a single factorized duality $T_y^+$ or $T_z^+$. In these cases, we also recover the nilmanifold with $f$-flux, as expected. Altogether, we have nicely reproduced the first two T-dualities in the key example of \ref{ssec:non_geometric_backgrounds}, which gives us confidence in our proposal.
    
    \subsection{Half-geometric T-correspondences as principal \texorpdfstring{$\sfTD^{\text{$\tfrac12$geo}}_n$}{TDnhalfgeometric}-bundles}
    
    A second piece of evidence justifying our proposal is that it contains the principal 2-bundles used in~\cite{Nikolaus:2018qop} to capture the topological part of half-geometric T-dualities. We start with a brief review of this construction.
    
    We follow the nomenclature of~\cite{Nikolaus:2018qop} and call a T-correspondence involving a geometric T-background of type $F^1$ and a T-fold background \emph{half-geometric}. Topological T-backgrounds of type $F^1$ have been shown to correspond to functors represented by the 2-group $\sfTB^\text{F1}_n$~\cite{Nikolaus:2018qop}. This 2-group is given by the semidirect product of the strict 2-group $\sfTB^\text{F2}_n$ with the abelian subgroup of $\beta$-transformations $\fro(n;\IZ)\subset \sfGO(n,n;\IZ)$,
    \begin{equation}
        \sfTB^\text{F1}_n\coloneqq \fro(n;\IZ)\ltimes \sfTB^\text{F2}_n~.
    \end{equation}
    For the definition of the action and the resulting semidirect product, see~\cite[App.~A.4]{Nikolaus:2018qop}; the corresponding cocycles of principal $\sfTB^\text{F1}_n$-bundles (without connection) were also given in~\cite{Nikolaus:2018qop}.
    
    A half-geometric T-correspondence can then be described by a principal $\sfTD^{\text{$\tfrac12$geo}}_n$-bundle, where the structure 2-group is defined as 
    \begin{equation}\label{eq:TD_half_geo}
        \sfTD^{\text{$\tfrac12$geo}}_n\coloneqq \frso(n;\IZ)\ltimes \sfTD_n~.
    \end{equation}
    Note that this semidirect product makes sense because of \ref{rem:action_of_subgroups}. The left-leg projection is equivariant in a particular sense and induces a map $\hat p\colon\sfTD^{\text{$\tfrac12$geo}}_n\rightarrow \sfTB^\text{F1}_n$. The main result of~\cite{Nikolaus:2018qop} is that this map is a bijection and that every $F^1$-background is the image of the left-leg projection of a principal $\sfTD^{\text{$\tfrac12$geo}}_n$-bundle. We now observe the following.
    
    \begin{proposition}\label{prop:embed_half_geometric}
        The structure 2-group $\sfTD^{\text{$\tfrac12$geo}}_n$ is a sub-2-group of the 2-group $\underline{\widehat{\sfTD}_n}$ contained in the 2-groupoid $\scTD_n$, as explained in \ref{rem:contained_2-group}. A principal $\sfTD^{\text{$\tfrac12$geo}}_n$ is then a special principal $\scTD_n$-bundle (without differential refinement, in particular vanishing scalar field $q$) in the sense of~\ref{def:special_TD_n_bundle}.
    \end{proposition}
    \begin{proof}
        Both facts are obvious from a direct comparison between the 2-group structure of $\sfTD^{\text{$\tfrac12$geo}}_n$ as defined in~\cite{Nikolaus:2018qop} and the monoidal product on $\scTD_n$. The vector space $\frso(n;\IZ)$ should be identified with $\beta$-transformations in $\scGO(n,n;\IZ)$, and it is therefore clear that principal $\sfTD^{\text{$\tfrac12$geo}}_n$-bundles are special principal $\scTD_n$-bundles.
    \end{proof}
    
    Together with~\cite[Thm.~4.2.2]{Nikolaus:2018qop}, this shows that our construction provides the freedom to capture the topological part of any $F^1$-T-background, and there is a left-leg projection from principal $\sfTD^{\text{$\tfrac12$geo}}_n$-bundles to principal $\sfTB^\text{F1}_n$-bundles, cf.~again~\cite{Nikolaus:2018qop}. Conversely, however, it seems clear that a differential refinement of an $F^1$-background naturally induces scalar fields; in the presence of these, the 2-group $\sfTD^{\text{$\tfrac12$geo}}_n$ is not large enough to accommodate all required transformations.

    \section{T-duality with scalar fluxes}\label{sec:non_geometric}
    
    It is now tempting to take the evident next step and try to generalize our proposal to the case of fully non-geometric $R$-spaces. This, however, does not work, as our description is based on principal 2-group(oid) bundles, while T-dualities involving $R$-spaces should have pairs of nilmanifolds in the fibers that do not admit a group structure. 
    
    Still, there is an evident next step, namely the incorporation of scalar fluxes, arising from the complete Kaluza--Klein reduction of a 3-form $H$. We will explore this extension in this section.
    
    \subsection{Kaluza--Klein reduction leads to the tensor hierarchy}
    
    \paragraph{Tensor hierarchy.} A geometric T-background is the kinematical datum for a supergravity action. There are now a number of physical reasons why Kaluza--Klein reductions lead to gauged supergravities. In these theories, a subgroup $\sfK$ of the symmetry group $\sfG$ of the space $\sfG/\sfH$, in which the scalar fields take values (in our case, the quotient space $Q_n$), is ``gauged,'' i.e.~promoted from a global to a local symmetry by adding a principal $\sfK$-bundle with connection to the kinematical datum and assuming the fields to take values in certain associated vector bundles. The subgroup $\sfK$ as well as its action is defined by a datum called the \emph{embedding tensor} (see below for further details). 
    
    Usually, this construction does not end here, but higher form fields forming the connection of a higher principal bundle are incorporated. In the supergravity literature, cf.~\cite{Samtleben:2008pe}, this is usually justified by the fact that the Bianchi identities (and, correspondingly, gauge transformations) are not covariant and require iterative corrections by higher form fields. More precisely, the combination of the new gauge potential with the connection on the gerbe naturally leads to a differential graded Lie algebra, which in turn produces a particular higher Lie algebra equipped with an adjustment datum, cf.~\cite{Borsten:2021ljb}. 
    
    Importantly for us, the non-geometric fluxes essentially define (at least parts of) the embedding tensor~\cite{Aldazabal:2011nj,Geissbuhler:2011mx,Grana:2012rr}. A general tensor hierarchy for double field theory was then defined in~\cite{Hohm:2013nja}. We note that also~\cite{Alfonsi:2020nxu} argues that the gauge potential forms arising in a Kaluza--Klein interpretation of double field theory should be arranged into a tensor hierarchy. All of this strongly suggests that in order to introduce interesting scalar fluxes, we should incorporate the embedding tensor into our picture.
    
    \paragraph{Algebraic structure underlying tensor hierarchies.} Let us briefly summarize the algebraic structure underlying the tensor hierarchies of gauged supergravity~\cite{deWit:2002vt,Samtleben:2008pe} as explained in~\cite{Borsten:2021ljb,Saemann:2019dsl}. In its simplest form, we start from a global symmetry group $\sfG$ with Lie algebra $V_0=\frg$ together with a representation $V_{-1}$. The \emph{embedding tensor} is a linear map
    \begin{equation}
        \Theta\colon V_{-1}\rightarrow V_0
    \end{equation}
    satisfying
    \begin{equation}\label{eq:closure_constraint}
        [\Theta(v_1),\Theta(v_2)]=\Theta(\Theta(v_1)v_2)~,
    \end{equation}
    which identifies a Lie subalgebra $\frh=\rmim(\Theta)$. We can package these structures into a differential graded Lie algebra
    \begin{equation}
        V_\Theta=( V_{-1}\xrightarrow{~\Theta~}V_0)~.
    \end{equation}
    By~\cite[Prop.~4.1]{Borsten:2021ljb}, this differential graded Lie algebra can be promoted to a weak Lie 2-algebra, which allows for an adjustment by~\cite[Thm.~6.2]{Borsten:2021ljb}. We note that the representation $V_{-1}$ may be enlarged to a graded vector space, giving rise to more general tensor hierarchies with underlying adjusted weak Lie $n$-algebras. This graded vector space is the structure $L_\infty$-algebra of an underlying higher principal bundle with connection.
    
    Besides the \emph{quadratic closure constraint}~\eqref{eq:closure_constraint}, there is also a \emph{representation constraint} that one imposes on the embedding tensor. This additional constraint can be seen as a requirement for supersymmetry~\cite{deWit:2002vt,deWit:2005hv}, for the locality of the action~\cite{deWit:2005ub}, or for anomaly cancellation~\cite{DeRydt:2008hw}. While supersymmetry or chiral fields that could cause anomalies do not directly appear in the present paper, it will turn out that the representation constraint is nevertheless necessary here to obtain the correct set of $R$-fluxes.
    
    \subsection{Identifying interesting spaces of scalar fluxes}
    
    We now specialize the above general picture to our situation, i.e.~$\sfG=\sfO(n,n;\IR)$, the isometry group of the scalar manifold $Q_n$, and $V_{-1}=\IR^{2n}$, the space in which the gauge potential 1-forms take values.  Correspondingly, the embedding tensor is a map $\Theta\colon\IR^{2n}\rightarrow \fro(n,n;\IR)$, which exponentiates to a map
    \begin{equation}
        \bar\Theta\colon\IR^{2n}\rightarrow \sfO(n,n;\IR)~.
    \end{equation}
    The map $\Theta$ needs to satisfy the closure constraint~\eqref{eq:closure_constraint} as well as the representation constraint. We note that a generic map $\Theta$ is an element in the tensor product of the fundamental and the adjoint representation of $\fro(n,n;\IR)$, which decomposes as follows in terms of Young diagrams:
    \begin{equation}
        \tyng(1,1) \otimes \tyng(1) = \tyng(1)\oplus \tyng(1,1,1) \oplus \tyng(2,1)~.
    \end{equation}
    Usually, the representation constraint is selected by requiring supersymmetry. Here, we have several strong heuristic reasons to impose the condition
    \begin{equation}\label{eq:representation_constraint}
        \Theta\in \tyng(1,1,1)
    \end{equation}
    which we present now. First, recall that $R$-flux corresponds to the 3-form $H$-flux wrapped around three of the $2n$ directions in the $2n$ torus fiber directions, subject to additional constraints reflecting the fact that the $2n$ coordinates cannot be regarded as geometric simultaneously. We therefore expect the totally antisymmetric representation. 
    
    Furthermore, we expect the $R$-form fluxes for $(n+1)$-dimensional torus fibers to originate from the $R$- and $Q$-fluxes of $n$-dimensional torus fibers under KK reduction, and this branching rule essentially fixes~\eqref{eq:representation_constraint}. We then have the following association of fluxes to $\sfGO(n,n;\IR)$-representations:
    \begin{equation}
        \mbox{$f$-flux}~\leftrightarrow~\tyng(1)~,~~~\mbox{$Q$-flux}~\leftrightarrow~\tyng(1,1)~,~~~\mbox{$R$-flux}~\leftrightarrow~\tyng(1,1,1)~.
    \end{equation}
    Here, ``$f$-flux'' refers to the curvatures of $\hat A$ and $\check A$, taking values in $\IZ^{2n}$. It corresponds to $H$ wrapped around one of the $2n$ directions in the $2n$-torus fibers and forms the fundamental representation, while the $Q$-flux corresponds to $H$ wrapped around two of the $2n$ directions in the $2n$-torus fibers and, hence, to the adjoint representation, namely the linearization of the non-linear adjoint representation of $\sfO(n,n;\mathbb R)$ on itself.
    
    A third argument for our choice~\eqref{eq:representation_constraint} comes from considering the embedding of T-duality into U-duality, for which supersymmetric arguments fully determine the representations. For the embedding $\sfE_{7(7)}\supset\sfSO(6,6)\times\sfSL(2)$, the representation $\mathbf{912}$ of the embedding tensor for $\sfE_{7(7)}$ decomposes as
    \begin{equation}
        \textbf{912} \to (\textbf{12},\textbf2)\oplus (\textbf{220},\textbf2)\oplus \dotsb~,
    \end{equation}
    and we only find the representations
    \begin{equation}
        \tyng(1,1,1)=\textbf{220}\eand \tyng(1)=\textbf{12}
    \end{equation}
    of $\sfSO(6,6)$, but not
    \begin{equation}
        \tyng(2,1)=\textbf{560}~.
    \end{equation}
    
    There are two more conditions that we need to impose on the exponentiated embedding tensor $\bar\Theta$. First, the $B$-field part of the connection of our principal $\scTD_n$-bundles takes values in the representation space $V_{-2}=\sfU(1)\times \IZ^{2n}$, and compatibility with this representation requires that the images of integer vectors should be a symmetry of the gauge 2-groupoid $\scTD_{n}$. In particular, we need to impose that
    \begin{equation}\label{eq:integrality_constraint}
        \bar\Theta(\IZ^{2n})\subset \sfGO(n,n;\IZ)~.
    \end{equation}
    
    Second, the fact that the exponentiated embedding tensor $\bar \Theta$ arises from exponentiating the actual embedding tensor $\Theta$ implies that the image of $\IZ^{2n}$ under $\bar\Theta$ lies in $\sfSO^+(n,n;\IZ)$, the intersection of $\sfGO(n,n;\IZ)$ with the connected component of $\sfGO(n,n;\IR)$ containing the identity. 
    
    We can then distill all these conditions into the following definition.
    \begin{definition}
        The \uline{space of general embedding tensors} $\bar R_n$ is the set of exponentiated embedding tensors with $\bar \Theta(\IZ^{2n})\subset \sfSO^+(n,n;\IZ)$ and satisfying the quadratic closure constraint~\eqref{eq:closure_constraint} as well as the linear representation constraint~\eqref{eq:representation_constraint}. The \uline{space of integral embedding tensors} $R_n\subset \bar R_n$ is given by those elements that also satisfy the further integrality condition~\eqref{eq:integrality_constraint}.
    \end{definition}

    Note that we merely used the framework of gauged supergravity to identify an interesting space of scalar fluxes. Compatibility with the $B$-field~\eqref{eq:integrality_constraint} renders the additional gauge symmetry contribution to $\scTD_n$ discrete, and therefore there are no additional gauge potentials. In particular, the image of the embedding tensor $\Theta$ is an abelian Lie algebra. Hence, in the following, $\Theta$ and $\bar\Theta$ will always be Lie algebra and Lie group homomorphisms, respectively, with domains the abelian Lie algebra $\IR^{2n}$ and Lie group $\IR^{2n}$, respectively.
    
    \subsection{Scalar fluxes and T-dualities along few directions}
    
    We have identified the scalar fluxes contained in the discrete moduli in $\bar\Theta$. In the following, we show that these moduli match the expectations on $R$-fluxes from the classifications of backgrounds. That is, for a geometric T-background with $n$ compact directions, there are no $R$-fluxes for $n\le2$, but $R$-fluxes exist for $n\geq 3$.
    
    For $n=0$ and $n=1$, there are unique group homomorphisms \(\bar\Theta\colon\IZ^0\to\sfSO^+(0,0;\IZ)\cong1\) and $\bar\Theta\colon\IZ\to\sfSO^+(1,1;\IZ)\cong1$, respectively. Hence, there are no $R$-fluxes in either case.
    
    For $n=2$, there exist group homomorphisms $\bar\Theta\colon\IZ^4\to\sfSO^+(2,2;\IZ)$, for which we must check the representation constraint. Infinitesimally, $\Theta\colon\IR^4\to\fro(2,2;\IR)$ forms a Lie algebra homomorphism. The image of this linearization is an abelian Lie subalgebra of \(\fro(2,2;\IR)\), whose dimension can be at most $2$. If the dimension of the image is \(0\), this corresponds to trivial $R$-charge. If the dimension is \(1\), it is straightforward to check that the representation constraint fails. If the dimension is \(2\), the image corresponds to a pair of mutually commuting translations in \(\IR^{2,2}\); the requirement that the exponentiated rotations be integral implies that one of them can be taken to be along a space-like 2-plane and the other along a time-like 2-plane. Thus, the putative $R$-flux can be put in a standard form, and the representation constraint can then be easily checked to fail. Hence, for \(n=2\) there are no non-trivial $R$-fluxes either.
    
    In dimensions $n\ge3$, however, non-trivial $R$-fluxes exist. One sufficient ansatz is to consider group homomorphisms
    \begin{equation}
        \IZ^n\to\fro(n;\IZ)~,
    \end{equation}
    where \(\fro(n;\IZ)\) is the abelian group of \(n\times n\) antisymmetric integer matrices, that are defined by pairing with an \(n\)-dimensional totally antisymmetric integer 3-tensor, i.e.\ an element of
    \begin{equation}
        \{f\colon\{1,\dotsc,n\}^3\to\mathbb Z\mid f_{ijk}=-f_{jik}=-f_{ikj}\}\cong\IZ^{\binom n3}~,	
    \end{equation}
    in which each of the \(n\) generators maps to linearly independent elements of \(\fro(n;\IZ)\); this then defines a group homomorphism
    \begin{equation}
        \IZ^{2n}\to\fro(n;\IZ)\subset\sfGO(n,n;\IZ)
    \end{equation}
    in which elements of the other \(\IZ^n\) simply map to zero, and which clearly satisfies the representation constraint.
    Thus in $n$ dimensions, the set of  $R$-fluxes contains at least \(\IZ^{\binom n3}\), which corresponds to the geometric cohomology 3-classes. Of course, there are many possible choices of the embedding $\fro(n;\IZ)\hookrightarrow\sfGO(n,n;\IZ)$ by conjugations; these correspond to the T-duality orbits of the geometric cohomology 3-classes.
    
    \subsection{The complete groupoid of T-duality}
    
    It remains to incorporate the data $R_n$ and $\bar R_n$ we have argued above to be interesting into our picture. To this end, we have to generalize $\scTD_n$ to an augmented quasi-groupoid, as we will argue now.
    
    Recall that for $n\le2$, the dimensional reduction of the Kalb--Ramond $B$-field and its field strength $H$ creates 1-forms and scalars with corresponding curvature 2- and 1-forms. These are accounted for in the gauge Lie 2-groupoid $\scTD_n$. As evident from~\eqref{eq:direct_product_cohomology_decomposition}, dimensional reduction of $H$ with $n\ge3$ will produce (global) $0$-forms, which clearly cannot be seen as curvatures of non-existing $(-1)$-forms. 
    
    The groupoid picture, however, suggests a natural place for these global $0$-forms. We note that 2-forms and their 3-form curvatures are essentially encoded by the 2-cells of $\scTD_n$, while the 1-forms and their 2-form curvatures correspond to the 1-cells. The scalars are then encoded by the $0$-cells. We can regard the Lie 2-groupoid $\scTD_n$ as a Lie 2-quasi-groupoid, i.e.~a simplicial manifold satisfying the relevant Kan condition, cf.~\cref{app:quasi-groupoids}. In this context, there is the notion of augmented groupoid, which allows us to include $(-1)$-cells as the image of a single face map. Such an augmentation is quite natural: consider for example the Čech groupoid of a surjective submersion $\sigma\colon Y\rightarrow M$, cf.~\ref{app:higher_principal_bundles}. We can augment the nerve of the Čech groupoid by $M$ and obtain the augmented quasi-groupoid
    \begin{equation}\label{eq:augmented_Cech_nerve}
        \check\scC_\text{aug}(Y\rightarrow M)\ \coloneqq\ \left(\begin{tikzcd}
            \ldots \arrow[r,shift left,shift left,shift left] 
            \arrow[r,shift right,shift right,shift right]
            \arrow[r,shift left] 
            \arrow[r,shift right] & Y^{[3]} \arrow[r,shift left,shift left] 
            \arrow[r,shift right,shift right]
            \arrow[r] &
            Y^{[2]}\arrow[r,shift left] 
            \arrow[r,shift right] & Y \arrow[r] & M
        \end{tikzcd}\right)~.
    \end{equation}
    
    In order to capture all aspects of non-geometric T-duality, we evidently have to augment the Lie 2-groupoid $\scTD_n$ by $\bar R_n$, resulting in the augmented Lie 2-quasi-groupoid $\scTD^\text{aug}_n$. 

    We note that the sequence of reductions indeed terminates here: $0$-form curvatures do not reduce any further. Thus, the picture of augmented Lie groupoids is indeed sufficient for arbitrary $n$.

    Our construction of $\scTD^\text{aug}_n$ starts from an enlarged version\footnote{Similar to $\scTD_n$, this 2-groupoid can be thought of as a bundle of 2-groups, with fiber $\sfTD_n$, on the disconnected orbifold $(Q_n\times R_n)/\sfGO(n,n;\IZ)$.} of $\scTD_n$.
    \begin{definition}
        The Lie 2-groupoid $\widehat{\scTD}_n$ is obtained from the groupoid $\scTD_n$ by replacing $Q_n$ with $Q_n\times R_n$ and the action of $\sfGO(n,n;\IZ)\ltimes \IR^{2n}$ on $Q_n$ with the following action on $Q_n\times R_n$:
        \begin{equation}\label{eq:replacement_action}
            \begin{aligned}
                \acton\colon(\sfGO(n,n;\IZ)\ltimes \IR^{2n})\times (Q_n\times R_n)&\rightarrow (Q_n\times R_n)~,
                \\
                (g,\xi;q,r)&\mapsto\big((g\acton r)(\xi)gq,g\acton r\big)=\big(gr(g^{-1}\xi)q,g\acton r\big)
            \end{aligned}
        \end{equation}
        with 
        \begin{equation}\label{eq:action_on_r}
            (g\acton r)(\zeta)\coloneqq gr(g^{-1}\zeta)g^{-1}
        \end{equation}
        for \(\zeta\in\IZ^{2n}\).
    \end{definition}
    Note that~\eqref{eq:replacement_action} is indeed a group action as one readily verifies by direct computation:
    \begin{equation}
        \begin{aligned}
            (g_1,\xi_1)\acton ((g_2,\xi_2)\acton (q,r))&=(g_1,\xi_1)\acton(g_2r(g_2^{-1}\xi_2)q,g_2\acton r)
            \\
            &=(g_1(g_2\acton r)(g_1^{-1}\xi_1)g_2r(g_2^{-1}\xi_2)q,g_1g_2\acton r)
            \\
            &=(g_1g_2r(g_2^{-1}g_1^{-1}\xi_1)r(g_2^{-1}\xi_2)q,g_1g_2\acton r)
            \\
            &=\Big(g_1g_2r\big((g_1g_2)^{-1}(\xi_1+g_1\xi_2)\big)q,g_1g_2\acton r\Big)
            \\
            &=((g_1,\xi_1)(g_2,\xi_2))\acton(q,r)~,
        \end{aligned}
    \end{equation}
    where we used the fact that $r$ is a group homomorphism.
    
    Physically speaking, the Lie 2-groupoid $\widehat{\scTD}_n$ is already large enough to encode the local $0$-form field strengths of the scalar fluxes, but it is too general; it remains to ensure that the local scalar fluxes glue together to form a global object.
    
    Consider now the Lie 2-quasi-groupoid that is the Duskin nerve~\cite{Duskin02simplicialmatrices} $\caN(\widehat{\scTD}_n)$ of $\widehat{\scTD}_n$.
    \begin{definition}
        The augmented Lie 2-quasi-groupoid $\scTD^\text{aug}_n$ is a Kan (augmented) simplicial manifold
        \begin{equation}
            \scTD^\text{aug}_n\coloneqq\ \left(\begin{tikzcd}
                \ldots \arrow[r,shift left,shift left,shift left] 
                \arrow[r,shift right,shift right,shift right]
                \arrow[r,shift left] 
                \arrow[r,shift right] & (\scTD^\text{aug}_n)_2\arrow[r,shift left,shift left] 
                \arrow[r,shift right,shift right]
                \arrow[r] &
                (\scTD^\text{aug}_n)_1\arrow[r,shift left] 
                \arrow[r,shift right] & (\scTD^\text{aug}_n)_0 \arrow[r] & (\scTD^\text{aug}_n)_{-1}
            \end{tikzcd}\right)
        \end{equation}
        with
        \begin{equation}
            (\scTD^\text{aug}_n)_0\coloneqq Q_n\times R_n
            \eand
            (\scTD^\text{aug}_n)_{-1}\coloneqq \bar R_n~,
        \end{equation}
        such that the forgetful functor to Lie 2-quasi-groupoids applied to $\scTD^\text{aug}_n$ yields $\caN(\widehat{\scTD}_n)$ and the final face map
        \begin{equation}
			\sff^0_{0}: (\scTD^\text{aug}_n)_0 \rightarrow (\scTD^\text{aug}_n)_{-1}
		\end{equation}
        is the evident projection $(Q_n\times R_n)\rightarrow R_n$ composed with the embedding $R_n\hookrightarrow \bar R_n$.
    \end{definition}
    
    Writing out the explicit form of $\scTD^\text{aug}_n$ becomes rather technical, and it is not further illuminating; fortunately, the above picture is sufficient for our purposes.
    
    \subsection{Generalized T-duality correspondences with scalar fluxes}\label{ssec:TDnhat-bundles}
    
    Because of the simplicity of the augmentation, we can directly extend the Čech cocycles of $\scTD_n$-bundles to cocycles of principal $\scTD^\text{aug}_n$-bundles. These are encoded in augmented simplicial maps from the augmented Čech 2-quasi-groupoid to the augmented Lie 2-quasi-groupoid $\scTD^\text{aug}_n$. Explicitly, we have the data
    \begin{subequations}
        \begin{equation}
            \begin{aligned}
                (g,z,\xi,m,\phi,q,r)
                &\in C^\infty(Y^{[3]},\sfGO(n,n;\IZ)\times \IZ^{2n}\times\IR^{2n}\times \IZ^{2n}\times \sfU(1)\times Q_n\times R_n)~,
                \\
                (g,\xi,q,r)&\in C^\infty(Y^{[2]},\sfGO(n,n;\IZ)\times \IR^{2n}\times Q_n\times R_n)~,
                \\
                (q,r)&\in C^\infty(Y,Q_n\times R_n)~,
                \\
                r&\in C^\infty(M,\bar R_n)~,
            \end{aligned}
        \end{equation}
        which satisfy the relations
        \begin{equation}\label{eq:augmented_cocycle}
            \begin{aligned}
                r_i&=r~,
                \\
                (q_j,r_j)&=(g_{ij}^{-1},-g_{ij}^{-1}\xi_{ij})\acton(q_i,r_i)~,~~~(q_{ijk},r_{ijk})=(q_{ij},r_{ij})=(q_i,r_i)~,
                \\
                g_{ik}&=g_{ij}g_{jk}~,~~~g_{ijk}=g_{ik}~,
                \\
                \xi_{ik}&=\xi_{ij}+g_{ij}\xi_{jk}+m_{ijk}~,~~~\xi_{ijk}=\xi_{ik}~,
                \\
                m_{ijk}+m_{ikl} &= m_{ijl}+g_{ij}m_{jkl}~,
                \\
                \phi_{ijk} + \phi_{ikl} &=\phi_{ijl} + (-1)^{|g_{ij}|}\phi_{jkl}-\langle \xi_{ij}, g_{ij}m_{jkl}\rangle+
                m_{jkl}^\rmT \rho_L(g_{ij})\xi_{jl} +
                \xi_{jk}^\rmT \rho_L(g_{ij})g_{jk}\xi_{kl}
                \\
                &\hspace{0.5cm}+\tfrac12 \xi_{kl}^\rmT \sigma_L(g_{ij},g_{jk})\xi_{kl}~,
            \end{aligned}
        \end{equation}
        as well as the following condition required for adjustment:
        \begin{equation}\label{eq:z-rel-2}
            z_{ijk}=\frac{(-1)^{|g_{ik}|}}{2}g_{ik}\eta\,\rmdiag(\sigma_L(g_{ij},g_{jk}))~.
        \end{equation}
        
        Similarly, a differential refinement is easily found because we have only added discrete structures to our gauge Lie 2-groupoid that do not affect the continuous cocycles. The differential refinement (excluding the additional scalar fields\footnote{It is arguable whether the scalar fields should be counted as part of the topological information or the differential refinement.}) is therefore the same as that of $\scTD_n$. That is, we have 1-~and 2-forms
        \begin{equation}
            \Lambda\in \Omega^1(Y^{[2]})~,~~~
            A\in \Omega^1(Y,\IR^{2n})~,~~~
            B\in \Omega^2(Y)~,
        \end{equation}
        satisfying the relations~\eqref{eq:diff_refined_cocycles2}.
    \end{subequations}    
    
    We thus propose that T-dualities with certain scalar fluxes on potentially non-geometric backgrounds are described by differentially refined, principal $\scTD^\text{aug}_n$-bundles. If the source of the T-duality is a geometric T-background (or a T-fold), then the cocycle is of a special form, and we can reconstruct the background globally (or locally) from the cocycle of the principal $\scTD^\text{aug}_n$-bundle as explained in previous sections. 
    
    The T-dual partner is obtained by applying the flip morphism~\eqref{eq:flip-plain} to the cocycle data. The action here is the same as that on $\scTD_n$, with an additional action on $\bar R_n$ and $R_n$ induced by the evident action~\eqref{eq:replacement_action}. Again, if the target of the T-duality is geometric (or locally geometric), we can reconstruct the background (locally).
    
    In the general case, however, it is not possible to reconstruct individual sources or targets of the T-duality. This is hardly surprising given the intrinsically non-geometric nature of T-folds and spaces with scalar fluxes, in particular $R$-spaces. Only a total description of the T-duality exists as is familiar e.g.~from double field theory.
    
    \subsection{Remarks on our construction}
    
    A few remarks on our proposal for T-dualities with scalar fluxes are in order.
    
    \paragraph{Embeddings of T-correspondences.} We first note that the augmented Lie quasi-groupoid $\scTD^\text{aug}_n$ naturally restricts to the higher Lie groupoids and higher Lie groups relevant to T-duality correspondences without $R$-fluxes or without $Q$- and $R$-fluxes, as expected. In particular, if the scalar fields are fixed to be constant, then $\scTD^\text{aug}_n$-bundles effectively reduce to $\sfTD^\ltimes_n$-bundles. We thus have embeddings of augmented 2-groupoids
    \begin{equation}\label{eq:embeddings1}
		\sfTD_n~\hookrightarrow~\scTD_n~\hookrightarrow~\scTD^\text{aug}_n
	\end{equation}
    together with embeddings of the corresponding higher principal bundles, which is also evident from the explicit cocycle descriptions of these bundles. Note the further embedding of 2-groupoids
    \begin{equation}
		\sfTD_n~\hookrightarrow~\sfTD^{\text{$\tfrac12$geo}}_n\hookrightarrow~\scTD_n
	\end{equation}
    that demonstrates that our proposal extends the half-geometric T-duality correspondences of~\cite{Nikolaus:2018qop}, cf.~\ref{prop:embed_half_geometric}.
    
    \paragraph{Compatibility of $Q$- and scalar-fluxes.} Notice that, even though the $(-1)$-form potentials $r$ are a~priori valued in the smooth space $\bar R_n$ (similar to all other potentials), they are constrained to be quantized as elements of $R_n$ by the cocycle condition $r=r_i\in R_n$. This accords with the fact that $(-1)$-form potentials do not encode independent local degrees of freedom, unlike potentials of higher form degrees.
    
    The condition~\eqref{eq:augmented_cocycle} implies a compatibility condition between the $Q$-flux and the scalar flux: a generic $Q$-flux cannot coexist with a generic scalar flux. If the $Q$-flux is given by $g_{ij}$ taking values in some subgroup $\Gamma\subset\sfGO(n,n;\IZ)$, then the scalar fluxes $r$ are constrained to take values in the stabilizer subgroup of $\Gamma$ under the group action~\eqref{eq:action_on_r}. In particular, a vanishing $Q$-flux is compatible with an arbitrary scalar flux in $R_n$, whereas a generic $Q$-flux is compatible only with the trivial scalar flux that is the constant map in $R_n$.
    
    This has the following physical interpretation. The scalar flux $r$, regarded as the pointwise embedding tensors $r_i\colon Y\to R_n\subset(\IR^{2n}\to\sfGO(n,n;\IR))$, identifies the abelian gauge group \(\IR^{2n}\) of the 1-forms with a subgroup of \(\sfGO(n,n;\IR)\) given by the image of $r$. In the presence of non-trivial Q-flux, as specified by $g_{ij}\in\sfGO(n,n;\IZ)$, however, this identification holds only locally, since \(g_{ij}\) acts non-trivially on the $r_i$. In order for this identification to be globally well-defined, one requires that $r$ be equivariant under the action of $g_{ij}$, which is implied by~\eqref{eq:augmented_cocycle}. This renders the scalar flux globally well-defined for each connected component of space-time (in the complement of domain walls).
    
    It is illustrative to examine the compatibility between $Q$-flux and scalar flux in the purely geometric case, i.e.\ when all fluxes correspond to the Kalb--Ramond 3-form flux $H$ wrapped around the $n$ geometric directions in an $n$-torus. The $Q$-flux being geometric in this sense corresponds to the ansatz $g_{ij}\in\fro(n;\IZ)\subset\sfGO(n,n;\IZ)$, and similarly $r_i$ must be a map $\IZ^{2n}\to\sfGO(n,n;\IZ)$ whose image lies in $\fro(n;\IZ)$. In such a case, since the adjoint action of $\fro(n;\IZ)$ on itself is trivial, the condition that $g_{ij}^{-1}\acton r = r$ corresponds to the condition that $r$ be constant on the orbits of the $\fro(n;\IZ)$-action on the domain $\IZ^{2n}$, i.e.
    \begin{equation}
        r(\hat a,\check a)=r(\hat a,\check a+g\hat a)
    \end{equation}
    for $(\hat a,\check a)\in\IZ^{2n}$ and arbitrary $g\in\fro(n;\IZ)$. The classification of such orbits is non-trivial. However, it is sufficient that $r$ satisfy the stronger condition
    \begin{equation}
        r(\hat a,\check a)=\hat r(\hat a)
    \end{equation}
    for some function $\hat r\colon\IZ^n\to\fro(n;\IZ)$. After additionally imposing the representation constraint, such scalar fluxes correspond to elements of $\IZ^{\binom n3}$. Hence, we see that the purely geometric case is consistent with the constraints imposed by~\eqref{eq:augmented_cocycle}. 
    
    \paragraph{T-correspondences without $H$- and $f$-fluxes.} Above in~\eqref{eq:embeddings1}, we considered restrictions of the $(-1)$- and $0$-cells in $\scTD_n^\text{aug}$. At the other extreme, one may demand that the $2$- and $1$-cells be trivial. The resulting principal bundles form the kinematic data of a sigma model on the Narain moduli space $\sfGO(n,n;\IZ)\setminus(Q_n\times R_n)$ with constraints on the superselection sectors coming from the augmentation that we are going to discuss below; if we further turn off the scalar fluxes completely, this reduces to a sigma model on $GM_n=\sfGO(n,n;\IZ)\setminus Q_n$ subject to the minor constraint~\eqref{eq:g-even} on discrete moduli.
    
    \paragraph{Classification of branes.} The augmented Lie 2-quasi-groupoid $\scTD^\text{aug}_n$ that we obtained also leads to a natural classification of branes that appear in toroidal compactifications of string theory, and we briefly comment on this in the following.
    
    In general, a codimension~\(k\) brane can be stable if it couples to a \((k-2)\)-form potential magnetically, so that branes can be classified by classifying the corresponding \((k-2)\)-form potentials; a codimension~\(k\) brane can also be stable if it carries a non-trivial topological charge, i.e.\ one in which the scalar field exhibits a non-trivial monodromy around the \((k-1)\)-sphere around the brane; such codimension~\(k\) branes are classified by the homotopy group \(\pi_{k-1}(\Sigma)\) of the manifold (or orbifold) that the scalars take values in. A higher gauge groupoid \(\scG\), describing both \(p\)-forms for \(p>0\) as well as scalars, unifies both of these conditions, such that codimension~\(k\) branes are uniformly classified by \(\pi_{k-1}(\scG)\); the fact that \(\pi_{k-1}(-)\) is abelian for \(k\ge3\) corresponds to the fact that three codimensions suffice to exclude anyonic statistics and, hence, non-abelian charges.
    
    In the case of $\scTD^\text{aug}_n$, we have the following result:
    \begin{equation}
        \begin{aligned}
            \pi_0(\scTD^\text{aug}_n) &= R_n~,
            \\
            \pi_1(\scTD^\text{aug}_n) &= \sfGO(n,n;\IZ)~,
            \\
            \pi_2(\scTD^\text{aug}_n) &= \IZ^{2n} \times \IZ^{2n}~,
            \\
            \pi_3(\scTD^\text{aug}_n) &= \IZ ~.
        \end{aligned}
    \end{equation}
    These deviate from the expectations somewhat due to the complications of adjustment, and we comment on each case in the following.
    
    Codimension~$1$ branes, or domain walls~\cite{Bergshoeff:2012pm}, are labeled by the change in scalar flux across them, which is thus labeled by an element of $R_n$. When the scalar flux belongs to $\fro(n;\IZ)\subset\sfGO(n,n;\IZ)$, the domain wall corresponds to an NS-brane wrapped around $n-3$ directions of the $n$-torus. Note that the presence of generic domain walls may be incompatible with the presence of generic defect branes as explained above.
    
    Codimension~$2$ branes, or defect branes~\cite{Bergshoeff:2011se}, are labeled by a non-trivial monodromy of the $Q_n/\sfGO(n,n;\IZ)$-valued scalar field around it, hence by $\sfGO(n,n;\IZ)$. Upon dimensional uplift, these correspond to NS-branes wrapped around $n-2$ directions of the $n$-torus, or to KK-branes, or to bound states of both, depending on the element of $\sfGO(n,n;\IZ)$. Somewhat unexpectedly, the condition for the existence of an adjustment implies that the monodromy $g$ must satisfy, for every pair of integers $m_1,m_2$, the condition that
    \begin{equation}
        g^{m_1+m_2}\eta\,\rmdiag(\sigma_L(g^{m_1},g^{m_2}))\in2\IZ^{2n}~,
    \end{equation}
    which is the special case of~\eqref{eq:g-even} for a codimension~2 brane. This condition always holds for $n=1$, but it fails for generic $g\in\sfGO(n,n;\IZ)$ for $n\ge2$. However, in case $g$ belongs to one of the special subgroups discussed in \cref{ssec:2-group_action} --- namely, the $\sfGL(n;\IZ)$ subgroup of $A$-transformations, the $\fro(n;\IZ)$ subgroup of $B$-transformations, the $\fro(n;\IZ)$ subgroup of $\beta$-transformations, the subgroup of factorized dualities, or the $\IZ_2\times\IZ_2$ subgroup generated by $\rmdiag(s_1,\dotsc,s_1,-s_2,\dotsc,-s_2)$ for $s_1,s_2=\pm1$ --- the condition always holds. Thus, such ``ordinary'' codimension~2 branes do exist. The fact that the permitted $Q$-fluxes do not form a subgroup of $\sfGO(n,n;\IZ)$ means that such defect branes are mutually non-local: the presence of one defect brane may forbid the presence of another defect brane somewhere else.
    
    One expects a $2n$-plet of codimension~3 branes, corresponding to a single $\sfGO(n,n;\IZ)$ orbit consisting of NS-branes wrapped around $n-1$ directions and KK-branes. Hence, the presence of the additional copy of $\IZ^{2n}$, which ultimately comes from the non-trivial 2-group structure of $\scGO(n,n;\IZ)$ in~\eqref{eq:def_scGOnnZ}, is somewhat unexpected. However, the adjustment condition~\eqref{eq:z-rel} requires that this spurious charge be fixed by the $Q$-fluxes, such that the actual possible set of codimension~3 brane charges is simply labeled by $\IZ^{2n}$ as expected.
    
    The unique codimension~4 brane corresponds to an NS-brane fully wrapped around the $n$-torus, coupling magnetically to the Kalb--Ramond field.
    
    \subsection{Example of T-duality correspondence with scalar fluxes}
    
    A generic T-fold with scalar flux is described as (part of) a full $\scTD^\text{aug}_n$-bundle over $X$. To render this data manageable, we can restrict ourselves to a configuration in which all fields are set to zero except for the scalar fields $(q,r)\in C^\infty(Y,Q_n\times R_n)$, the monodromies $(g,\xi)\in C^\infty(Y^{[2]},\sfGO(n,n;\IZ)\times\IR^{2n})$, and $z\in C^\infty(Y^{[3]},\IZ^{2n})$. By~\eqref{eq:z-rel}, the \(g_{ij}\) define a \(\sfGO(n,n;\IZ)\)-principal bundle that is even in the sense of~\eqref{eq:g-even}.
    By the same equation, the $g_{ij}$ also fix the higher monodromies $z_{ijk}$. Note that we can set the connection data \((\Lambda,A,B)\) consistently to zero. In this case, the data \((g_{ij},q_i,r_i)\) are equivalent (modulo the condition~\eqref{eq:g-even}) to that defining a cocycle of the action groupoid of $\sfGO(n,n;\IZ)$ on $Q_n\times R_n$.
    
    Thus, each connected component of the base manifold $X$ is associated to an element of $R_n$ --- the $R$-flux. Within each connected component, then, one has additional $Q$-flux around non-contractible cycles; if the $R$-flux vanishes for a connected component $X_i$, then the $Q$-flux is simply a group homomorphism
    \(\pi_1(X_i)\to\sfGO(n,n;\IZ)\) satisfying~\eqref{eq:g-even}.
    
    Concretely, in the case of the key example of an abelian gerbe with curvature $H$ on $T^3$, we T-dualize along all cycles in the base. Correspondingly, the base space $X$ is merely a point. It follows that we can set all cocycle data to zero except for $q$ and $r$, and the latter specifies the expected $R$-flux $r\in R_n$.
    
    \section*{Acknowledgments}
    
    We would like to thank Leron Borsten and Chris Hull for helpful discussion as well as Konrad Waldorf for pointing out the problem with adding connections in the constructions of~\cite{Nikolaus:2018qop}, which eventually led to this paper. This work was supported by the Leverhulme Research Project Grant RPG--2018--329 ``The Mathematics of M5-Branes.''
    
    \section*{Data availability statement}
    
    No additional research data beyond the data presented and cited in this work are needed to validate the research findings in this work. 

    \appendices
    
    \subsection{Lie 2-groupoid basics}\label{app:2-groupoid_basics}
    
    We will follow the conventions of~\cite{Jurco:2014mva} regarding 2-categories and speak of weak and strict 2-categories. The former are also known as bicategories; see~\cite{Benabou:1967:1} for the original definitions as well as~\cite[Chap.~7]{Borceux:1994aa} for a textbook account.
    
    \paragraph{Weak 2-categories.} A \underline{weak 2-category} $\scB$ consists of a collection of \uline{objects} or \uline{0-cells} $\scB_0$ and a category of morphisms $\scB(a,b)$ for every pair of objects $a,b\in \scB_0$. Objects and morphisms in these categories are known as \uline{1-} and \uline{2-cells}, respectively. For each object $a\in \scB_0$, there is an \uline{identity 1-cell} $\sfid_a$. Composition of 2-cells is denoted by $\circ$ and called \uline{vertical composition}. \uline{Horizontal composition}, on the other hand, is a collection of bifunctors $\scB(a,b)\otimes \scB(b,c)\rightarrow \scB(a,c)$ for all $a,b,c\in \scB_0$. Horizontal composition is not strict and comes with a set of natural isomorphisms known as left and right unitors,
    \begin{subequations}
        \begin{equation}
            \sfl\colon x\otimes \sfid_\sfs(x)\xRightarrow{~\cong~} x
            \eand 
            \sfr\colon\sfid_\sfs(x)\otimes x\xRightarrow{~\cong~} x~,
        \end{equation}
        as well as an associator, 
        \begin{equation}
            \sfa\colon(x\otimes y)\otimes z\xRightarrow{~\cong~} x\otimes (y\otimes z)~,
        \end{equation}
        for all 1-cells $x,y,z$. These morphisms satisfy coherence conditions known as the pentagon and triangle identities, see~\cite{Jurco:2014mva}. We will exclusively work with ``unital'' weak 2-categories that come with unital horizontal composition, reducing the coherence conditions to the pentagon identity for the associator:
        \begin{equation}
            \begin{tikzcd}
                & ((x\otimes y)\otimes u)\otimes v \arrow[rd,Rightarrow,"\sfa"] \arrow[ld,Rightarrow,"\sfa\otimes \sfid",swap] & 
                \\
                (x\otimes(y\otimes u))\otimes v  \arrow[d,Rightarrow,"\sfa",swap] & & (x\otimes y)\otimes (u\otimes v) \arrow[d,Rightarrow,"\sfa"]
                \\
                x\otimes((y\otimes u)\otimes v) \arrow[rr,Rightarrow,"\sfid\otimes \sfa"] & & x\otimes(y\otimes(u\otimes v))
            \end{tikzcd}
        \end{equation}
    \end{subequations}    
    
    \paragraph{2-functors.} Given two weak 2-categories $\scB$ and $\tilde \scB$, a \uline{unital lax 2-functor} $\Phi\colon\scB\rightarrow \tilde \scB$ consists of a function
    \begin{subequations}
        \begin{equation}
            \Phi_0\colon\scB_0\rightarrow \tilde \scB_0~,
        \end{equation}
        a collection of functors
        \begin{equation}
            \Phi_1^{ab}\colon\scB(a,b)\rightarrow \tilde \scB(\Phi_0(a),\Phi_0(b))~,
        \end{equation}
        and a collection of natural transformations
        \begin{equation}
            \Phi_2^{abc}\colon\Phi_1^{ab}(-)\tildeotimes \Phi_1^{bc}(-)~\xRightarrow{~~}\Phi_1^{ac}(-\otimes -)
        \end{equation}
        for all $a,b,c\in \scB_0$. The latter satisfy a coherence condition amounting to the commutative diagram
        \begin{equation}\label{eq:2-functor_associator_coherence}
            \begin{tikzcd}
                & \Phi_1^{ac}(x\otimes y)\,\tildeotimes\, \Phi_1^{cd}(z) \arrow[dr,Rightarrow,"\Phi_2^{acd}"] & 
                \\
                (\Phi_1^{ab}(x)\,\tildeotimes\,\Phi_1^{bc}(y))\,\tildeotimes\, \Phi_1^{cd}(z) \arrow[ur,Rightarrow,"\Phi_2^{abc}\otimes \sfid"]  \arrow[d,Rightarrow,"\tilde \sfa"] & & \Phi_1^{ad}((x\otimes y)\otimes z) \arrow[d,Rightarrow,"\Phi_1^{ad}(\sfa)"]
                \\
                \Phi_1^{ab}(x)\,\tildeotimes\,(\Phi_1^{bc}(y)\,\tildeotimes\, \Phi_1^{cd}(z))\arrow[dr,Rightarrow,"\sfid\otimes \Phi_2^{bcd}"] &  & \Phi_1^{ad}(x\otimes(y\otimes z))
                \\
                & \Phi_1^{ab}(x)\,\tildeotimes\,\Phi_1^{bd}(y\otimes z)\arrow[ur,Rightarrow,"\Phi_2^{abd}"] & 
            \end{tikzcd}
        \end{equation}
    \end{subequations}    
    If the natural transformations are natural isomorphisms, we call it a \uline{weak 2-functor}.
    
    We note that 2-functors $\Psi\colon\scB_1\rightarrow \scB_2$ and $\Phi\colon\scB_2\rightarrow \scB_3$ compose as 
    \begin{equation}\label{eq:composition_2_functors}
        \begin{gathered}
            \Xi=\Phi\circ \Psi~,
            \\
            \Xi_0=\Phi_0\circ \Psi_0~,~~~
            \Xi^{ab}_1=\Phi^{\tilde a\tilde b}_1\circ \Psi^{ab}_1~,
            \\
            \Xi^{abc}_2(x,y)=\Phi^{\tilde a\tilde c}_1(\Psi^{abc}_2(x,y))\circ \Phi^{\tilde a\tilde b\tilde c}_2(\Psi^{ab}_1(x),\Psi^{bc}_1(y))~,
        \end{gathered}
    \end{equation}
    where $\tilde a=\Psi_0(a)$, etc.
    
    \paragraph{Lie 2-groupoids.} A \uline{(weak) 2-groupoid} is a weak 2-category in which all morphisms are equivalences. That is, all 2-cells are strictly invertible, and all 1-cells are invertible up to isomorphism. A \uline{Lie 2-groupoid} is then a 2-groupoid internal to a suitable category of smooth manifolds.\footnote{Note that the naive choice of smooth manifolds and smooth maps between these does not contain all pullbacks, which leads to problems in the composition. This is a well-known technicality, which can be resolved by working with diffeological spaces and which we ignore here.}
    
    \subsection{Higher groups}\label{app:higher_groups}
    
    \paragraph{2-groups.} A 2-group is a categorified group. In its most general form, a \uline{(weak) 2-group} is a weak monoidal small category in which all morphisms are invertible and all objects are weakly invertible, cf.~e.g.~\cite{Baez:0307200}. Equivalently, it can be regarded as a monoidal category of morphisms contained in a pointed\footnote{Although the pointing is unique, one should use pointed (2-)functors between pointed (2-)groupoids in order to get the correct automorphisms, cf.~the $n$Lab page \url{https://ncatlab.org/nlab/show/looping}}  Lie 2-groupoid with a single 0-cell.
    
    If the associator in a 2-group is trivial, then we obtain a \uline{strict 2-group}, which can be regarded as a group object internal to $\CatCat$, the category of small categories. As shown in~\cite{Brown:1976:296-302}, strict 2-groups are equivalent to crossed modules of groups. 
    
    \paragraph{Crossed modules of groups.} A \uline{crossed module of groups} $\caG=(\sfH\xrightarrow{~\sft~}\sfG,\acton)$ is a pair of groups $\sfG,\sfH$ together with a group homomorphism $\sft\colon\sfH\rightarrow \sfG$ and an action of $\sfG$ on $\sfH$ by automorphisms $\acton\colon\sfG\times \sfH\rightarrow \sfH$ such that, for all $g\in\sfG$ and $h_{1,2}\in\sfH$, we have
    \begin{equation}
        \sft(g\acton h_1)\ =\ g\sft(h_1)g^{-1}
        \eand
        \sft(h_1)\acton h_2\ =\ h_1h_2h_1^{-1}~.
    \end{equation}
    This admits an evident specialization to crossed modules of Lie groups.
    
    We will use the conventions of~\cite{Jurco:2014mva}, under which the strict 2-group $\underline{\caG}$ corresponding to a crossed module $\caG=(\sfH\xrightarrow{~\sft~}\sfG,\acton)$ is defined as follows:
    \begin{equation}
        \begin{aligned}
            \begin{tikzcd}
                \sfG\ltimes \sfH \arrow[r,shift left] 
                \arrow[r,shift right] & \sfG
            \end{tikzcd}~,~&~~
            \begin{tikzcd}[column sep=2.0cm,row sep=large]
                \phantom{\sft(h)} g & \sft(h^{-1})g\arrow[l,bend left,swap,out=-20,in=200]{}{(g,h)}
            \end{tikzcd}~,
            \\
            (g_1,h_1)\circ (\sft(h_1^{-1})g_1,h_2)&\coloneqq(g_1,h_1h_2)~,
            \\
            (g_1,h_1)\otimes (g_2,h_2)&\coloneqq(g_1g_2,(g_1\acton h_2)h_1)~,
            \\
            \sfinv(g_1,h_1)&\coloneqq (g_1^{-1},g_1^{-1}\acton h_1^{-1})~.
        \end{aligned}    
    \end{equation}    
    
    \paragraph{Morphisms.} A \uline{strict morphism} of crossed modules of groups $\Phi\colon\caG\rightarrow \tilde \caG$ is simply a map
    \begin{equation}
        \Phi~~:~~(\sfH\xrightarrow{~\sft~}\sfG,\acton)~~\longrightarrow~~(\tilde \sfH\xrightarrow{~\tilde \sft~}\tilde \sfG,\tilde \acton)
    \end{equation}
    consisting of a pair of group homomorphisms $\Phi_0\colon\sfG\rightarrow \tilde \sfG$ and $\Phi_1\colon\sfH\rightarrow \tilde \sfH$ that are compatible with $\sft$ and $\acton$ in evident ways.
    
    A \uline{very weak morphism} of crossed modules of groups $\Phi\colon\caG\rightarrow \tilde \caG$, also known as a \uline{butterfly}, cf.~\cite{Aldrovandi:0808.3627}, is a commutative diagram of groups
    \begin{equation}\label{eq:butterfly}
        \begin{tikzcd}
            \sfH_1 \arrow[dd,"\sft_1",swap] \arrow[dr,"\lambda_1"]& & \sfH_2 \arrow[dl,"\lambda_2",swap] \arrow[dd,"\sft_2"]
            \\
            & \sfE \arrow[dl,"\gamma_1",swap] \arrow[dr,"\gamma_2"]& 
            \\
            \sfG_1 & & \sfG_2
        \end{tikzcd}
    \end{equation}
    where $\sfE$ is a group, $\lambda_{1,2}$ and $\gamma_{1,2}$ are group homomorphisms, the NE--SW diagonal is a short exact sequence (i.e.~a group extension), and the NW--SE diagonal is a complex.
    
    Between these two notions lies the notion of a \uline{weak morphism} of crossed modules, which is induced by a lax 2-functor of the corresponding two one-object Lie 2-groupoids whose hom-categories (i.e.~category of morphisms) are $\underline{\caG}$ and $\underline{\tilde{\caG}}$, respectively, cf.~\cite{Jurco:2014mva}, as defined in~\ref{app:2-groupoid_basics}.\footnote{In this paper, we refrain from using the notion of crossed intertwiners developed in~\cite{Nikolaus:2018qop} for practical reasons.} Such a morphism $\Phi$ is thus encoded by a functor and a natural transformation,
    \begin{subequations}\label{eq:weak_morphism}
        \begin{equation}
            \Phi_1\colon\underline{\caG}\rightarrow \underline{\tilde{\caG}}
            \eand 
            \Phi_2\colon\Phi_1(-)\tildeotimes \Phi_1(-)\Rightarrow \Phi_1(-\otimes -)~.
        \end{equation}
        Besides the naturality condition 
        \begin{equation}
            \Phi_2(g_1,g_2)\tilde \circ (\Phi_1(g_1,h_1)\tildeotimes \Phi_1(g_2,h_2))=\Phi_1((g_1,h_1)\otimes (g_2,h_2))\tilde \circ \Phi_2(\sft(h_1^{-1})g_1,\sft(h_2^{-1})g_2)
        \end{equation}
        for all $g_{1,2}\in \sfG$ and $h_{1,2}\in \sfH$, we have the coherence condition~\eqref{eq:2-functor_associator_coherence} with trivial associators, resulting in
        \begin{equation}
            \Phi_2(g_1\otimes g_2,g_3)\tilde \circ(\Phi_2(g_1,g_2)\tildeotimes \sfid_{\Phi_1(g_3)})
            =
            \Phi_2(g_1,g_2\otimes g_3)\tilde \circ (\sfid_{\Phi_1(g_1)}\tildeotimes \Phi_2(g_2,g_3))~.
        \end{equation}
        Strict morphisms are evidently included here as weak morphisms with $\Phi_2$ trivial.
    \end{subequations}
    Two weak morphisms of Lie 2-groups $\Psi\colon\underline{\caG}_1\rightarrow \underline{\caG}_2$ and $\Phi\colon\underline{\caG}_2\rightarrow \underline{\caG}_3$ compose into a morphism $\Xi=\Phi\circ \Psi$ with
    \begin{equation}
        \begin{aligned}
            \Xi_1(g,h)&=\Phi_1(\Psi_1(g,h))~,
            \\
            \Xi_2(g_1,g_2)&=\Phi_1(\Psi_2(g_1,g_2))\circ \Phi_2(\Psi_1(g_1),\Psi_1(g_2))
        \end{aligned}
    \end{equation}
    for all $(g,h)$ in the morphisms of $\underline{\caG}_1$, cf.~\eqref{eq:composition_2_functors}.
    
    Weak morphisms of crossed modules are particularly useful for our discussion as they can be readily postcomposed with the lax 2-functors defining principal 2-bundles, cf.~\ref{app:higher_principal_bundles}.
    
    \paragraph{2-group actions.} Any 2-group $\scH$ comes with a 2-group of automorphisms (or equivalences) $\sfAut(\scH)$, having 2-group endomorphisms that are equivalences of categories as its objects and natural 2-transformations between these as its morphisms. An action of a (weak) 2-group $\scG$ on another 2-group $\scH$ is then readily defined as a homomorphism of 2-groups $\Phi\colon\scG\rightarrow \sfAut(\scH)$~\cite{Breen:1992:465-514,Carrasco:1996:4059-4112}. 
    
    Here, we will use the reformulation of~\cite[Prop.~3.2]{Garzn:2001aa} for unital such actions. That is, a unital action of a 2-group $\scG$ on a 2-group $\scH$ is given by a bifunctor
    \begin{subequations}
        \begin{equation}
            \acton\colon\scG\times \scH\rightarrow \scH
        \end{equation}
        and natural isomorphisms
        \begin{equation}
            \begin{aligned}
                \Upsilon_\scG\colon (g_1\otimes g_2)\acton h \xrightarrow{~\cong~} g_1\acton(g_2\acton h)~,
                \\
                \Upsilon_\scH\colon g\acton(h_1\otimes h_2)\xrightarrow{~\cong~} (g\acton h_1)\otimes (g\acton h_2)
            \end{aligned}
        \end{equation}
    \end{subequations}    
    for all objects $g,g_{1,2}\in \scG$ and $h,h_{1,2}\in \scH$. These natural isomorphisms have to satisfy the coherence conditions listed in~\cite[Prop.~3.2]{Garzn:2001aa}. We write $\scG\curvearrowright \scH$ for such an action. We also note that the proof of this proposition gives a helpful definition of the bifunctor $\acton$ in terms of the homomorphism $\scG\rightarrow \sfAut(\scH)$.
    
    \paragraph{Semidirect products.} We further take~\cite[Def.~3.4]{Garzn:2001aa} as our definition of a semidirect product of 2-groups. Given two weak 2-groups $\scG$ and $\scH$ with a unital action $\scG\curvearrowright\scH$, we define the semidirect product $\scG\ltimes \scH$ as the 2-group with underlying Lie groupoid $\scG\times \scH$ and monoidal product
    \begin{subequations}
        \begin{equation}
            (G_1,H_1)\otimes (G_2,H_2) \coloneqq (G_1\otimes G_2,H_1\otimes (G_1\acton H_2))
        \end{equation}
        for all morphisms $G_{1,2}$ in $\scG$ and $H_{1,2}$ in $\scH$. The unit object is 
        \begin{equation}
            \unit_{\scG\ltimes \scH}=(\unit_\scG,\unit_\scH)~,
        \end{equation}
        and the associator is given by
        \begin{equation}\label{eq:associator_semidirect_product}
            \begin{aligned}
                &\sfa(g_1,h_1;g_2,h_2;g_3,h_3)\coloneqq
                \\
                &\hspace{1cm}\Big(\sfa(g_1,g_2,g_3),\big(\sfid_{h_1}\otimes \Upsilon_\scH^{-1}(g_1,h_2,g_2\acton h_3)\big)\circ \big(\sfid_{h_1}\otimes (\sfid_{g_1\acton h_2}\otimes \Upsilon_\scG(g_1,g_2,h_3))\big)\\
                &\hspace{8cm}\circ\sfa(h_1,g_1\acton h_2,(g_1\otimes g_2)\acton h_3)\Big)
            \end{aligned}
        \end{equation}
    \end{subequations}    
    for all objects $g_{1,2,3}\in \scG$ and $h_{1,2,3}\in \scH$, where the inverse of $\Upsilon_\scH$ is with respect to vertical composition. We note that the definitions of group action and semidirect product we use here subsume those used in~\cite{Nikolaus:2018qop}. 
    
    \subsection{Higher Lie algebras}\label{app:higher_Lie_algebras}
    
    \paragraph{$L_\infty$-algebras.} An \uline{$L_\infty$-algebra} $\frL$ is a $\IZ$-graded vector space $\frL=\bigoplus_{i\in\IZ} \frL_i$ together with totally antisymmetric multilinear maps $\mu_k\colon\frL^{\wedge k}\rightarrow \frL$ of degree $|\mu_k|=2-k$ satisfying the homotopy Jacobi identity
    \begin{equation}\label{eq:hom_Jac_rel}
        \sum_{i+j=n}\sum_{\sigma\in \bar S_{i|j}}\chi(\sigma;\ell_1,\ldots,\ell_{n})(-1)^{j}\mu_{j+1}(\mu_i(\ell_{\sigma(1)},\ldots,\ell_{\sigma(i)}),\ell_{\sigma(i+1)},\ldots,\ell_{\sigma(n)})=0~,
    \end{equation}
    where the sum runs over all $(i,j)$-unshuffles and $\chi$ denotes the (graded) Koszul sign of the permutation of the arguments. An $L_\infty$-algebra concentrated in non-positive degrees (i.e.~non-trivial only there) is a model for a \uline{semistrict higher Lie algebra}, and we use these two terms interchangeably. If all maps $\mu_k$ with $k>2$ are trivial, we call the $L_\infty$-algebra~\uline{strict}.
    
    \paragraph{Semistrict Lie~2-algebras.} We will be particularly interested in the case of $L_\infty$-algebras concentrated in degrees $-1$ and $0$ that form models for semistrict Lie 2-algebras. They consist of two vector spaces $\frL=\frL_{-1}\oplus \frL_0$ together with maps
    \begin{equation}
        \begin{gathered}
            \mu_1\colon\frL_{-1}\rightarrow \frL_0~,
            \\
            \mu_2\colon\frL_{0}\wedge \frL_0\rightarrow \frL_0~,~~~            
            \mu_2\colon\frL_{-1}\wedge \frL_0\rightarrow \frL_{-1}~,~~~            
            \mu_2\colon\frL_{0}\wedge \frL_{-1}\rightarrow \frL_{-1}~,
            \\
            \mu_3\colon\frL_0\wedge \frL_0\wedge \frL_0\rightarrow \frL_{-1}
        \end{gathered}
    \end{equation}
    satisfying~\eqref{eq:hom_Jac_rel}.
    
    A morphism $\phi\colon\frL\rightarrow \tilde \frL$ is given by linear maps
    \begin{equation}
        \phi_0\colon\frL_0\rightarrow \tilde \frL_0~,~~~\phi_1\colon\frL_{-1}\rightarrow \tilde \frL_{-1}~,~~~\phi_2\colon\frL_{0}\wedge \frL_{0}\rightarrow \tilde \frL_{-1}
    \end{equation}
    such that
    \begin{equation}\label{eq:Lie_2_algebra_morph}
        \begin{aligned}
            0 &= \phi_1(\mu_1(v_1)) - \tilde\mu_1(\phi_1(v_1))~,\\
            0 &= \phi_1(\mu_2(w_1,w_2)) - \tilde\mu_1(\phi_2(w_1,w_2))-\tilde\mu_2(\phi_1(w_1),\phi_1(w_2))~,\\
            0 &= \phi_1(\mu_2(w_1,v_1)) +\phi_2(\mu_1(v_1),w_1) - \tilde\mu_2(\phi_1(w_1),\phi_1(v_1))~,\\
            0 &= \phi_1(\mu_3(w_1,w_2,w_3)) -\phi_2(\mu_2(w_1,w_2),w_3) + \phi_2(\mu_2(w_1,w_3),w_2)\\
            &\phantom{{}={}} - \phi_2(\mu_2(w_2,w_3),w_1) - \tilde\mu_3(\phi_1(w_1),\phi_1(w_2),\phi_1(w_3)) \\
            &\phantom{{}={}} + \tilde\mu_2(\phi_1(w_1),\phi_2(w_2,w_3))- \tilde\mu_2(\phi_1(w_2),\phi_2(w_1,w_3))\\
            &\phantom{{}={}}+\tilde\mu_2(\phi_1(w_3),\phi_2(w_1,w_2))
        \end{aligned}
    \end{equation}
    for all $v_1\in \frL_{-1}$ and $w_{1,2,3}\in \frL_0$.
    
    \paragraph{Crossed modules of Lie algebras.} Applying the tangent functor to a crossed module of Lie groups $\caG=(\sfH\xrightarrow{~\sft~}\sfG,\acton)$, we obtain a crossed module of Lie algebras $\frG=(\frh\xrightarrow{~\sft~}\frg,\acton)$, where $\frg$ and $\frh$ are the Lie algebras of $\sfG$ and $\sfH$, respectively. Such a crossed module is equivalent to a strict 2-term $L_\infty$-algebra $\frL=\frL_{-1}\oplus \frL_0$ under the relation
    \begin{equation}
        \begin{gathered}
            \frg= \frL_0~,~~~\frh= \frL_{-1}~,
            \\
            [w_1,w_2]=\mu_2(w_1,w_2)~,~~~w\acton v=\mu_2(w,v)~,~~~[v_1,v_2]=\mu_2(\mu_1(v_1),v_2)~,
        \end{gathered}
    \end{equation}    
    cf.~also~\cite{Baez:2003aa}. Applying the tangent functor to a morphism of Lie 2-groups yields a morphism of strict 2-term $L_\infty$-algebras where $\phi_0$, $\phi_1$ and $\phi_2$ are the linearizations or differentials of $\Phi_0$, $\Phi_1$, and $\Phi_2$. The required properties of $\phi$ follow from those of $\Phi$.
    
    \subsection{Principal 2-bundles with adjusted connection}\label{app:higher_principal_bundles}
    
    \paragraph{Čech groupoid.} Consider a surjective submersion $\sigma\colon Y\rightarrow M$, which defines the \uline{Čech groupoid} 
    \begin{subequations}
        \begin{equation}\label{eq:def_Cech_groupoid}
            \check\scC(Y\rightarrow M)\ \coloneqq\ \left(\begin{tikzcd}
                Y^{[2]}\arrow[r,shift left] 
                \arrow[r,shift right] & Y
            \end{tikzcd}\right)~,~~~
            \begin{tikzcd}[column sep=2.0cm,row sep=large]
                y_1 & y_2\arrow[l,bend left,swap,out=-20,in=200]{}{(y_1,y_2)}~,
            \end{tikzcd}
        \end{equation}
        where $Y^{[2]}$ is the fibered product
        \begin{equation}
            Y^{[2]}=\{(y_1,y_2)\in Y\times Y~|~\sigma(y_1)=\sigma(y_2)\}~.
        \end{equation}
    \end{subequations}    
    This groupoid trivially extends to a higher Lie $n$-groupoid with trivial $k$-morphisms for $k\geq 2$. For most purposes, one may restrict $Y$ to be an ordinary cover given in terms of open subsets of $\IR^n$. In certain cases, it is more convenient to replace the Čech groupoid by a more general, higher groupoid giving rise to hypercovers, cf.~e.g.~\cite{Aldrovandi:0808.3627}.
    
    \paragraph{Cocycle description.} The cocycles of a higher principal bundle with higher structure group $\scG$ subordinate to the submersion $\sigma$ are then given by a higher functor from $\check\scC(Y\rightarrow M)$ to the one-object groupoid $\sfB\scG$ whose (higher) category of morphisms is $\scG$. These cocycles can be differentially refined to allow for a connection~\cite{Jurco:2014mva}.
    
    The introduction of a general connection\footnote{We simply use the term connection to refer to what is sometimes in the abelian gerbe literature called a connective structure and a curving.} requires a so-called \uline{adjustment}, cf.~\cite{Sati:2009ic,Saemann:2019dsl,Kim:2019owc} and in particular~\cite{Rist:2022hci} for details. For $\scG$ a 2-group given by a crossed module of Lie groups $\caG\coloneqq (\sfH\xrightarrow{~\sft~}\sfG,\acton)$ with corresponding Lie 2-algebra given by the crossed module of Lie algebras $\frG\coloneqq (\frh\xrightarrow{~\sft~}\frg,\acton)$, an adjustment is a map~\cite{Rist:2022hci}
    \begin{subequations}\label{eq:adjustmentCondition}
        \begin{equation}\label{eq:general_adjustment}
            \kappa:\sfG\times\frg\rightarrow\frh~,
        \end{equation}
        which is unital in $\sfG$, linear in $\frg$, and satisfies
        \begin{align}
            \kappa(\sft(h),X)&=h(X\acton h^{-1})~,
            \\
            \kappa(g_2g_1,X)&=g_2\acton\kappa(g_1,X)+\kappa\big(g_2,g_1Xg^{-1}_1-\sft(\kappa(g_1,X))\big)
        \end{align}
        for all $g_{1,2}\in\sfG$, $h\in\sfH$, and $X\in\frg$.
    \end{subequations}
    
    The differential cocycles for a principal $\scG$-bundle with connection then are given by the data~\cite{Rist:2022hci}
    \begin{subequations}\label{eq:adjusted_cocycles}
        \begin{equation}
            \begin{gathered}
                h\in C^\infty(Y^{[3]},\sfH)~,
                \\
                (g,\Lambda)\in C^\infty(Y^{[2]},\sfG)\oplus\Omega^1(Y^{[2]},\frh)~,
                \\
                (A,B)\in \Omega^1(Y,\frg)\oplus\Omega^2(Y,\frh)
            \end{gathered}
        \end{equation}
        such that\footnote{We note that our conventions are related to those in~\cite{Saemann:2012uq} by the map $B\mapsto -B$ and $H\mapsto -H$.}
        \begin{equation}
            \begin{aligned}
                h_{ikl}h_{ijk}&=h_{ijl}(g_{ij}\acton h_{jkl})~,
                \\
                g_{ik}&=\sft(h_{ijk})g_{ij}g_{jk}~,
                \\
                \Lambda_{ik}&=\Lambda_{jk}+g_{jk}^{-1}\acton\Lambda_{ij}-g_{ik}^{-1}\acton(h_{ijk}\nabla_i h_{ijk}^{-1})~,
                \\
                A_j&=g^{-1}_{ij}A_ig_{ij}+g^{-1}_{ij}\rmd g_{ij}-\sft(\Lambda_{ij})~,
                \\
                B_j&=g^{-1}_{ij}\acton B_i+\rmd\Lambda_{ij}+A_j\acton \Lambda_{ij}+\tfrac12[\Lambda_{ij},\Lambda_{ij}]-\kappa(g_{ij},F_i)
            \end{aligned}
        \end{equation}
        for all appropriate $(i,j,\ldots)\in Y^{[n]}$. The curvature of this principal 2-bundle is the sum of a 2-form $F$ and a 3-form $H$ and is given by
        \begin{equation}
            \begin{aligned}
                F&\coloneqq\rmd A+\tfrac12[A,A]+\sft(B)\in \Omega^2(Y,\frg)~,
                \\
                H&\coloneqq\rmd B+A\acton B-\kappa(A,F)\in \Omega^3(Y,\frh)~.
            \end{aligned}
        \end{equation}
    \end{subequations}
    Important examples of such adjusted connections are found in~\cite{Rist:2022hci} and (in infinitesimal form and a more general context) in~\cite{Saemann:2019dsl}. 
        
    In the first definition of non-abelian gerbes in~\cite{Breen:math0106083,Aschieri:2003mw}, the cocycle data was enlarged by a $\delta\in \Omega^2(Y^{[2]},\frh)$, which was not expected from the local description. The adjustment fixes $\delta$ in terms of the remaining data. The choice $\kappa=0$ contradicts the adjustment conditions~\eqref{eq:adjustmentCondition}. Imposing it anyway essentially reduces the cocycles to the fake flat ones satisfying $F=0$. 
    
    We note that the idea of modifying the $B$-field gauge transformations has appeared in other contexts. The first mention we are aware of is in the context of heterotic supergravity~\cite{Bergshoeff:1981um,Chapline:1982ww}. The connections introduced here are adjusted connections for the special case of a minimal model of the string Lie 2-algebra. In the context of higher gauge theory, a special class of adjusted connections was identified in~\cite{Sati:2008eg}, comprising those of heterotic supergravity.
    
    Furthermore, there is the work of~\cite{Ho:2012nt}, in which an additional datum is introduced that allows for a modification of the infinitesimal $B$-field gauge transformations to avoid the problem of fake flatness. This datum is somewhat related, but not identical, to that underlying the special adjustments of~\cite{Rist:2022hci}, and also the $B$-field deformation is related, but ultimately different. It is unclear to us how to describe the most general case of adjusted connections in the framework of~\cite{Ho:2012nt}. 
    
    In the context of the special case of certain higher gauge algebras constructed from Leibniz algebras, a $B$-field modification was proposed in~\cite{Strobl:2019mgf}. For a more general discussion of Leibniz algebras, see also~\cite{Borsten:2021ljb}. 
    
    The ambiguity in deforming the infinitesimal $B$-field gauge transformations as well as the curvature by terms proportional to the 2-form curvature $F$ was also identified in~\cite{Gruetzmann:2014ica} for a general semistrict Lie 2-algebra. Two particular examples of adjustments are then given.
    
    Finally, we note that a version of adjusted connections had been observed earlier in the context of gauge-matter theories with general gauge algebroids~\cite{Strobl:2004im}. The integrated version of these, together with a much more general framework for adjusted connections, was recently discussed in~\cite{Fischer:2024vak}.        
    
    \paragraph{Bundle isomorphisms.} Coboundaries are encoded in natural isomorphisms between two functors given in terms of the above cocycle data. They are encoded in maps
    \begin{subequations}\label{eq:adjusted_coboundaries}
        \begin{equation}
            \begin{gathered}
                b\in  C^\infty(Y^{[2]},\sfH)~,
                \\
                (a,\lambda)\in  C^\infty(Y,\sfG)\oplus\Omega^1(Y,\frh)~,
            \end{gathered}
        \end{equation}
        and two cocycles $(h,g,\Lambda,A,B)$ and $(\tilde h,\tilde g,\tilde \Lambda,\tilde A,\tilde B)$ are equivalent if
        \begin{equation}
            \begin{aligned}
                \tilde h_{ijk}&=a_i^{-1}\acton(b_{ik}h_{ijk}(g_{ij}\acton b_{jk}^{-1})b_{ij}^{-1})~,
                \\
                \tilde g_{ij}&=a_i^{-1}\sft(b_{ij})g_{ij}a_j~,
                \\
                \tilde \Lambda_{ij}&=a^{-1}_j\acton\Lambda_{ij}+\lambda_j-\tilde{g}^{-1}_{ij}\acton\lambda_i+(a_j^{-1}g_{ij}^{-1})\acton(b_{ij}^{-1}\nabla_ib_{ij})~,                    
                \\
                \tilde{A}_i&=a_i^{-1}A_ia_i+a_i^{-1}\rmd a_i-\sft(\lambda_i)~,
                \\
                \tilde{B}_i&=a_i^{-1}\acton B_i+\rmd\lambda_i+\tilde{A}_i\acton\lambda_i+\tfrac12[\lambda_i,\lambda_i]-\kappa(a_i,F_i)~.
            \end{aligned}
        \end{equation}
    \end{subequations}
    
    \paragraph{Postcomposition with 2-group morphisms.} Consider two crossed modules $\caG=(\sfH\xrightarrow{~\sft~}\sfG,\acton)$ and $\tilde\caG=(\tilde\sfH\xrightarrow{~\tilde\sft~}\tilde\sfG,\tilde\acton)$ as well as a principal $\caG$-bundle $\scP$ with cocycles $(g,h)$. Then a 2-group morphism $\Phi\colon\underline{\caG}\rightarrow \underline{\tilde \caG}$ yields a principal $\tilde \caG$-bundle with cocycles given by 
    \begin{subequations}\label{eq:cocycles_under_morphisms}
        \begin{equation}
            \tilde g_{ij}=\Phi_1(g_{ij})\eand \tilde h_{ijk}=\Phi^{\tilde \sfH}_1(h_{ijk},m_{ij},m_{jk})\Phi^{\tilde \sfH}_2(g_{ij},g_{jk})~,
        \end{equation}
        where $\Phi^{\tilde \sfH}_1(h_{ijk})$ is the component of $\Phi_1$ in $\tilde \sfH$. This follows abstractly from the interpretation of $\scP$ and $\Phi$ as weak 2-functors and the fact that 2-functors compose nicely, cf.~\eqref{eq:composition_2_functors}.
        
        For the connection part, we can generalize the discussion in~\cite{Rist:2022hci} to apply the $L_\infty$-algebra morphism induced by the morphism $\Phi$ to the local connection forms:
        \begin{equation}
            \begin{aligned}
                A_i\mapsto \tilde A_i=\phi_0(A)~,~~~B_i\mapsto \tilde B_i=\phi_1(B)+\tfrac12\phi_2(A,A)~.    
            \end{aligned}
        \end{equation}
    \end{subequations}    This construction is familiar from homotopy Maurer--Cartan theory, cf.~e.g.~\cite{Jurco:2018sby} for more details.
    
    Interestingly, as observed in~\cite{Rist:2022hci}, this fully defines $\sft(\Lambda_{ij})$ via the cocycle condition~\eqref{eq:adjusted_cocycles} relating $A_i$ to $A_j$. It remains to lift this to a full map $\Lambda_{ij}\colon\Omega^1(Y^{[2]},\sfH)$, which is best done on a case-by-case basis.
    
    Note that, in the case of morphisms $\Phi\colon\caG\rightarrow \tilde \caG$ that are given by butterflies, the induced morphism of principal 2-bundles is more complicated and may require a refinement of the cover, cf.~\cite[\S~4.2]{Aldrovandi:0909.3350}. One such example is discussed in detail in~\cite{Rist:2022hci}. 
    
    \subsection{Quasi-groupoids and augmentation}\label{app:quasi-groupoids}
    
    In the following, we briefly summarize some basic material on quasi-groupoids; for a detailed review in our conventions, see~\cite{Jurco:2016qwv} and references therein.
    
    \paragraph{Simplicial manifolds.} Recall that a \uline{simplicial object} in a category $\scC$ is a $\scC$-valued presheaf $X\colon\triangle^{\operatorname{op}}\to\scC$ on the simplex category $\triangle$, which is (the skeleton of) the 1-category of finite non-empty well-ordered sets and order-preserving functions. Every finite well-ordered set is isomorphic to the ordinal $\underline n=\{0,1,\dotsc,n-1\}$, and the image of $\underline n$ is the set of $(n-1)$-simplices: $X_n\coloneqq X(\underline{n+1})$. The images under $X$ of injective order-preserving maps $\underline n\to\underline{n+1}$ give the \uline{face maps} $\sff^n_0,\dotsc,\sff^n_n\colon X_n\to X_{n-1}$, and the images under $X$ of surjective order-preserving maps $\underline{n+2}\to\underline{n+1}$ give the \uline{degeneracy maps} $\sfd^n_0,\dotsc,\sfd^n_n\colon X_n\to X_{n+1}$.
    
    A \uline{simplicial set} is then simply a simplicial object in $\sfSet$, and a \uline{simplicial manifold} is a simplicial object in a suitable category of smooth manifolds. Notice that every simplicial set can be trivially regarded as a discrete simplicial manifold. The \uline{standard simplicial $n$-simplex} $\Delta^n$ is the simplicial set $\triangle^\text{op}\rightarrow \sfSet$ represented by $\underline{n+1}$.
    
    \paragraph{Lie quasi-groupoids.} An \uline{$(n,i)$-horn} in $\Delta^n$ is the union of all faces of $\Delta^n$ except for the $i$th one. The $(n,i)$-horns of a simplicial manifold $\scM$ are the images of the $(n,i)$-horns of $\Delta^n$ in $\scM$. If every horn in $\scM$ can be filled to a simplex, and the face maps from the simplices in $\scM$ to the horns in $\scM$ are surjective submersions, we say that $\scM$ is a Kan simplicial manifold or a \uline{Lie quasi-groupoid}. If all the fillers for $(m,i)$-horns are unique for every $m>n$ and $0<i<m$, then we say that $\scM$ is a \uline{Lie $n$-quasi-groupoid}.
    
    \paragraph{From (higher) groupoids to quasi-groupoids.} The nerve of any groupoid $\scG$, i.e.~the simplicial manifold whose 0-simplices are the objects of $\scG$, whose 1-simplices are the morphisms of $\scG$, and whose 2-simplices are the pairs of composable morphisms of $\scG$, etc., is a 1-quasi-groupoid. In the case of 2-groupoids, the definition of a nerve is slightly more involved but similarly feasible. A common choice is the \uline{Duskin nerve}~\cite{Duskin02simplicialmatrices}, which also forms a 2-quasi-groupoid. A problem in this treatment of 2-groupoids is that the explicit expression of horizontal composition is replaced by the existence of a set of horn fillers. This, together with a vast redundancy of information in quasi-groupoids, is the reason for us not working with quasi-groupoids from the outset.
    
    \paragraph{Augmentation.} In the definition of simplicial objects, we have restricted ourselves to non-empty well-ordered sets, but we can naturally extend this to all finite well-ordered sets, including $\underline{-1}=\varnothing$; we denote the resulting category by $\triangle_+$. An \uline{augmented simplicial object} in $\scC$ is then a $\scC$-valued presheaf on $\triangle_+$. Concretely, an augmented simplicial object $X^+_\bullet$ is equivalently a triple $(X_\bullet,X_{-1},\sff^{-1}_0)$ where $X_\bullet$ is an (unaugmented) simplicial object, $X_{-1}$ is an object in $\scC$, and $\sff^0_0\colon X_0\to X_{-1}$ is a morphism in $\scC$ such that
    \begin{equation}
        \sff^0_0\circ \sff^1_0=\sff^0_0\circ \sff^1_1~.
    \end{equation}
    We define an \uline{augmented quasi-groupoid} in a corresponding way. A particularly useful example is that of the augmented Čech groupoid~\eqref{eq:augmented_Cech_nerve}.
    
    \bibliographystyle{latexeu}
    
    \bibliography{bigone}
    
\end{document}